\documentclass{article}  

\usepackage{graphicx}

\usepackage{geometry}
\usepackage{graphicx}
\usepackage{natbib}
\usepackage{color}
\usepackage{amssymb}
\usepackage{lineno}
\usepackage{txfonts}
\usepackage{amstext}
\usepackage{mathabx}
\usepackage{multirow}
\usepackage{ulem}
\usepackage{lscape}
\usepackage{txfonts}
\usepackage{subfigure}
\usepackage{float}
\usepackage{ctable} 
\usepackage[bottom]{footmisc}
\usepackage{caption}  
\usepackage[colorlinks=true, linkcolor=blue, citecolor=blue, urlcolor=blue]{hyperref}

\usepackage{etoolbox}
\makeatletter
\patchcmd{\@afterheading}{\@nobreaktrue}{}{}{}
\makeatother

\newcommand{\jgr}{J. Geophys. Res. }

\newcommand{\icarus}{Icarus }
\newcommand{\aap}{Astron. Astrophys. }
\newcommand{\apj}{Astrophys. J. }

\newcommand{\planss}{Planet. Space Sci. }

\newcommand{\ssr}{Space Sci. Rev. }

\newcommand{\nat}{Nature }
\newcommand{\science}{Science }

\newcommand{\pasp}{Publ. Astron. Soc. Pac. }

\newcommand{\natastron}{Nat. Astron. }
\newcommand{\natcommun}{Nat. Commun.}

\newcommand{\psj}{Plan. Sci. J. }

\newcommand{\arcsec}{\hbox{$^{\prime\prime}$}}

\newcommand{\tab}[1]{Table~\ref{#1}}

\newcommand{\degre}{\ensuremath{^\circ}}

\begin{document} 


\begin{center}
\textbf{\LARGE Observations of the temporal evolution of Saturn's stratosphere following the Great Storm of 2010-2011}\\
\vspace{0.4cm}
\textbf{\Large I. Temporal evolution of the water abundance in Saturn's hot vortex of 2011-2013}
\end{center}
\vspace{0.4cm}
\normalsize

\large\noindent C. Lefour$^{1}$, T. Cavali\'e$^{1,2}$, H. Feuchtgruber$^{3}$, R. Moreno$^{2}$, L. N. Fletcher$^{4}$, T. Fouchet$^{2}$, E. Lellouch$^{2}$, E. Barth$^{5}$, P. Hartogh$^{6}$\normalsize\\
\vspace{0.2cm}

\noindent$^1$Laboratoire d'Astrophysique de Bordeaux, Univ. Bordeaux, CNRS, B18N, all\'ee Geoffroy Saint-Hilaire, 33615 Pessac, France (ORCID: 0009-0004-6922-9388, email: camille.lefour@u-bordeaux.fr)\\
$^2$LIRA, Observatoire de Paris, Université PSL, CNRS, Sorbonne Universit\'e, Universit\'e Paris Cité, 5 place Jules Janssen, 92195 Meudon, France\\
$^3$Max Planck Institute for Extraterrestrial Physics, Garching, Germany\\
$^4$School of Physics and Astronomy, University of Leicester, Leicester, UK\\
$^5$Southwest Research Institute, Boulder, CO 80302, USA\\
$^6$Max-Planck-Institute for Solar System Research, G\"ottingen, Germany\\

\vspace{0.2cm}
\noindent\textbf{Received:} 15 November 2024\\
\noindent\textbf{Accepted:} 14 April 2025\\
\noindent\textbf{Published:} ??????\\
\vspace{0.2cm}

\noindent\textbf{DOI:} ??????\\
\vspace{0.5cm}

\section*{Abstract}
Water vapour is delivered to Saturn's stratosphere by Enceladus' plumes and subsequent diffusion in the planet system. It is expected to condense into a haze in the middle stratosphere. The hot stratospheric vortex (the `beacon') that formed as an aftermath of Saturn's Great Storm of 2010 significantly altered the temperature, composition, and circulation in Saturn's northern stratosphere. Previous photochemical models suggested haze sublimation and vertical winds as processes likely to increase the water vapour column density in the beacon. We aim to quantify the temporal evolution of stratospheric water vapour in the beacon during the storm. We mapped Saturn at 66.44 and 67.09\,$\mu$m on seven occasions from July 2011 to February 2013 with the PACS instrument of the \textit{Herschel} Space Observatory\footnote{\textit{Herschel} is an ESA space observatory with science instruments provided by European-led Principal Investigator consortia and with important participation from NASA.}. The observations probe the millibar levels, at which the water condensation region was altered by the warmer temperatures in the beacon. Using radiative transfer modelling, we tested several empirical and physically based models to constrain the cause of the enhanced water emission found in the beacon. The observations show an increased emission in the beacon that cannot be reproduced only accounting for the warmer temperatures reported in the beacon. An additional source of water vapour is thus needed. We find a factor (7.5$\pm$1.6) increase in the water column in the beacon compared to pre-storm conditions using empirical models. Combining our results with a cloud formation model for July 2011, we evaluate the sublimation contribution to 45-85\% of the extra column derived from the water emission increase in the beacon. The observations confirm that the storm conditions enhanced the water abundance at the millibar levels because of haze sublimation and vertical winds in the beacon. Future work on the haze temporal evolution during the storm will help to better constrain the sublimation contribution over time.

\section{Introduction}

The unexpected detection of water vapour and carbon dioxide in the stratospheres of the giant planets and Titan \citep{Feuchtgruber1997,Coustenis1998,Lellouch2002,Burgdorf2006} was the starting point for identifying the external sources of oxygen to these planets. Indeed, H$_2$O (and CO$_2$ in Uranus and Neptune) cannot be supplied to the upper layers of these atmospheres from their oxygen-rich interiors \citep{Li2022,Cavalie2024a,Venot2020} because condensation around the tropopause acts as a transport barrier. Several types of sources have been proposed, including ablation of interplanetary dust particles (IDPs, \citealt{Prather1978}), material expelled from their icy moons and rings \citep{Strobel1979}, and large comet impacts \citep{Lellouch1995}. In the past two decades, observations with \textit{Herschel}, \textit{Cassini}, ALMA, IRAM, etc., have resulted in constraints on most of the dominant sources of oxygen in the giant planets and Titan: comet Shoemaker-Levy 9 in Jupiter \citep{Lellouch2002,Bezard2002,Moses2000b,Moses2005,Cavalie2013}, IDP for H$_2$O in Uranus and Neptune \citep{Moses2017,Teanby2022}, and an ancient comet for CO in Neptune \citep{Lellouch2005,Moreno2017} and possibly in Uranus too \citep{Cavalie2014}. At Saturn, while CO may come from an ancient comet impact \citep{Cavalie2009,Cavalie2010}, \citet{Cavalie2019} showed using spatially resolved H$_2$O maps obtained with \textit{Herschel} that stratospheric H$_2$O is sourced from the geysers of Enceladus \citep{Waite2006,Porco2006,Hansen2006}, the subsequently formed H$_2$O torus \citep{Hartogh2011a} which is transported towards the rings and the planet atmosphere \citep{Cassidy2010,Waite2018,Hsu2018,Mitchell2018}. The Enceladus source is also compatible with the H$_2$O observations at Titan \citep{Moreno2012,Lara2014,Dobrijevic2014}. The meridional distribution of the column density of stratospheric H$_2$O in Saturn, observed with \textit{Herschel}, can be represented by a Gaussian centred at the equator \citep{Cavalie2019}. After entering the atmosphere of the planet, water vapour is mixed to lower altitudes (higher pressures) until it reaches the millibar level in the middle stratosphere, where it condenses into ice particles and is expected to form a stratospheric water haze as has been predicted by models \citep{Moses2000b,Ollivier2000}.

An unexpected planetary-scale storm appeared in December 2010 in the atmosphere of Saturn at around 35\degre N \citep{Sanchez-Lavega2011,Fischer2011,Fletcher2011}. Such storms, referred to as Great White Spots, had been observed six times between 1876 and 1990, with a periodicity of about a Saturnian year \citep{Sanchez-Lavega2018}. The 2010 storm quickly spread longitudinally to form the biggest storm witnessed in Saturn's atmosphere in decades. \citet{Fletcher2011} showed that it significantly disrupted the slowly evolving seasonal cycle in the stratosphere between 20\degre N to 50\degre N in temperatures, winds, and composition, within 45 days after the start of the disturbance. The storm also had indirect consequences for Saturn's equatorial stratospheric oscillation, disrupting the pattern for a number of years \citep{Fletcher2017}. The original anticyclonic oval at $\sim$0.1\,bar was caused by the adiabatic cooling of an upwelling plume from the deep troposphere. The central tropospheric cool vortex spread over $\sim$10\degre~in latitude. Above this tropospheric vortex, on each side, subsidence of air parcels in the stratosphere at 1\,mbar (Saturn's stratosphere spans from 0.001 to 100\,mbar approximately) is supposed to be the cause of a dramatic increase in the infrared emission \citep{Fletcher2011}. Initially, a 16\,K difference was reported between these two warm stratospheric regions, referred to as `beacons', and the cool central tropospheric vortex. Data taken in May 2011 by \textit{Cassini}/CIRS showed that the two beacons had merged into a single hot spot. The temperature at 1\,mbar had reached as high as 220\,K \citep{Fletcher2012}, resulting in a region warmer than Jupiter's stratosphere. The beacon persisted for many months, drifting in longitude at 35\degre N and slowly cooling with time until it completely dissipated in 2014.

This unique storm and its stratospheric counterpart have not only disrupted the temperature field, but also the hydrocarbon chemistry inside the beacon \citep{Fletcher2012,Hesman2012,Cavalie2015,Moses2015}. \citet{Moses2015} especially demonstrated that vertical downwelling winds in the beacon could explain the local increase in C$_2$H$_2$ and C$_2$H$_6$ abundances in the beacon observed with \textit{Cassini}/CIRS. In addition, they anticipated changes in the vertical profile of water vapour, because the intense warming inside the beacon in the altitude range where stratospheric water usually condenses enabled its presence in the vapour phase. 

In this paper, we present \textit{Herschel} observations of water vapour in the stratosphere of Saturn in the two years that followed the outbreak of this storm with the aim of quantifying any changes in the water vapour vertical and horizontal distribution in the beacon. We detail the observations in Section \ref{part:Observations} and the radiative transfer modelling in Section \ref{part:Models}. The results are presented in Section \ref{part:Results}. We discuss the results and give our conclusions in Section \ref{part:Discussion}.

\section{Observations \label{Observations}}
\label{part:Observations}

 \subsection{Observation strategy}
 \label{part:Strategy}
   We used observations carried out with \textit{Herschel} \citep{Pilbratt2010}. The data were taken in the framework of the \textit{Herschel} Solar System Observations key programme (HssO, \citealt{Hartogh2009}) and from the Open Time programme OT2\_tcavalie\_7. While the HssO observations resulted in the first detection of the beacon with the observatory on July 12 2011 (i.e. after the merger event took place in May 2011, when the beacon was at its brightest and hottest), the subsequent OT2\_tcavalie\_7 programme was designed to monitor the water emission in the beacon until the end of the lifetime of \textit{Herschel} (i.e. May 2013). The observation constraints of \textit{Herschel} resulting from its L2-orbit and orientation with respect to the Sun offered the possibility to observe Solar System planets only beyond the Earth orbit and when in quadrature. As a result, the observations covered the evolution of the water emission in the beacon in four time windows: July 2011 (HssO programme), and February 2012, July-August 2012, and January-February 2013 (OT2\_tcavalie\_7 programme). By extrapolating from \textit{Cassini} and Very Large Telescope (VLT) observations of the beacon from \citet{Fletcher2012} and accounting for the beacon drift rate measured in the first months of the existence of the beacon of ($2.70\pm0.04$)\degre.day$^{-1}$ \citep{Fletcher2012}, the observation starting times were set such that the beacon would be close to the eastern planetary limb. Observing the beacon at the limb allowed us to optimize the signal-to-noise ratio (S/N), taking advantage of limb darkening in the continuum and limb brightening in the line.
 
   These data, all taken with the 5$\times$5 spatial pixels ($\sim$50\arcsec $\times$ 50\arcsec \, on sky) of the Photodetector Array Camera and Spectrometer (PACS, \citealt{Poglitsch2010}) set in line spectroscopy mode, consist in 3$\times$3 raster maps (i.e. 9 positions) with 3\arcsec~separation between each of the 9 raster positions, at 66.44 and 67.09\,$\mu$m, two spectral lines of H$_2$O. Each of the seven H$_2$O maps thus contains 225 spectra at 66.44 or 67.09\,$\mu$m, as we have a matrix of 25 pixels at each of the 9 raster map positions, and each pixel includes a full spectrum. The combination of pixel array and raster mapping results in oversampling the planet, given the limited 9.4\arcsec~spatial resolution at these wavelengths. The spectral resolving power is $\frac{\nu}{\Delta\nu}\sim$2500-3000. It results in a spectral resolution $\Delta\nu$ a factor $\sim$100 greater than the natural line width. The observation of Saturn, a bright continuum source at these wavelengths, required us to use an engineering mode in which the readout time of the spectrometer electronics was set to its minimal value (1/32 of a second) to avoid detector saturation. Additional details, like dates, Observation Identification numbers, and integration times, are given in \tab{tab:Obs_list}.

 \begin{table*}
 \centering
       \begin{tabular}{llllll}
       \specialrule{.2em}{.1em}{.1em} 
       \multicolumn{6}{c}{\textit{Window 1: July 2011}} \\
       \hline 
        Date              & OD    & DOY2011 & Obs. ID            & $\Delta t$ [s] & $\lambda$ \\
       \hline                    
       2011-07-12 12:40:53  & 789  & 192 & 1342224015 & 5347 & 67.09\,$\mu$m \\
       \specialrule{.2em}{.1em}{.1em} 
       \multicolumn{6}{c}{\textit{Window 2: February 2012}} \\
       \hline   
       Date              & OD    & DOY2011 & Obs. ID            & $\Delta t$ [s] & $\lambda$ \\
       \hline   
       2012-02-02 00:05:21  & 994  & 397 & 1342238584 & 3259 & 66.44\,$\mu$m  \\
       2012-02-02 21:33:59  & 995  & 398 & 1342238616 & 3259 & 67.09\,$\mu$m  \\
       \specialrule{.2em}{.1em}{.1em} 
       \multicolumn{6}{c}{\textit{Window 3: July-August 2012}} \\
       \hline   
       Date              & OD    & DOY2011 & Obs. ID            & $\Delta t$ [s] & $\lambda$ \\
       \hline   
       2012-07-29 16:58:22  & 1173 & 576 & 1342248742 & 3259 & 66.44\,$\mu$m    \\
       2012-08-10 17:50:26  & 1185 & 588 & 1342249394 & 3259 & 67.09\,$\mu$m    \\
       \specialrule{.2em}{.1em}{.1em} 
       \multicolumn{6}{c}{\textit{Window 4: January-February 2013}} \\
       \hline   
       Date              & OD    & DOY2011 & Obs. ID            & $\Delta t$ [s] & $\lambda$ \\
       \hline   
       2013-01-30 03:19:45  & 1357 & 760 &  1342262558  & 3259 & 66.44\,$\mu$m  \\
       2013-02-02 06:10:01  & 1360 & 763 &  1342262796  & 3259 & 67.09\,$\mu$m   \\
       \specialrule{.2em}{.1em}{.1em} 
       \end{tabular}
        \caption{Summary of the \textit{Herschel}-PACS observations of Saturn. The given time corresponds to the start of the observation. OD means operational day, DOY2011 is the day number since 2011-01-01, $\Delta t$ is the total integration time and $\lambda$ the water line central wavelength.}
        \label{tab:Obs_list}
   \end{table*}

 \subsection{Data reduction}
 \label{part:Data_reduction}
 
\begin{figure}[!ht]
\centering
\includegraphics[width=0.7\textwidth]{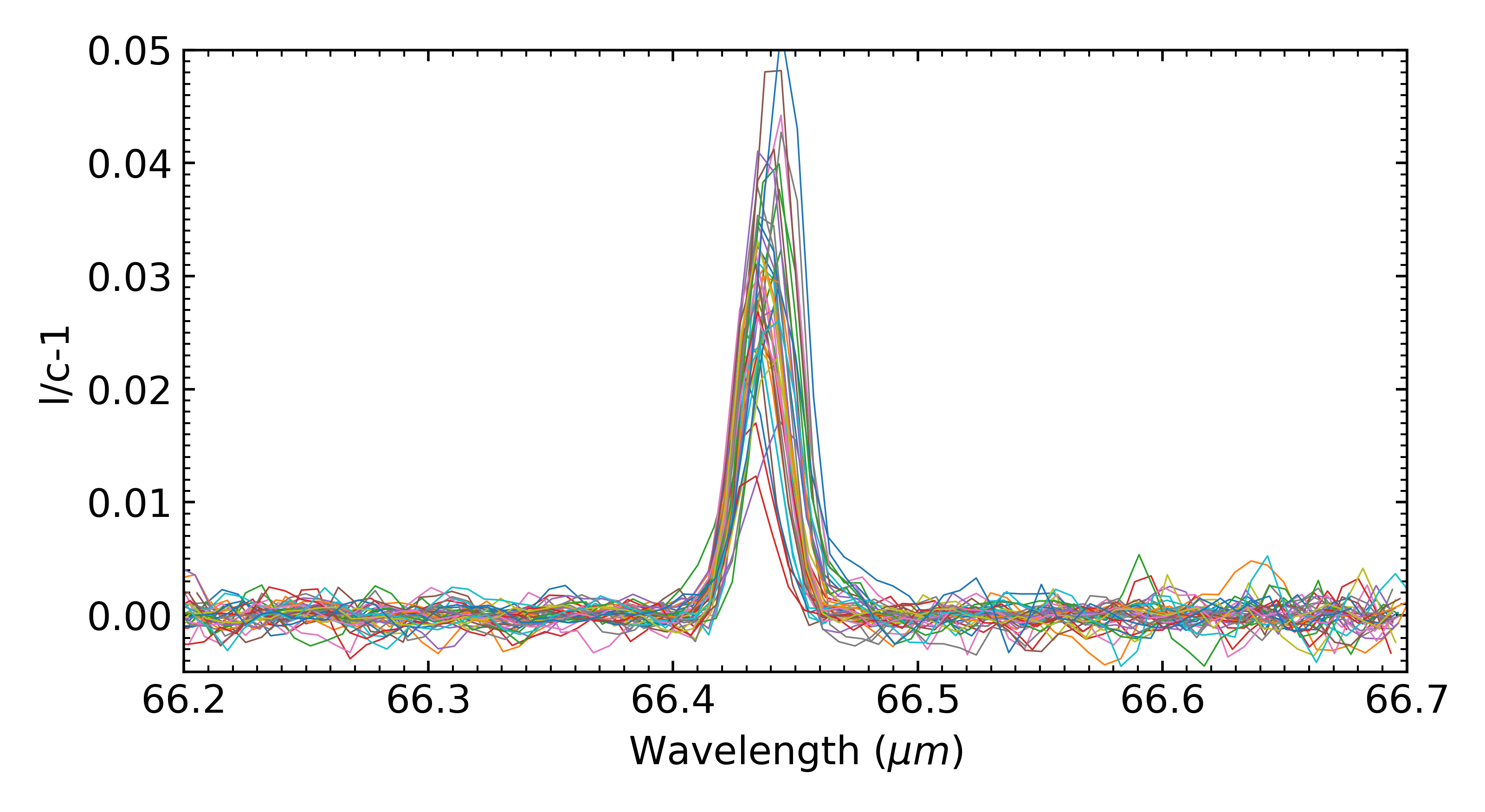}
\includegraphics[width=0.7\textwidth]{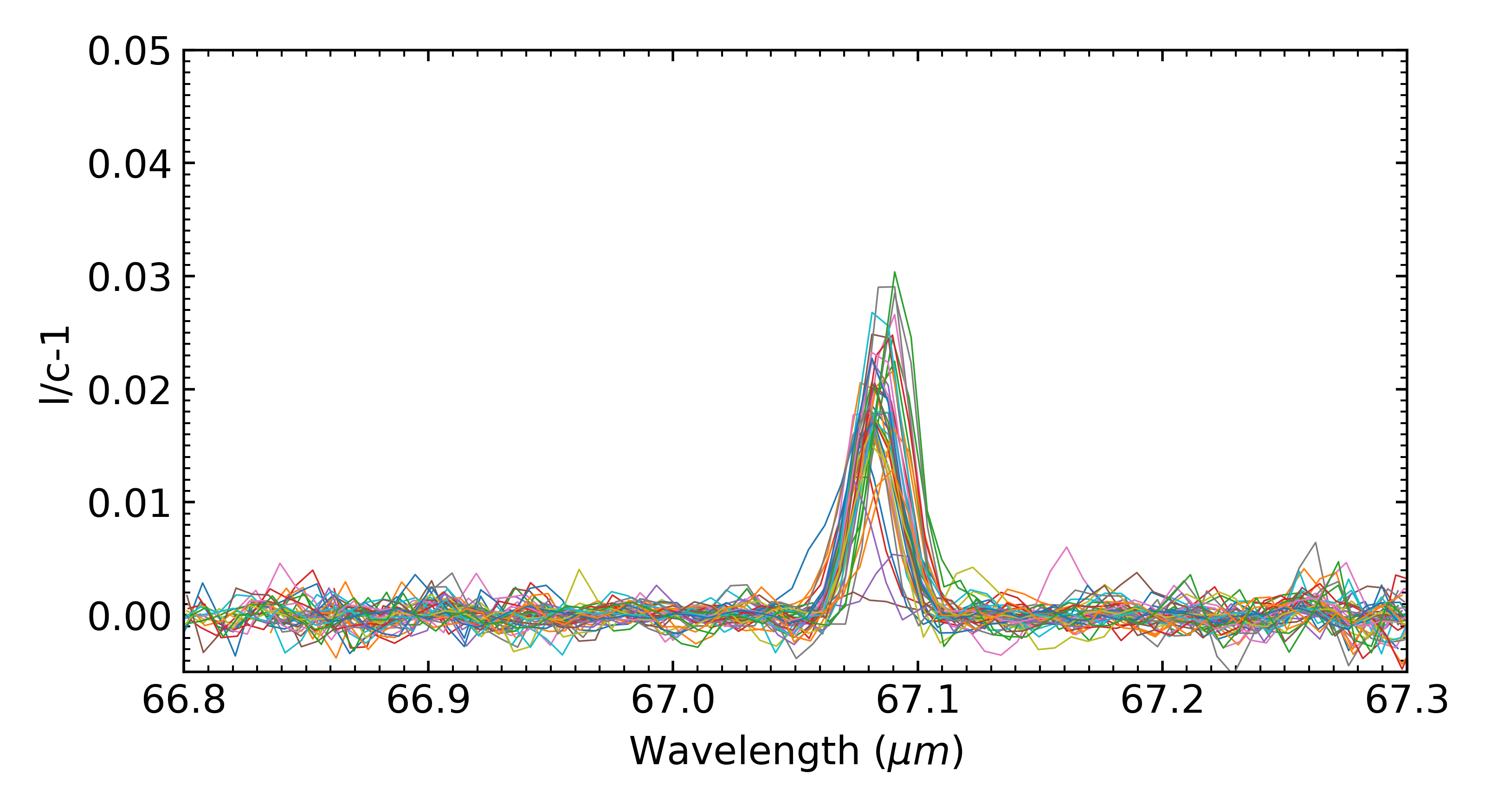}
\caption{Observed spectra after baseline removal for the second observation window (February 2012) at 66.44 $\mu$m (\textit{top}) and 67.09 $\mu$m (\textit{bottom}). The spectra are presented as line-to-continuum ratio minus one. The scatter in the peak position of the line is mainly due to Doppler shift induced by Saturn's rapid rotation. For both observations, all the plotted spectra correspond to the pointings selected within the planetary disc, as shown with the black dots on Fig. \ref{fig:Line_area_maps} second row (i.e. second observation window).}
\label{fig:Examples_lines}
\end{figure}

   We reduced the PACS data similarly to those obtained in January 2011 and presented in \citet{Cavalie2019}. After initial reduction in HIPE 8.0 (\textit{Herschel} Interactive Processing Environment, \citealt{Ott2010}) and additional processing (flat-fielding, outlier removal, and rebinning), we calibrated the central position of the seven maps using the continuum data. The planet surface in the continuum data is centred using the ephemerides extracted from JPL/Horizons for each of our seven observations. An offset of the order of 2\arcsec \, was applied in the RA-dec plane to centre the data. A pointing uncertainty of the order of 1/5 of the beam size (approximately 2\arcsec) is expected from the telescope pointing precision. We only selected at this stage spectra that show detections of the water emission; that is, the spectra within the planetary disc or in the direct vicinity of the atmospheric limb. Since the line at 66.44 $\mu$m is stronger than the line at 67.09 $\mu$m (by about 30\%), the S/N is greater at 66.44 $\mu$m. We then subtracted instrumental baselines caused by the strong continuum emission of the planet using Lomb periodograms or polynomial fits. The total uncertainty on line amplitude from the baseline removal is 10-15\% at 66.44 $\mu$m and 15-20\% at 67.09 $\mu$m. Examples of processed spectra from the second observation window are presented in Fig. \ref{fig:Examples_lines} for both wavelengths. A shift in the position of the line peak is observed and is due to two effects: 1) because of Saturn's rapid rotation, the lines are blue-shifted at the western limb and red-shifted at the eastern limb, and 2) \citet{Poglitsch2010} demonstrated that a line shift is induced at some raster positions because of the non-uniform illumination of the instrument slit. 

 As in other giant planet observations using PACS \citep{Cavalie2013,Cavalie2019}, reasonable uncertainties cannot be estimated on an absolute flux calibration and we instead express the spectra in terms of the line-to-contiuuum ratio. Because the lines are spectrally unresolved and reflect the Gaussian response of the instrument, all the spectral information is contained in the line area. We consequently analysed the water maps in terms of line area for those pixels on the planetary disc or in the direct vicinity of the limb. We performed a Gaussian fitting because of the Gaussian shape of the spectral resolution and then computed the line area using the parameters given by the fit. This step produces an additional uncertainty from the noise level of a few percent. The overall uncertainty on the line area is between 10 to 20\%, or a 1$\sigma$ root-mean-square (\textit{rms}) value of about 0.01 (in units of $\mu$m $\times$ \% continuum).

 \subsection{Observation results}
 \label{part:Observation_results}
 
\begin{figure*}
\centering
\includegraphics[width=0.99\textwidth]{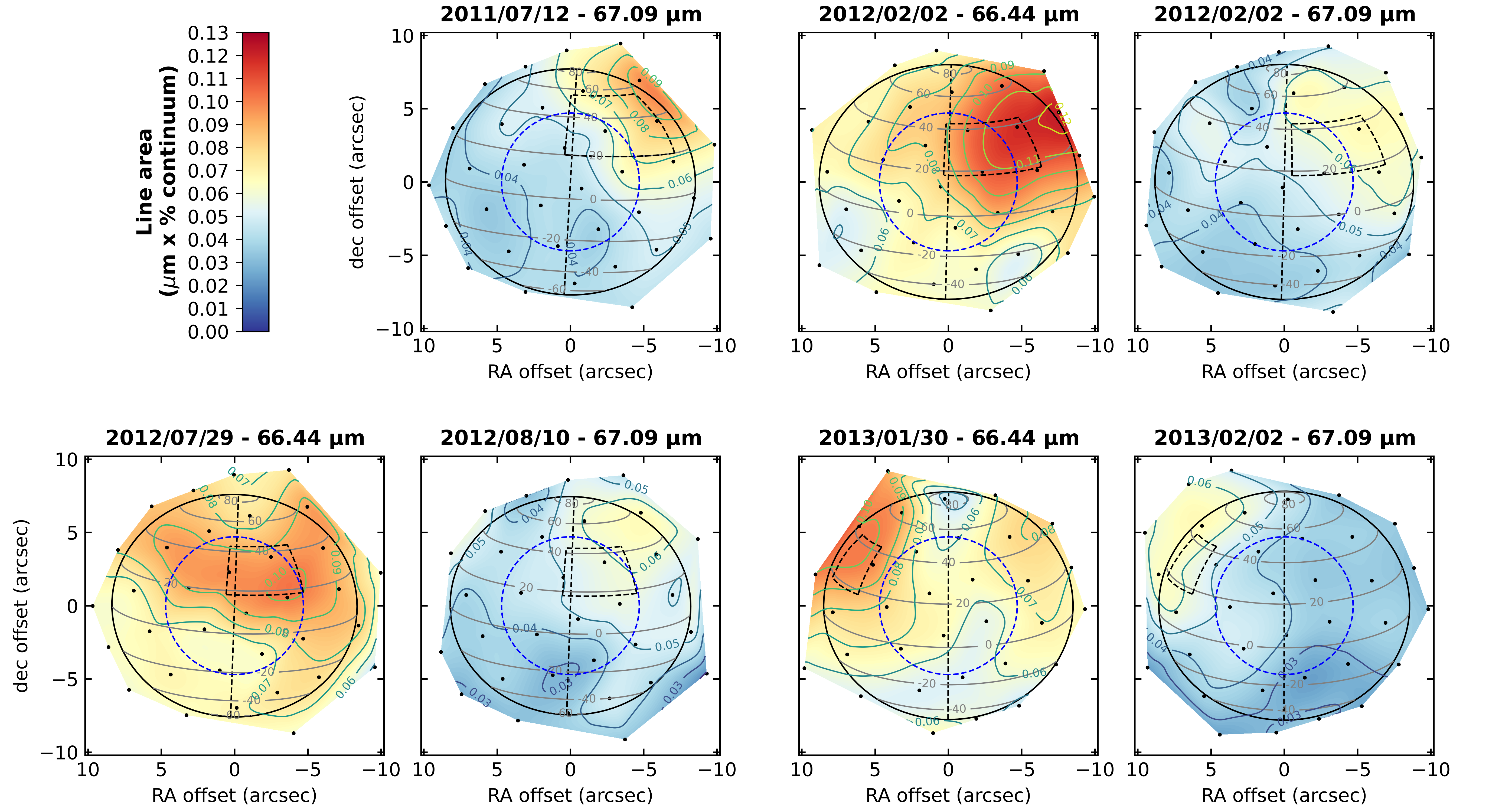}
\caption{Water line area maps expressed in units of $\mu$m$\times$\% of continuum. The four observation time windows are July 2011 at 67.09 $\mu$m, February 2012 at 66.44 and 67.09 $\mu$m, July-August 2012 at 66.44 and 67.09 $\mu$m, and January-February 2013 at 66.44 and 67.09 $\mu$m. Saturn's 1-bar level is represented by the black ellipse, with the north pole facing upwards. A selection of iso-latitudes are plotted with grey lines. The rotation axis corresponds to the thin dashed black line. The beam spatial extent is illustrated by the circle in a dashed blue line. The black dots correspond to the central position of the spatial pixels. The position and extension of the beacon (see Table \ref{tab:Beacon_data}) at mid-observation time is represented with the thick dashed black lines. A cubic interpolation was applied to the data.}
\label{fig:Line_area_maps}
\end{figure*}

   The final water line area maps are presented in Fig. \ref{fig:Line_area_maps} for our four observation windows. Water emission is detected all across the planet on each map, at 66.44 and 67.09 $\mu$m. Water emission is strongly enhanced in the northern hemisphere at the eastern limb in July 2011 and in February 2012. However, it is not the case for the last two windows. In July-August 2012, the maximum of line emission is located in-between the eastern limb and the central meridian, and it is surprisingly on the western limb in January-February 2013. This prompted us to revise a posteriori the drift rate of the beacon determined by \citet{Fletcher2012}. This is detailed in Section \ref{part:Beacon_drift}.
 
   Fig. \ref{fig:Line_area_maps} shows the position of the beacon at mid-integration time after revising its drift rate (see the areas delimited by the thick dashed black lines). It is noteworthy that the strongest water emission is not located at the central position of the beacon, most importantly because the emission is enhanced at the atmospheric limb compared to nadir geometry for geometrical reasons. The emission is also spread out in longitude during the integration time (1.5 hours for the first window, i.e. a rotation of 51\degre, and 54 minutes for the other windows, i.e. 31\degre). Finally, the beam size covering half of the planet dilutes the emission.

   The 66.44 $\mu$m and 67.09 $\mu$m lines have a line of sight beam-convolved opacity of about 6 and 2, respectively, at the line centre. They are thus sensitive both to temperature and water abundance. The temperature was monitored by \textit{Cassini}/CIRS over the whole period covered by our observations, and we present the adopted thermal fields in Section \ref{part:Temperature}. The water abundance, especially in the beacon, can therefore be derived from the \textit{Herschel} observations. We explore different scenarios with various water abundance fields in Section \ref{part:Composition} to explain the water emission in the beacon region. During the storm, the exceptional increase in temperature in the beacon resulted in a shift of the usual condensation level at $\sim$2 mbar (at 35\degre N) to deeper pressure levels, up to 12 mbar in July 2011, corresponding to a shift of about 100 km in altitude.
 
In Section \ref{part:Step_1}, we first evaluate whether the observed maps can be reproduced solely by the measured increase in temperature while maintaining the nominal water distribution from \citet{Cavalie2019}. In Section \ref{part:Step_2}, we test two profiles derived from the photochemical model of \citet{Moses2015}: one accounting only for haze sublimation and the other incorporating both sublimation and downward winds. \citet{Moses2015} showed that these winds are essential to explain the localized increase in C$_2$H$_2$ and C$_2$H$_6$ abundances at the millibar level, but additionally showed that the sublimation of stratospheric water haze is expected due to the shift of the condensation line within the beacon. Finally, we adopt a simpler empirical approach in which the water vapour mole fraction in the beacon is treated as a free parameter, allowing us to determine the required water column density to match the observed water line maps. This third scenario is developed in Section \ref{part:Step_3}.

\section{Modelling}
\label{part:Models}

We modelled the spectroscopic observations with the line-by-line radiative transfer model described in \citet{Cavalie2019}, which accounts for the 3D ellipsoidal geometry of the planet and the emission/absorption of the rings. Saturn's rotation (9.87 km.s$^{-1}$ at the equator) induces longitudinal smoothing of the beacon emission over 51\degre and 31\degre, respectively, during the 1.5 hour integration time for the first window and 54 minutes for the other windows. It is taken into account by averaging the radiative transfer results at the start, mid and end observation time. Several inputs are necessary in the model: the 3D temperature and background composition fields, the geometry and orientation of the planet with respect to \textit{Herschel}, the position of the beacon at the time of the observations, and the water spatial distribution.

 \subsection{Temperature fields}
 \label{part:Temperature}

\subsubsection{Background temperature field}
\label{part:Background_temperature}

For the background thermal field (i.e. without the beacon), we used the altitude-latitude seasonal field derived by \citet{Fletcher2018b} from the entire set of low-spectral-resolution (15 cm$^{-1}$) \textit{Cassini}/CIRS observations. For each of our observations, we extracted the field at the relevant date. Each zonal field was then re-gridded on a regularly spaced latitude-longitude grid (5\degre$\times$5\degre) at each pressure level using linear interpolation.

\subsubsection{Temperatures in the beacon region}
\label{part:Temperature_beacon}

    In the latitudinal and longitudinal region where the presence of the beacon alters the temperature field, we replaced the data of our background thermal field by relevant retrievals performed as close as possible in time to our observations. We interpolated linearly at the latitude and longitude boundaries the temperatures between the background and beacon retrievals to smooth the whole thermal field for each date.

\begin{figure*}[!ht]
\centering
\includegraphics[width=0.47\textwidth]{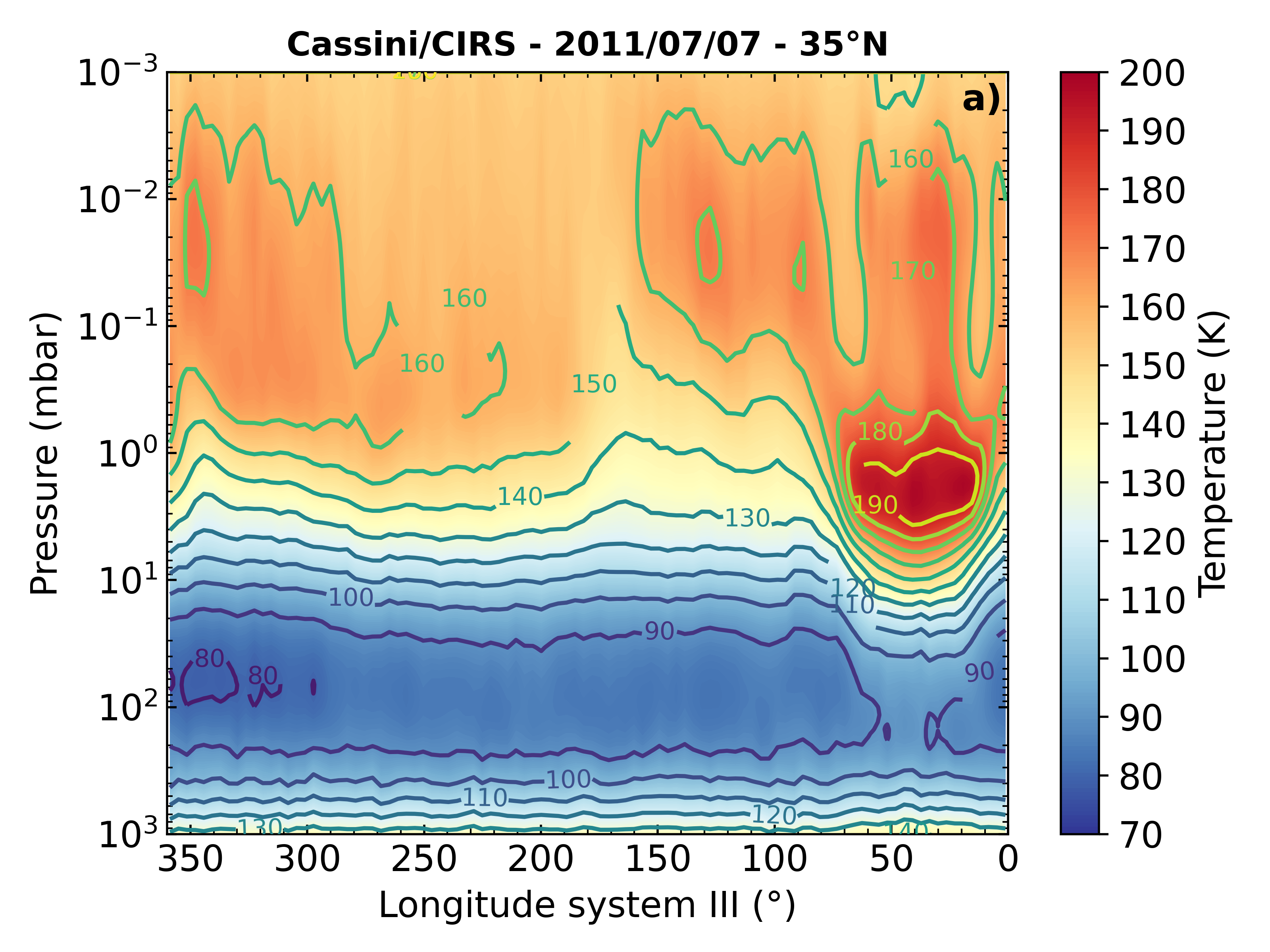}
\includegraphics[width=0.47\textwidth]{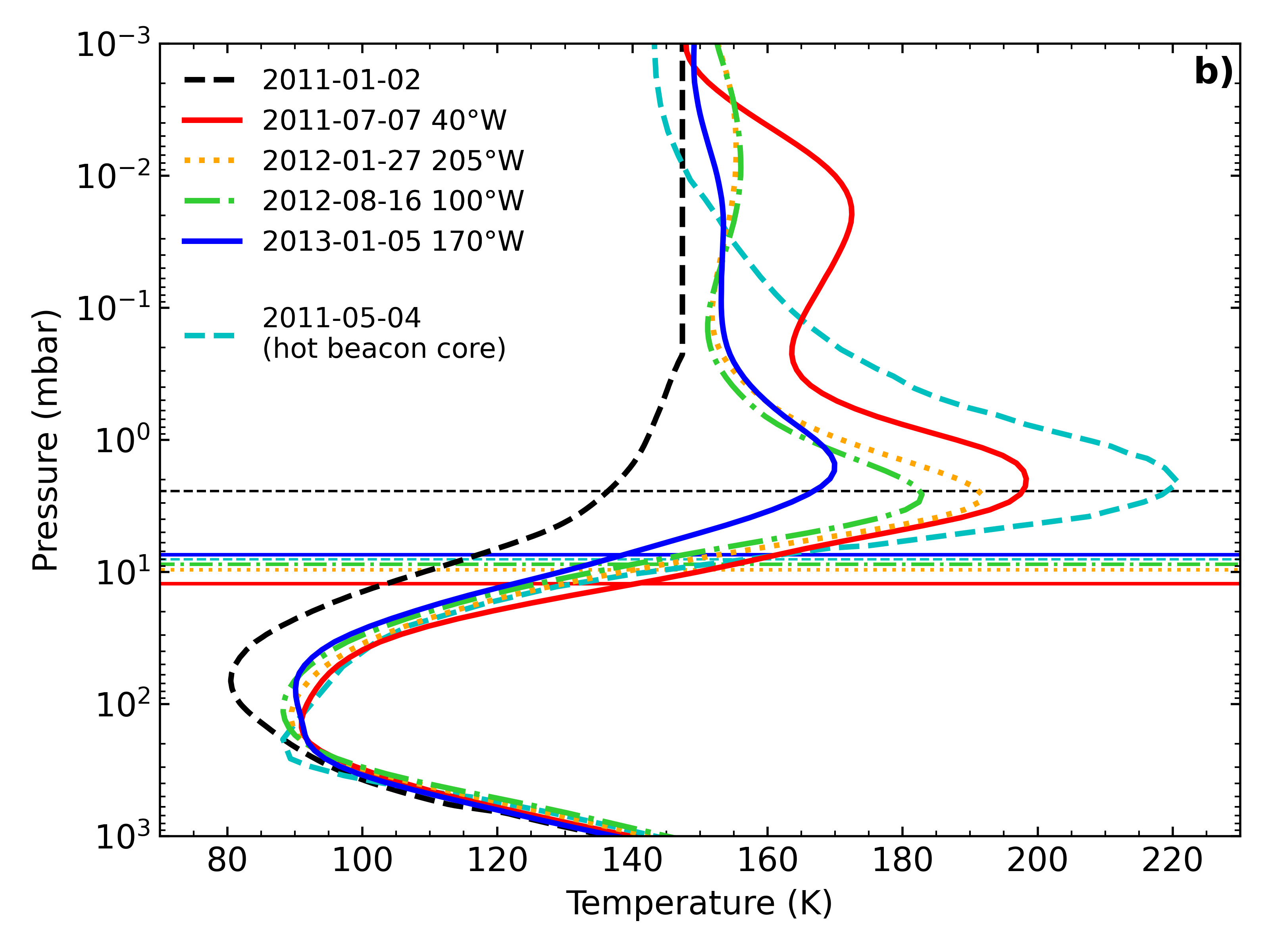} \\
\includegraphics[width=0.47\textwidth]{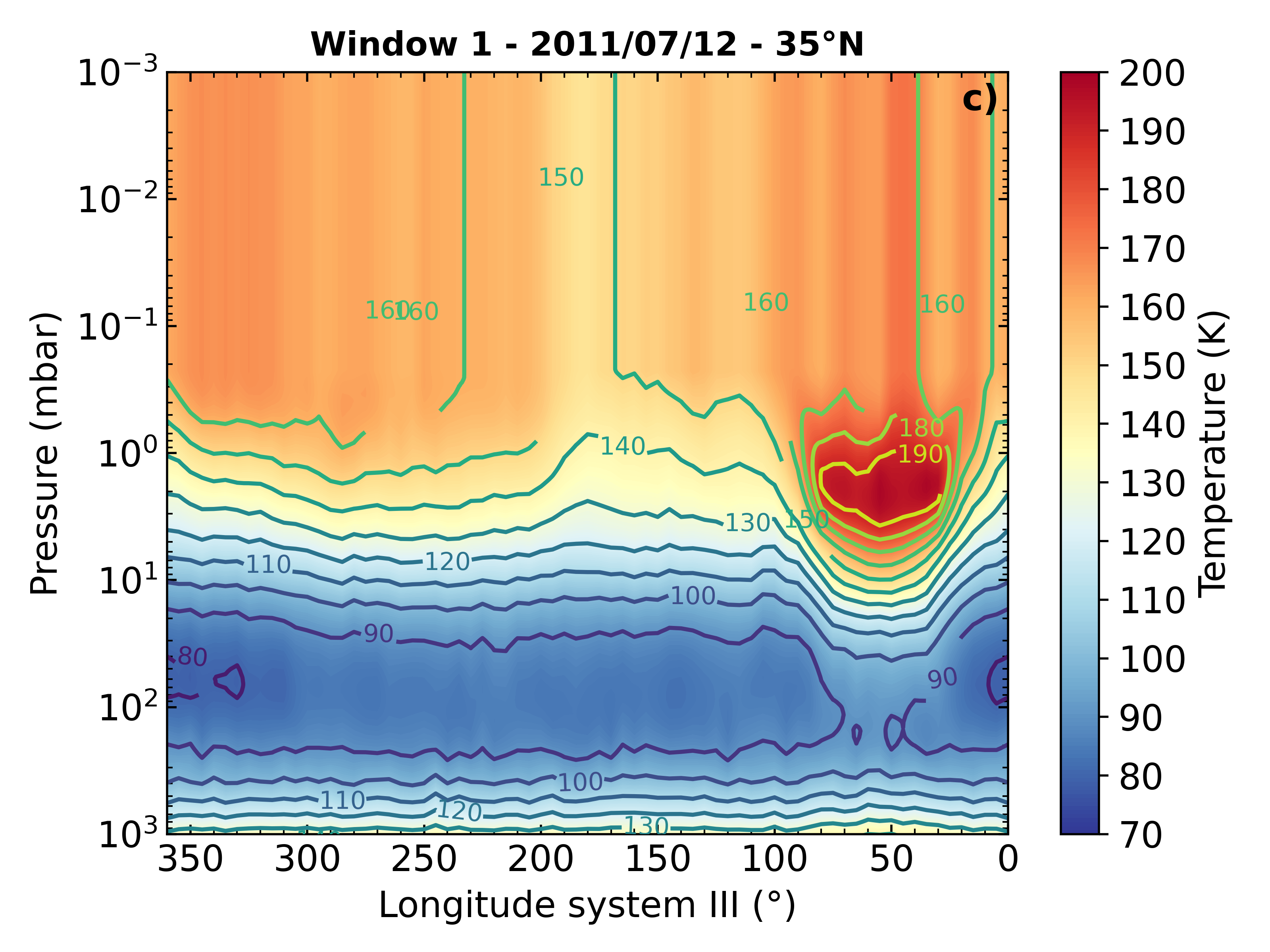}
\includegraphics[width=0.47\textwidth]{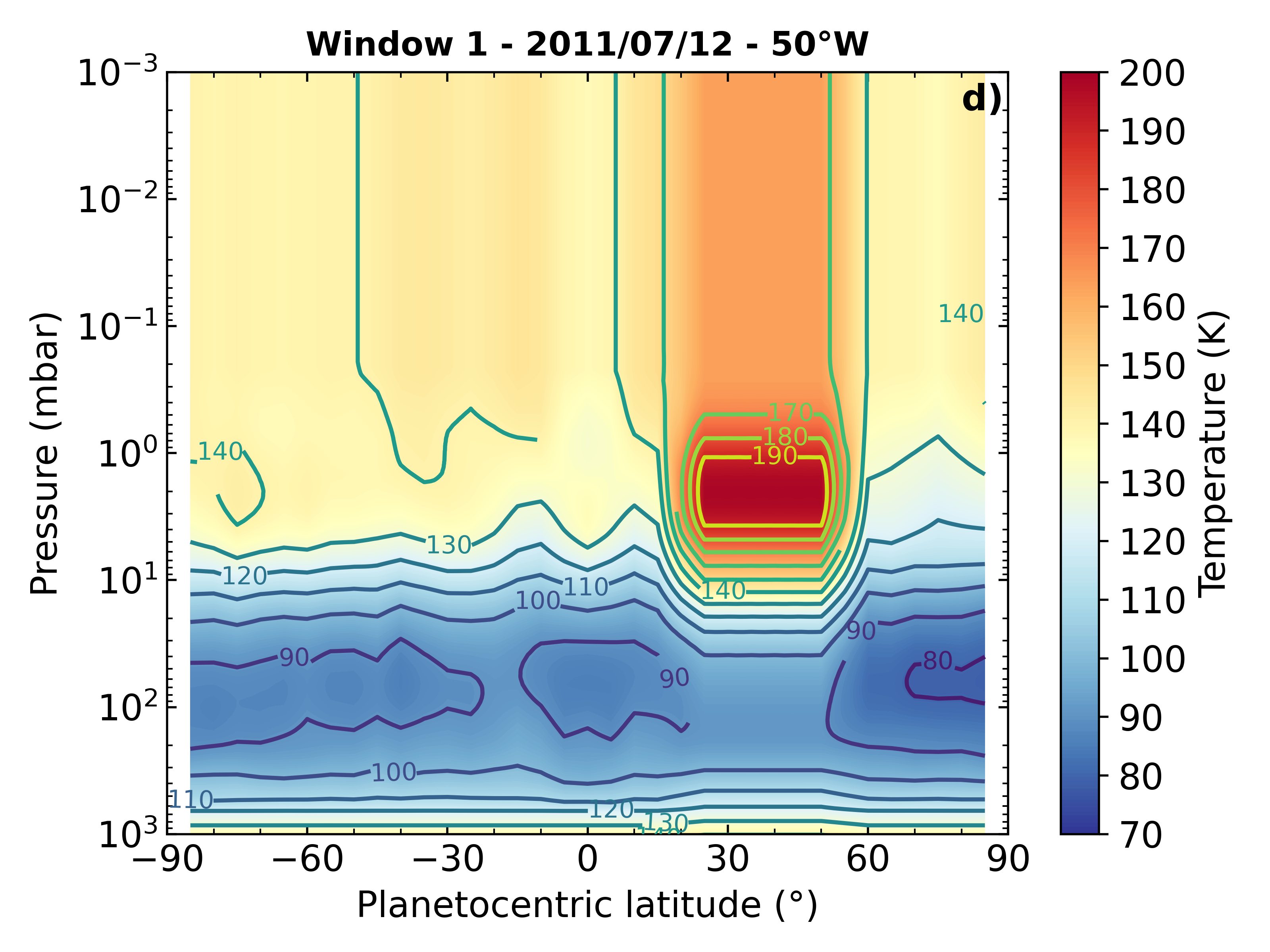}
\caption{\textbf{a) and b)} Temperature fields at an averaged latitude of 35\degre N retrieved from \textit{Cassini}/CIRS observations by \citet{Fletcher2012}. \textbf{a)} Pressure-longitude cut for July 7 2011. \textbf{b)} Vertical profiles extracted from the July 7 2011, January 27 2012, August 16 2012, and January 5 2013 thermal fields (see Fig. \ref{fig:Thermal_fields_profiles_CIRS_ALL}). The coloured lines are taken at the central longitude of the beacon. The central longitude is defined following the beacon drift presented in Section \ref{part:Beacon_drift}. The dashed black line represents the quiescent conditions (taken here on January 2 2011) at the early stages of the storm. The May 4 2011 hot beacon core temperature profile is depicted by the dashed sky-blue line and was used in the photochemical model of \citet{Moses2015}. The horizontal lines represent the water condensation level calculated using \citet{Fray2009} for each temperature profile (same colour and line code for a given date). \textbf{c) and d)} Final thermal fields used in our radiative transfer calculations for the first window (July 12 2011), constructed from the above \textit{Cassini}/CIRS data and the background seasonal data from \citet{Fletcher2018b}. An isothermal extrapolation from 0.2 mbar to the top of the atmosphere was performed. \textbf{c)} Similar to a) at 35\degre N, after the beacon drift was accounted for. \textbf{d)} Pressure-latitude constructed field at 50\degre W.}
\label{fig:Thermal_fields_profiles_CIRS_et_final}
\end{figure*}

    For our first two observation windows (July 2011 and February 2012), we took the beacon temperatures retrieved by \citet{Fletcher2012} from July 7 2011 and January 27 2012 \textit{Cassini}/CIRS observations. These are the closest dates to our observations and, given the long radiative timescales of several weeks in Saturn's stratosphere, we assumed no temperature changes within the beacon over the short difference in time between the \textit{Cassini}/CIRS and the \textit{Herschel}/PACS observations (i.e. $\approx$ a week). The July 7 2011 temperature field consists of a longitudinal section centred at the beacon latitude (see Fig. \ref{fig:Thermal_fields_profiles_CIRS_et_final} a). The altitudinal coverage encompasses the upper troposphere and the stratosphere. We deduced the latitudinal extension of the beacon from the IRTF/TEXES observations of \citet{Fouchet2016}. The January 27 2012 temperature field consists of latitudinal and longitudinal sections centred on the beacon (see Fig. \ref{fig:Thermal_fields_profiles_CIRS_ALL}, second row).

    For our last two windows (July-August 2012 and January-February 2013), we used temperature retrievals from similar, as-of-yet unpublished, \textit{Cassini}/CIRS data obtained on August 16 2012 and January 5 2013. As opposed to the other dates, we have no information regarding the latitudinal extent of the beacon in January 2013, since no latitudinal coverage was performed with \textit{Cassini}/CIRS between August 2012 and October 2014. We assumed the same latitudinal extent as in the 2012 data. All the beacon region thermal fields that we used to reconstruct a full 3D field for each observation are presented in Fig. \ref{fig:Thermal_fields_profiles_CIRS_ALL}.
    
    CIRS spectra were latitudinally averaged between 25-44\degre N (planetocentric) to increase the S/N in the inversions. These averaged spectra were then used to retrieve the four longitude-pressure fields taken on July 7 2011, January 27 2012, August 16 2012, and January 5 2013. A similar approach was used for the two latitude-pressure fields taken on January 27 2012 and August 16 2012, with a longitudinal average around the beacon centre between 195-225\degre W and 90-110\degre W, respectively. The resulting thermal fields thus lack detailed information on the horizontal structure because of the latitudinal and/or longitudinal averaging. The initial information on the horizontal details cannot be recovered; nor can the uncertainty and fluctuations on the pre-average variations. The temperatures close to the beacon centre are underestimated, while the temperatures further away of the beacon centre are overestimated with the averaging. However, the overall effect should be compensated for, mostly because the Planck function varies linearly with the temperature in our frequency domain, and the thermal field is smoothed out. The beacon being smaller than the PACS beam size, we only look at the larger scales. Uncertainties on the temperature field of the order of 2 K are estimated at the pressure levels of interest (several millibar) from the temperature retrievals and from the fluctuations arising from the time gap between the CIRS data and our PACS data.

    The vertical profiles at the central latitude and longitude of the beacon, retrieved from the aforementioned \textit{Cassini}/CIRS observations of July 7 2011, January 27 2012, August 16 2012, and January 5 2013 are displayed in Fig. \ref{fig:Thermal_fields_profiles_CIRS_et_final} b. They are compared to a quiescent conditions profile obtained at the same latitude in January 2 2011 a few weeks after the onset of the storm. Fig. \ref{fig:Thermal_fields_profiles_CIRS_et_final} b additionally displays the warmest temperature profile measured by CIRS as it was used in the photochemical model of \citet{Moses2015}. This thermal profile was retrieved on May 4 2011 from \textit{Cassini}/CIRS spectra averaged around the beacon centre ($\pm$10\degre\, average in longitude and $\pm$5\degre\, in latitude, \citealt{Fletcher2012}).

    The temperature peak in the beacon corresponds to a temperature increase over quiescent conditions of 67 K, 62 K, 53 K, and 36 K, respectively for the four observation periods. It is obvious that the stratosphere had not returned to quiescent conditions even two years after the storm onset. We also note that we simply extrapolate isothermally the temperatures from the 0.2 mbar level up to the top of the stratosphere as shown in Fig. \ref{fig:Thermal_fields_profiles_CIRS_et_final} c, because these levels are not well constrained by \textit{Cassini}/CIRS, particularly using the low-spectral-resolution nadir observations of \citet{Fletcher2012}. The two water lines analysed in this paper are not sensitive to these high levels anyway (see the water contribution functions on Fig. \ref{fig:Fonctions_contributions}). 

    As the beacon travels at a given drift rate and the above fields are not measured at the same time as our \textit{Herschel} observations, the final step when building the thermal fields is to shift them in longitude accordingly (see next section). 

\subsubsection{Beacon drift rate}
\label{part:Beacon_drift}

    \citet{Fletcher2012} observed that the beacon drifted in longitude at a rate of ($2.70\pm0.04$)\degre.day$^{-1}$ (or $-26.4\pm0.4$ m.s$^{-1}$ at 35\degre) from July 2011 to March 2012. It is thus an important point to consider when building the thermal fields for our analysis, because the \textit{Cassini}/CIRS observations presented in Section \ref{part:Temperature_beacon} were not recorded at the same date as our \textit{Herschel}/PACS data. Consequently, we must account for this drift to have the beacon properly located in longitude in the thermal fields we use for the radiative transfer analysis. For the July 2011 and January 2012 thermal fields, we applied the beacon drift rate derived by \citet{Fletcher2012} which is valid for our first two windows, and corrected the beacon centre longitude. The resulting field for the July 2011 window is shown in Fig. \ref{fig:Thermal_fields_profiles_CIRS_et_final} c as an example. 

\begin{figure}[!ht]
\centering
\includegraphics[width=0.7\textwidth]{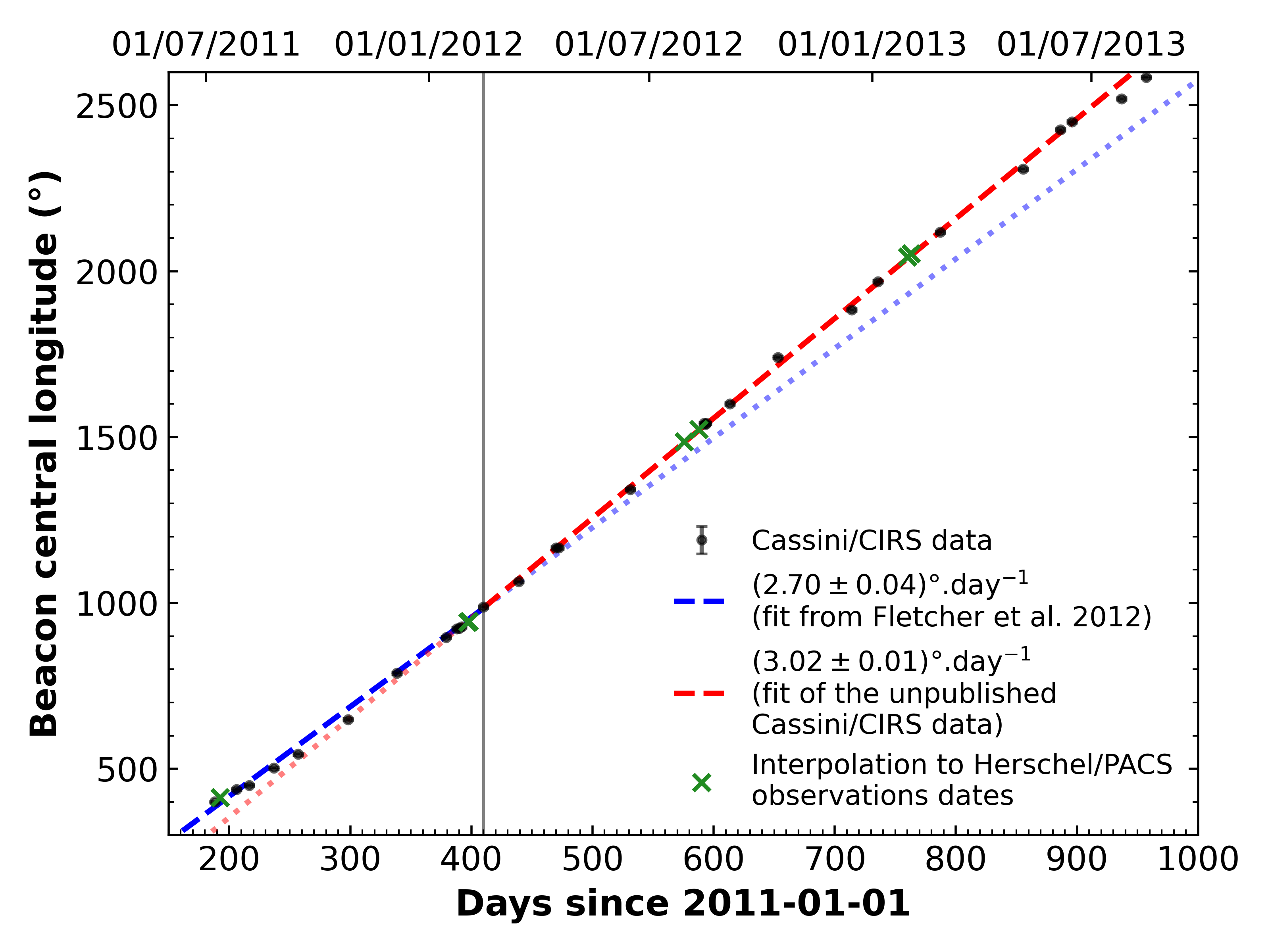}
\caption{Central longitude of the beacon through time. The black dots are derived from the \textit{Cassini}/CIRS thermal fields averaged on the 25-45\degre N latitudinal band, i.e. for an average latitude of 35\degre N, at the millibar level in the stratosphere. Data points are derived from the longitudinal centre position of the beacon, calculated as the middle of the longitude-temperature curve, which is rather symmetric with respect to the middle longitude. The dashed blue line represents the trend from \citet{Fletcher2012} between July 2011 and March 2012. The dashed red line shows the trend after March 2012 derived in this paper. The dotted lines correspond to the extrapolation of the two trends. The vertical grey line delimits the two temporal periods. The interpolation at our \textit{Herschel}/PACS observations dates are represented by the green x-shaped points.}
\label{fig:Beacon_position}
\end{figure}

    Beyond March 2012, the beacon drift rate remained unconstrained. We thus analysed unpublished CIRS thermal fields similar to those presented in \citet{Fletcher2012} to derive an updated drift rate over time. The entire \textit{Cassini}/CIRS dataset used is presented in Fig. \ref{fig:Donnees_beacon_position_extension} (top panel). The temperatures are derived from CH$_4$ emission averaged between 1280-1320 cm$^{-1}$ and therefore sensitive to stratospheric temperatures averaged over the millibar levels. The data correspond to a latitudinal average between 25-44\degre N, and thus cover the beacon latitudinal centre. They extend from 2010 (i.e. before the storm onset) to the end of 2014 (i.e. $\sim$2 years after our last \textit{Herschel}/PACS observation window), so we have an overall panorama of the entire storm event in the stratosphere at the millibar levels as a function of time and longitude. For each of the dates after the merging in April-May 2011 (we only look at the evolution of the merged beacon), we identified the central longitude of the beacon as the middle of the thermal bell-shaped curve, which is rather symmetrical with respect to the middle longitude. We did not use the maximum temperature peak in the beacon region as an estimator of the beacon centre because of the variability of the maximum location with time, and also because of the uncertainty on the temperature retrieval from inversion models (2-3 K, \citealt{Fletcher2012}). To illustrate this point, the determination of the beacon centre on July 7 2011 from the temperature peak (see Fig. \ref{fig:Thermal_fields_profiles_CIRS_et_final} a) would be complex as the region of maximal temperature at around 190 K is really extended with longitude (about 50\degre). The beacon central longitude over time estimated from the previous method is summed up in Fig. \ref{fig:Beacon_position}. We find a beacon drift velocity of ($3.02\pm0.01$)\degre.day$^{-1}$ (or $-29.4\pm0.1$ m.s$^{-1}$ at 35\degre N) beyond March 2012. The thermal fields of our last two \textit{Herschel}/PACS windows (July-August 2012 and January-February 2013) were corrected using this new drift rate. 

    The increase in the drift rate between the two periods is difficult to explain owing to the lack of wind field measurements in the stratosphere of Saturn over the lifetime of the beacon at the relevant pressure levels. We note, however, that the retrograde motion of the beacon is consistent with two independent direct wind measurements performed with ALMA in the millibar layers. \citet{Cavalie2024b} measure $-33\pm18$ m.s$^{-1}$ between 30\degre N and 45\degre N in data taken with ALMA in January 2012. Stratospheric wind measurements performed in May 2018, also with ALMA, at the beacon latitudes by \citet{Benmahi2022} indicate as well retrograde winds, with a wind velocity of -50$\pm$20 m.s$^{-1}$.

    With the knowledge of the beacon central position over time, we retrieved the longitudinal extension of the beacon over time (see Fig. \ref{fig:Donnees_beacon_position_extension} bottom panels) from the above data. We applied a longitudinal shift of the thermal fields of Fig. \ref{fig:Donnees_beacon_position_extension} (top panel), according to the results of Fig. \ref{fig:Beacon_position}, to align the beacon at 180\degre W over time. Fig. \ref{fig:Donnees_beacon_position_extension} bottom right panel results from a bilinear interpolation of the aligned data of the left panel and shows the beacon extension over time at the millibar levels. The thermal signature of the beacon is seen up to the end of 2013 at these levels and returns to quiescent conditions.

 \subsection{Observation geometry}
 \label{part:Geometry}
 
 We retrieved the observation geometry of Saturn from JPL/Horizons for the different observation times. These data, which are essential for extracting the appropriate temperature information from the full 3D thermal fields and then analysing the maps with radiative transfer calculations, are contained in \tab{tab:Obs_geometry}. In this paper, all latitudes ($\varphi$) are planetocentric, and all longitudes ($\lambda$) are System III longitudes.
 
   The beacon geometrical data considered in this paper (see \tab{tab:Beacon_data}) have been derived from the observations presented in \citet{Fletcher2012}, and which cover our first two observation windows. For the last two windows, we used the unpublished thermal mapping observations of August 16 2012 and January 5 2013 presented in Fig. \ref{fig:Thermal_fields_profiles_CIRS_ALL}, and accounting for the drift rate of Section \ref{part:Beacon_drift}. For the 2013 data, we assume a beacon latitudinal extent identical to that seen in the 2012 data as the latitude-pressure field was not measured on this date. 

 \begin{table*}
      \centering 
      \begin{tabular}{llllllll}
       \specialrule{.2em}{.1em}{.1em} 
       \multicolumn{8}{c}{\textit{Window 1: July 2011}} \\
       \hline 
       Date              & $\theta$ & $\lambda_\mathrm{obs}$ & $\varphi_\mathrm{obs}$ & $\lambda_\Sun$ & $\varphi_\Sun$ & NP & $L_S$ \\
       \hline                    
       2011-07-12 13:25:00  & 17.06 & 83.95 & 9.43 & 78.43 & 12.57 & 356.79 & 23.41 \\
       \specialrule{.2em}{.1em}{.1em} 
       \multicolumn{8}{c}{\textit{Window 2: February 2012}} \\
       \hline   
       Date              & Ang. diam. & $\lambda_\mathrm{obs}$ & $\varphi_\mathrm{obs}$ & $\lambda_\Sun$ & $\varphi_\Sun$ & NP & $L_S$ \\
       \hline                    
       2012-02-02 00:32:00  & 17.62 & 245.72 & 18.41 & 251.20 & 15.85 & 358.57 & 30.05 \\
       2012-02-02 22:01:00  & 17.65 & 251.56 & 18.41 & 257.02 & 15.86 & 358.57 & 30.08 \\
       \specialrule{.2em}{.1em}{.1em} 
       \multicolumn{8}{c}{\textit{Window 3: June-August 2012}} \\
       \hline   
       Date              & Ang. diam. & $\lambda_\mathrm{obs}$ & $\varphi_\mathrm{obs}$ & $\lambda_\Sun$ & $\varphi_\Sun$ & NP & $L_S$ \\
       \hline                    
       2012-07-29 17:25:00  & 16.73 &   60.70 & 15.84 &    55.10 & 18.51 & 357.96 & 35.79 \\
       2012-08-10 18:17:00  & 16.42 &   97.90 & 16.25 &    92.64 & 18.68 & 358.03 & 36.18 \\
       \specialrule{.2em}{.1em}{.1em} 
       \multicolumn{8}{c}{\textit{Window 4: January-February 2013}} \\
       \hline   
       Date              & Ang. diam. & $\lambda_\mathrm{obs}$ & $\varphi_\mathrm{obs}$ & $\lambda_\Sun$ & $\varphi_\Sun$ & NP & $L_S$ \\
      \hline                    
       2013-01-30 03:47:00  & 17.02 & 180.70 & 23.25 & 186.49 & 21.05 & 359.86 & 41.63 \\
       2013-02-02 06:37:00  & 17.11 & 188.94 & 23.27 & 194.74 & 21.09 & 359.87 & 41.76 \\
       \specialrule{.2em}{.1em}{.1em} 
     \end{tabular}
      \caption{Observation geometry of Saturn for the various epochs at mid-observation. Saturn's equatorial angular diameter, given in arcsec as seen from \textit{Herschel}, is represented by $\theta$. $\lambda_\mathrm{obs}$ and $\varphi_\mathrm{obs}$ are the longitude and latitude (respectively) of the sub-observer point. $\lambda_\Sun$ and $\varphi_\Sun$ are the longitude and latitude (respectively) of the sub-solar point. NP is the North Polar angle in degrees, and $L_S$ is Saturn's solar longitude in degrees. The different values are extracted from JPL/Horizons.}
      \label{tab:Obs_geometry}
   \end{table*}

 \begin{table}
     \centering
     \begin{tabular}{llllll}
       \specialrule{.2em}{.1em}{.1em} 
       \multicolumn{6}{c}{\textit{Window 1: July 2011}} \\
       \hline   
       Date              & $\lambda_\mathrm{b}$ & $\Delta\lambda_\mathrm{b}$ & $\varphi_\mathrm{b}$ & $\Delta\varphi_\mathrm{b}$ & $S_\mathrm{b}$ [cm$^{2}$]   \\
       \hline                    
       2011-07-12 13:25:00  & 52 & 65 & 35 & 35 & 2.8$\times 10^{19}$  \\   
       \specialrule{.2em}{.1em}{.1em}
       \multicolumn{6}{c}{\textit{Window 2: February 2012}} \\
       \hline   
       Date              & $\lambda_\mathrm{b}$ & $\Delta\lambda_\mathrm{b}$ & $\varphi_\mathrm{b}$ & $\Delta\varphi_\mathrm{b}$ & $S_\mathrm{b}$ [cm$^{2}$]  \\
       \hline                    
       2012-02-02 00:32:00  & 222 & 52 & 31 & 26  & 1.5$\times 10^{19}$  \\   
       2012-02-02 22:01:00  & 224 & 52 & 31 & 26  & 1.5$\times 10^{19}$  \\   
       \specialrule{.2em}{.1em}{.1em} 
       \multicolumn{6}{c}{\textit{Window 3: July-August 2012}} \\
       \hline   
       Date              & $\lambda_\mathrm{b}$ & $\Delta\lambda_\mathrm{b}$ & $\varphi_\mathrm{b}$ & $\Delta\varphi_\mathrm{b}$  & $S_\mathrm{b}$ [cm$^{2}$]  \\
       \hline                    
       2012-07-29 17:25:00  & 46 & 41 & 30 & 25   & 0.9$\times 10^{19}$  \\   
       2012-08-10 18:17:00  & 82 & 40 & 30 & 25   & 0.9$\times 10^{19}$  \\   
       \specialrule{.2em}{.1em}{.1em} 
       \multicolumn{6}{c}{\textit{Window 4: January-February 2013}} \\
       \hline   
       Date              & $\lambda_\mathrm{b}$ & $\Delta\lambda_\mathrm{b}$ & $\varphi_\mathrm{b}$ & $\Delta\varphi_\mathrm{b}$  & $S_\mathrm{b}$ [cm$^{2}$]   \\
       \hline                    
       2013-01-30 03:47:00  & 241 & 29 & 30? & 25?   & 0.6$\times 10^{19}$?  \\   
       2013-02-02 06:37:00  & 250 & 29 & 30? & 25?   & 0.6$\times 10^{19}$?  \\   
       \specialrule{.2em}{.1em}{.1em} 
     \end{tabular}
      \caption{Beacon geometry data at the time of the \textit{Herschel}/PACS observations. The beacon central longitudes ($\lambda_\mathrm{b}$) were estimated following the beacon drift rate presented in Section \ref{part:Beacon_drift}. The beacon longitudinal extensions ($\Delta\lambda_\mathrm{b}$) were calculated from the thermal contour at 160 K at the 2 mbar level from the \textit{Cassini}/CIRS longitude-pressure thermal fields presented in Section \ref{part:Temperature} and in Fig. \ref{fig:Thermal_fields_profiles_CIRS_ALL}. The beacon latitudinal extensions ($\Delta\lambda_\mathrm{b}$) were estimated in the same method for the two windows of 2012. For the first window, we deduced the beacon latitudinal extension from the IRTF/TEXES observations of \citet{Fouchet2016}. For the last window only (January-February 2013), as the latitude-pressure thermal field was not measured, the beacon latitudinal data are taken as upper limits from the previous window in July-August 2012. The lack of knowledge on this window is depicted by a `?' symbol. The beacon central latitudes ($\varphi_\mathrm{b}$) are deduced from the middle of the latitudinal extension. The beacon surface area $S_\mathrm{b}$ is defined from the thermal contour at 160 K at the 2 mbar level, from the final 3D thermal fields built in Section \ref{part:Temperature}.}
      \label{tab:Beacon_data}
   \end{table}

 \subsection{Composition}
 \label{part:Composition}
 
 \subsubsection{Background composition}
 \label{part:Background_composition}

\begin{figure}[!ht]
 \centering
 \includegraphics[width=0.7\textwidth]{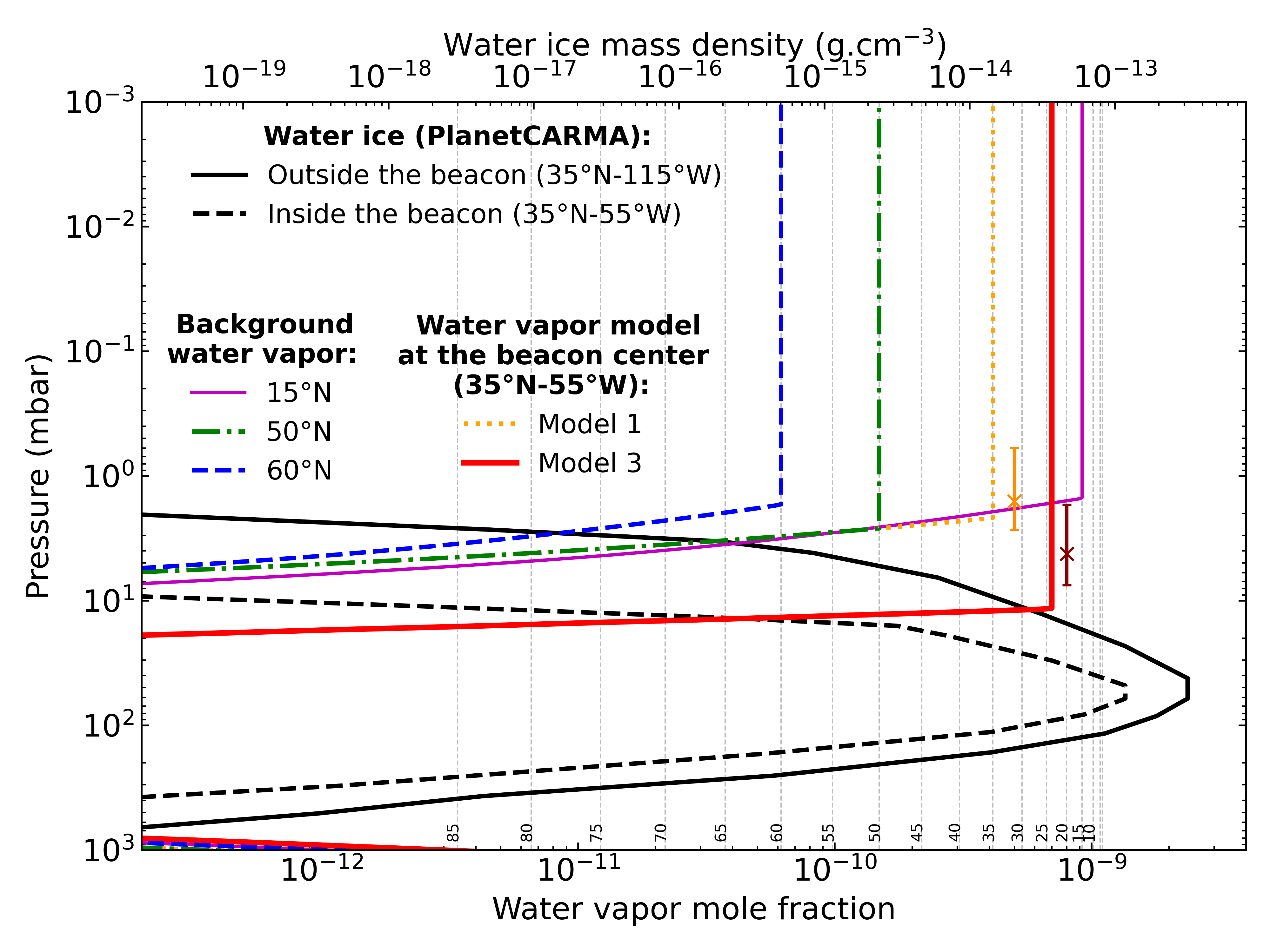}
 \caption{Profiles of the water vapour mole fraction (bottom x scale) and the water ice mass density (top x scale) as a function of pressure. The water vapour profiles are extracted from the water 3D field of July 12 2011. The solid purple line, the dashed-dotted green line and the dashed blue line depict the background field derived from \citet{Cavalie2019} at, respectively, 15, 50, and 60\degre N. The thin vertical dashed lines represent the values of the water vapour mole fraction at $\sim$1-2 mbar found in \citet{Cavalie2019} in quiescent conditions, as a function of latitude. The latitudes are labelled vertically at the bottom of the graph. Model 1 (dotted orange line) indicates the water profile of Step 1 (see Section \ref{part:Step_1}) in which we keep the pre-storm water field of \citet{Cavalie2019} to test the enhancement of the water line amplitude due to temperature increase only. The red line shows the best fit of Model 3 (i.e. Step 3, Section \ref{part:Step_3}) for July 2011, in which we vary the water mole fraction in the beacon. The orange and dark red bars represent the vertical sensitivity range from the water contribution functions, corresponding to the Step 1 and Step 3 models, respectively (see also Fig. \ref{fig:Fonctions_contributions}). The crosses correspond to the altitude of maximum contribution to the line. The water ice profiles were computed with the PlanetCARMA cloud model of \citet{Barth2020} adapted to the Saturn stratosphere with the July 12 2011 thermal profile outside the beacon at 35\degre N-115\degre W for the solid black line and inside the beacon at 35\degre N-55\degre W for the dashed black line.}
 \label{fig:Profils_eau}
 \end{figure}

 We started from the background atmospheric composition described in \citet{Cavalie2019} regarding H$_2$, He, CH$_4$, NH$_3$, and PH$_3$. We also adopted the empirical distribution for stratospheric H$_2$O they derived from \textit{Herschel}/PACS observations of January 2011. The H$_2$O mole fraction at the millibar level varies with latitude following a Gaussian distribution centred on the equator of Saturn, as follows:
 
\begin{equation}
 y_\text{H$_2$O}(\Phi) = y_\text{eq} \times \text{exp}{\left( \displaystyle -\frac{\Phi^2}{2 \sigma^2} \right)}
 \label{eq:gaussian}
\end{equation}

 \noindent with $y_\text{H$_2$O}$ the water mole fraction, $\Phi$ the planetocentric latitude, $y_\text{eq}=1.1$ ppb the water mole fraction at the equator, and $\sigma=25$\degre. At a given latitude, $y_\text{H$_2$O}$ was set to be constant for pressures lower than the water condensation level computed according to the thermal fields and the saturation law published by \citet{Fray2009}. Examples of this background field are illustrated in Fig. \ref{fig:Profils_eau} at several latitudes.

   This water vapour originates from an external flux coming from the torus of water vapour fed by the plumes of Enceladus \citep{Hansen2006,Porco2006,Waite2006,Hartogh2011a,Cassidy2010}, which diffuses vertically from the top of the atmosphere until it reaches the top of the water cold trap. The cold trap extends from 20 bar in the troposphere \citep{Cavalie2024a} to $\sim$2 mbar in the stratosphere in quiescent conditions (see Fig. \ref{fig:Thermal_fields_profiles_CIRS_et_final} b). This upper boundary of the cold trap varies with latitude because of the thermal field. At this level, the water vapour from the external flux condenses and is expected to form the stratospheric water haze \citep{Moses2000b,Ollivier2000}.

 \subsubsection{H$_2$O distribution in the beacon}
 \label{part:H2O_profiles}
 
   Within the beacon, the temperature dramatically increases during the storm in the pressure range where H$_2$O condensation usually occurs \citep{Fletcher2011,Fletcher2012}. This leads to a downward shift of the condensation level as shown in Fig. \ref{fig:Thermal_fields_profiles_CIRS_et_final} b. As a consequence, there is a whole new region in which H$_2$O could exist in its vapour phase during the lifetime of the beacon. In quiescent conditions, there is no water emission emanating from these layers because of condensation. Thus, the amount of water vapour in these layers had never been characterized before. 

\begin{figure*}[!ht]
 \centering
 \includegraphics[width=0.99\textwidth]{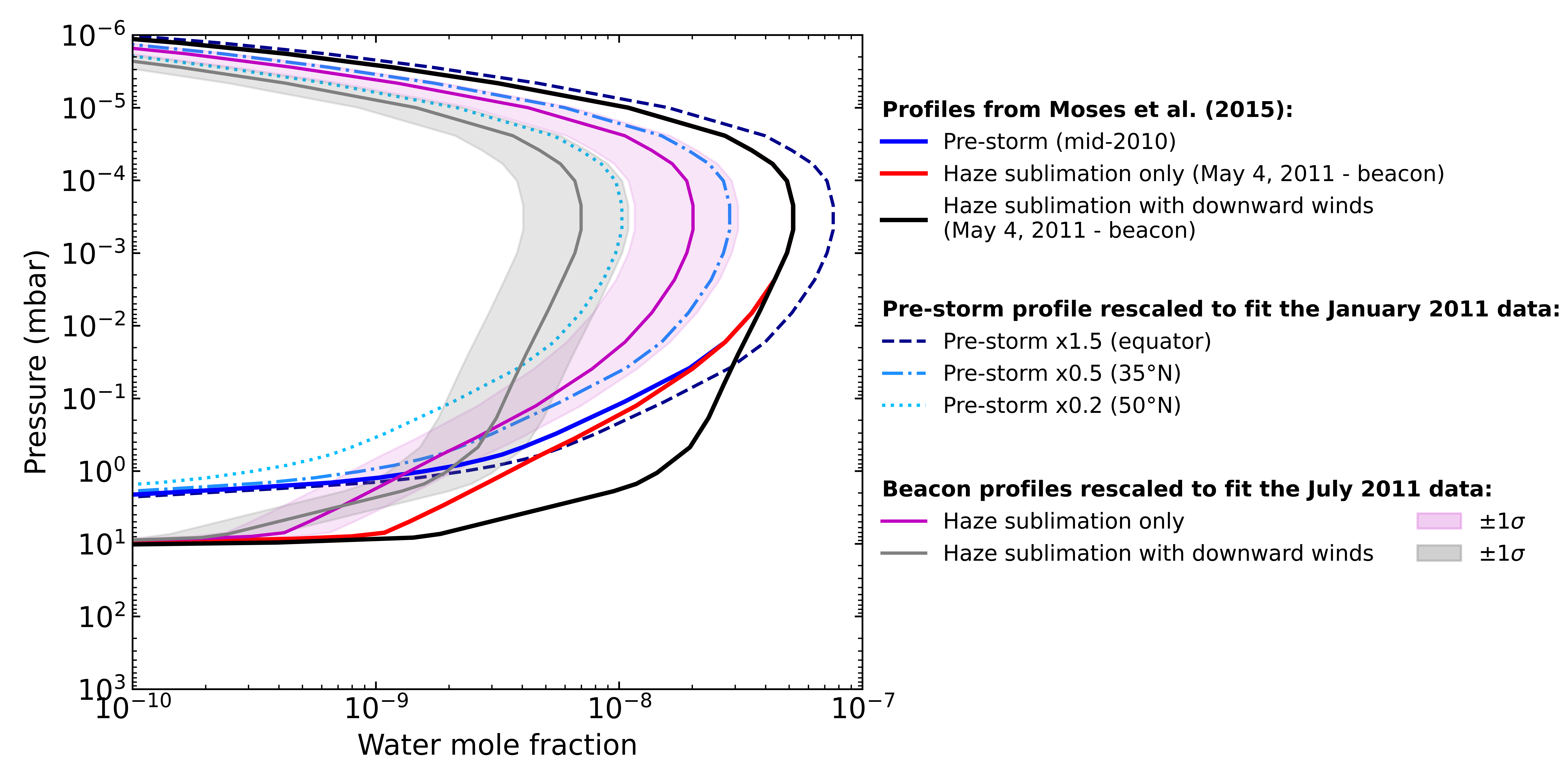}
 \caption{Water vertical profiles as a function of pressure derived from \citet{Moses2015} and from Moses, priv. comm. Their original photochemical profiles are depicted by the thick solid blue, red, and black lines, respectively, in the pre-storm conditions, in the haze sublimation model, and the vertical transport with haze sublimation model. They are derived from a disc-averaged value for the water influx at the top of the atmosphere. As these profiles originate from a disc-averaged observation of water in the stratosphere of Saturn, we first had to rescale the pre-storm profile as a function of latitude following the water distribution found by \citet{Cavalie2019} with an additional fitting factor to reproduce the January 2011 data. The resulting water profiles are plotted in blue at different latitudes. The purple and grey areas represent the best fit results for July 2011 with the two rescaled water models in the beacon.}
 \label{fig:Profils_Moses}
 \end{figure*}

   Given the long timescales related to eddy mixing in the 1-10 mbar layer ($\tau=H^2/K_{zz}\sim200-400$ years compared to the few weeks between the onset of the storm and the first \textit{Herschel}/PACS maps) in quiescent conditions, the effect of vertical mixing should be negligible even if local turbulence may have completely been altered in the beacon. \citet{Moses2015} found that downward winds could explain the abundance increases seen at 2 mbar in C$_2$H$_2$ and C$_2$H$_6$ with \textit{Cassini}/CIRS \citep{Fletcher2012}. Their best fit to the data was obtained with a Gaussian wind profile peaking at $\log_{10}(P~\text{mbar})=-0.5$ (i.e. 0.3 mbar) with a maximum velocity of -10 cm.s$^{-1}$ and a standard deviation of $\log_{10}(P~ \text{mbar})=0.8$ (i.e. from 0.05 to 2 mbar). It would then roughly take 3-4 weeks to transport H$_2$O from 0.1 mbar to 2 mbar; that is, to the warmer layers in the beacon. \citet{Moses2015} estimated that the H$_2$O column density would increase by a total factor of 8.5 above the condensation level in the beacon, in which a factor of 2.8 is due to those vertical winds and a factor of 3 accounting for the haze sublimation. The water vertical profiles in the beacon obtained in \citet{Moses2015} from photochemical modelling are illustrated in Fig. \ref{fig:Profils_Moses}. These profiles were computed to study the beacon at its hot core stage of May 4 2011.
 
\begin{figure*}[!ht]
 \centering
 \includegraphics[width=1\textwidth]{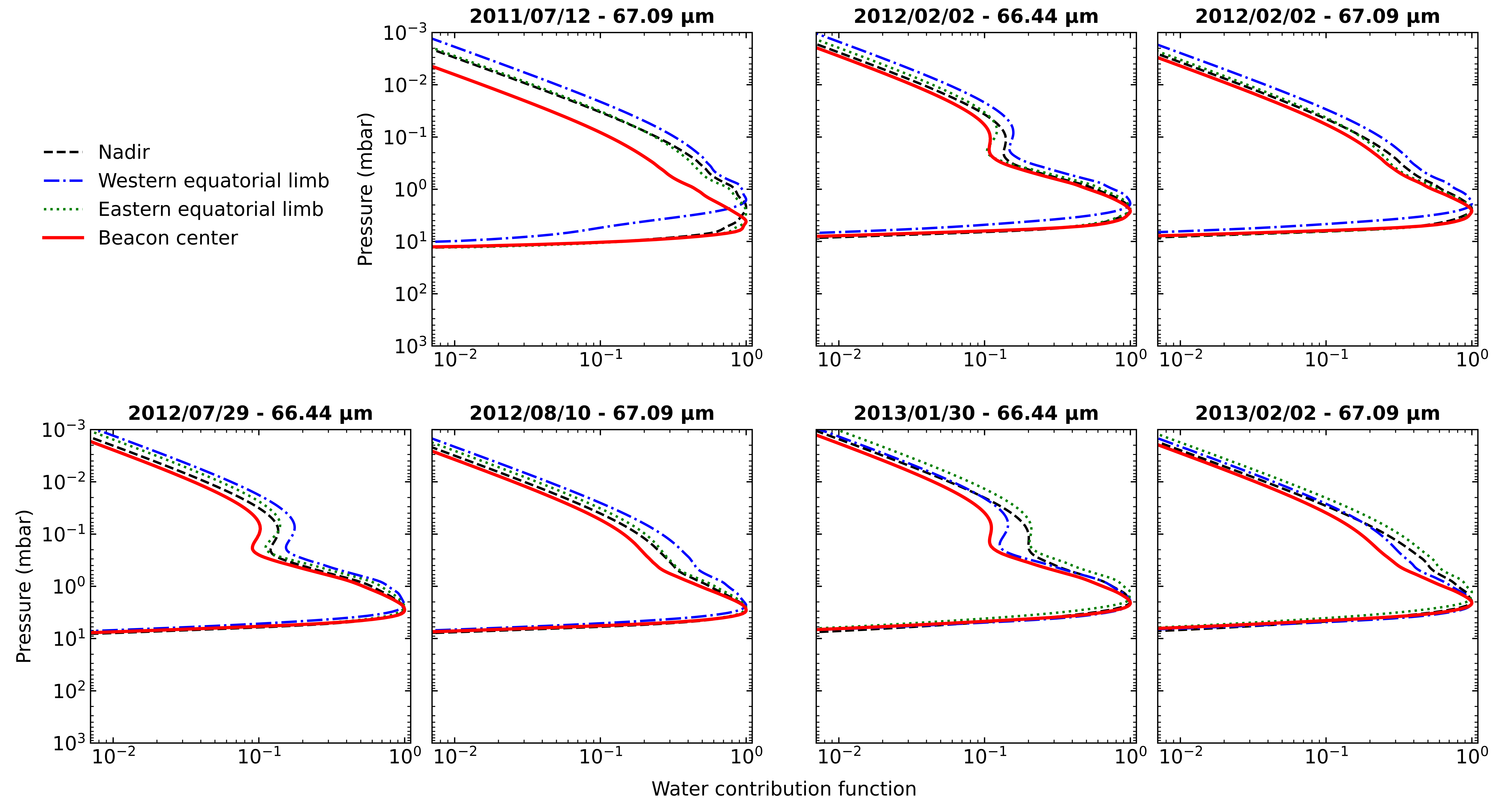}
 \caption{Water contribution functions for each of our seven observations, calculated with the best fit 3D water abundance field in the beacon from Step 3 (see Section \ref{part:Step_3}). Dashed black lines refer to a line of sight at the planet disc centre. Dashed-dotted blue lines and dotted green lines correspond to a line of sight at, respectively, the western and eastern equatorial limbs. The solid red lines depict a line of sight at the beacon centre.}
 \label{fig:Fonctions_contributions}
 \end{figure*}

   Our approach consisted of three steps: (1) we first assessed whether the temperature increase in the beacon, combined with the latitudinal distribution from \citet{Cavalie2019}, was sufficient to reproduce the observed water maps; (2) we then tested more physical vertical water profiles derived from the photochemical models of \citet{Moses2015}; and (3) finally, we determined the column density required to account for the beacon strong emission over time with empirical models. The water vapour contribution function calculations from Step 3 are presented in Fig. \ref{fig:Fonctions_contributions}. They indicate that, in the beacon, the enhanced H$_2$O line emission observed in Fig. \ref{fig:Line_area_maps} can be produced in the new water vapour existence region, within the Step 3 water models. We can thus use the H$_2$O maps to quantify the extra H$_2$O column density in this pressure range.

\section{Results}
\label{part:Results} 

 \subsection{Step 1: The effect of the temperature increase}
 \label{part:Step_1}

	A first model is to consider that no additional water vapour is present in the heated layers during the storm (i.e. no haze sublimation and/or no downwelling winds filling the layers). This model will be referred to as Step 1. The water abundance field in this model is the same as in the pre-storm conditions presented in Fig. \ref{fig:Profils_eau} and Section \ref{part:Background_composition}. With this model, we only account for the effect of the local temperature increase on water line emission for the layers above the typical condensation level (see Fig. \ref{fig:Thermal_fields_profiles_CIRS_et_final} b), coupled with the contribution to the line of the water abundance field in quiescent conditions found in \citet{Cavalie2019} in January 2 2011.

\begin{figure*}[!ht]
 \centering
 \includegraphics[width=0.47\textwidth]{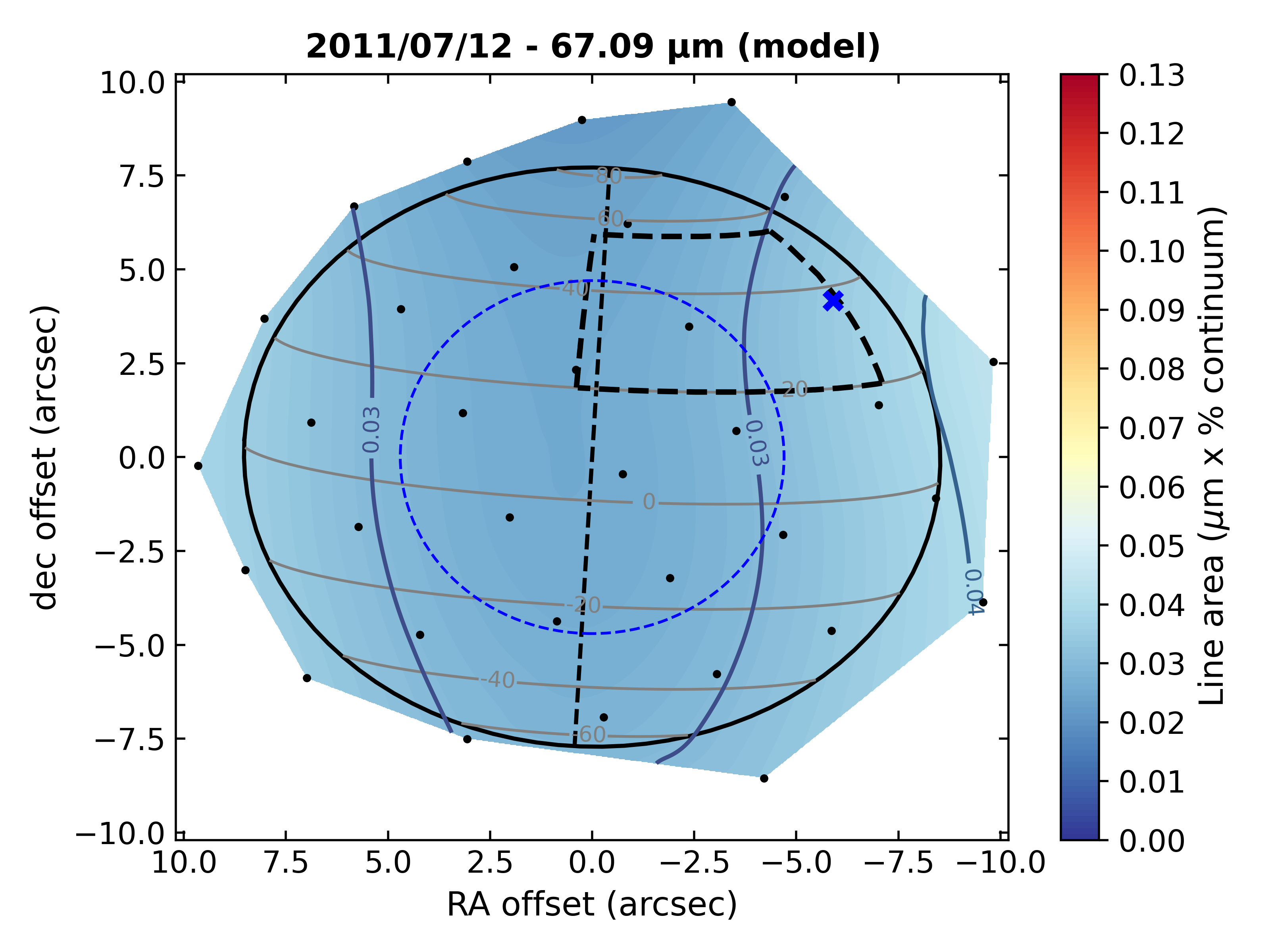}
 \includegraphics[width=0.47\textwidth]{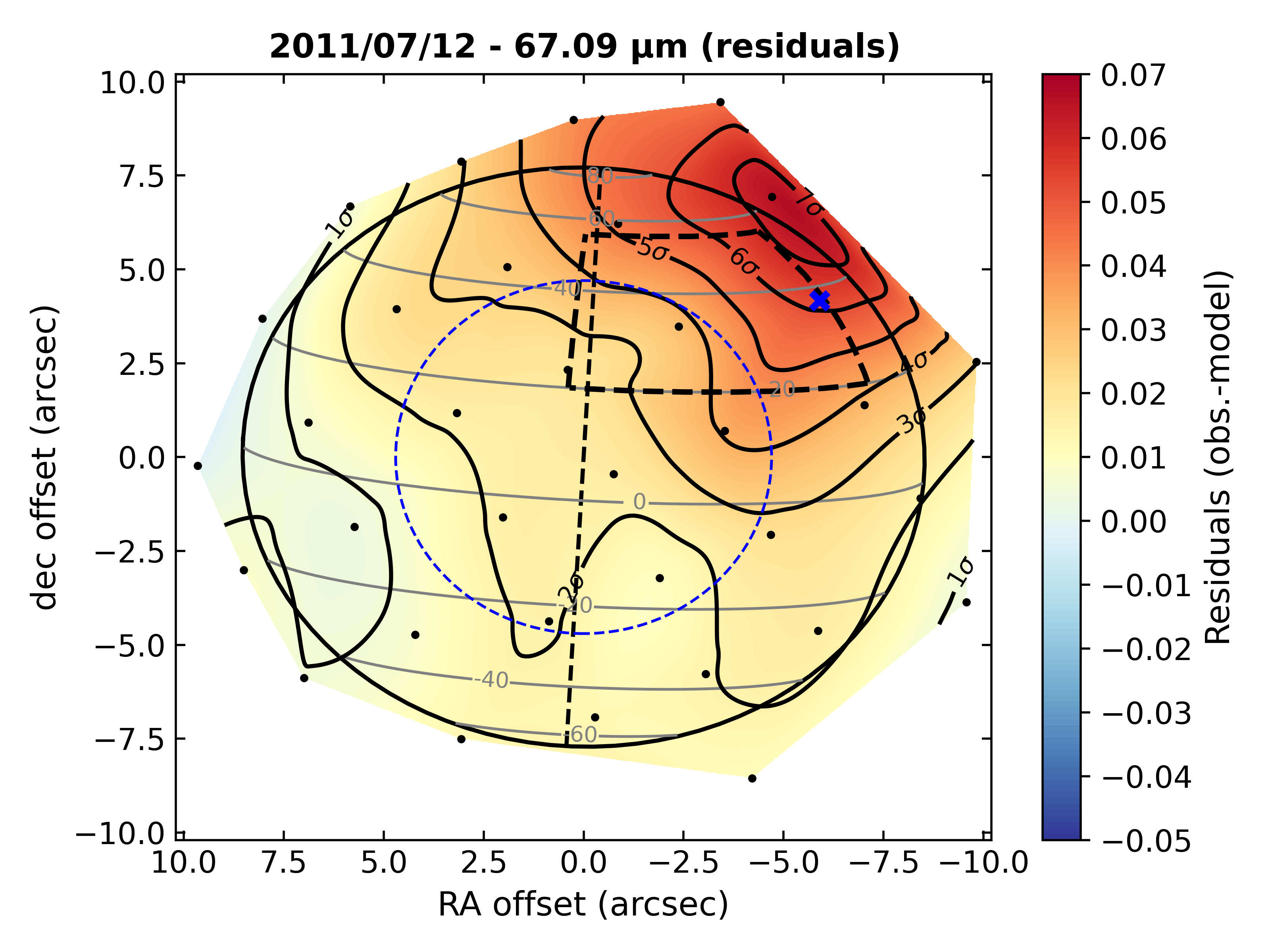}\\

 \includegraphics[width=0.47\textwidth]{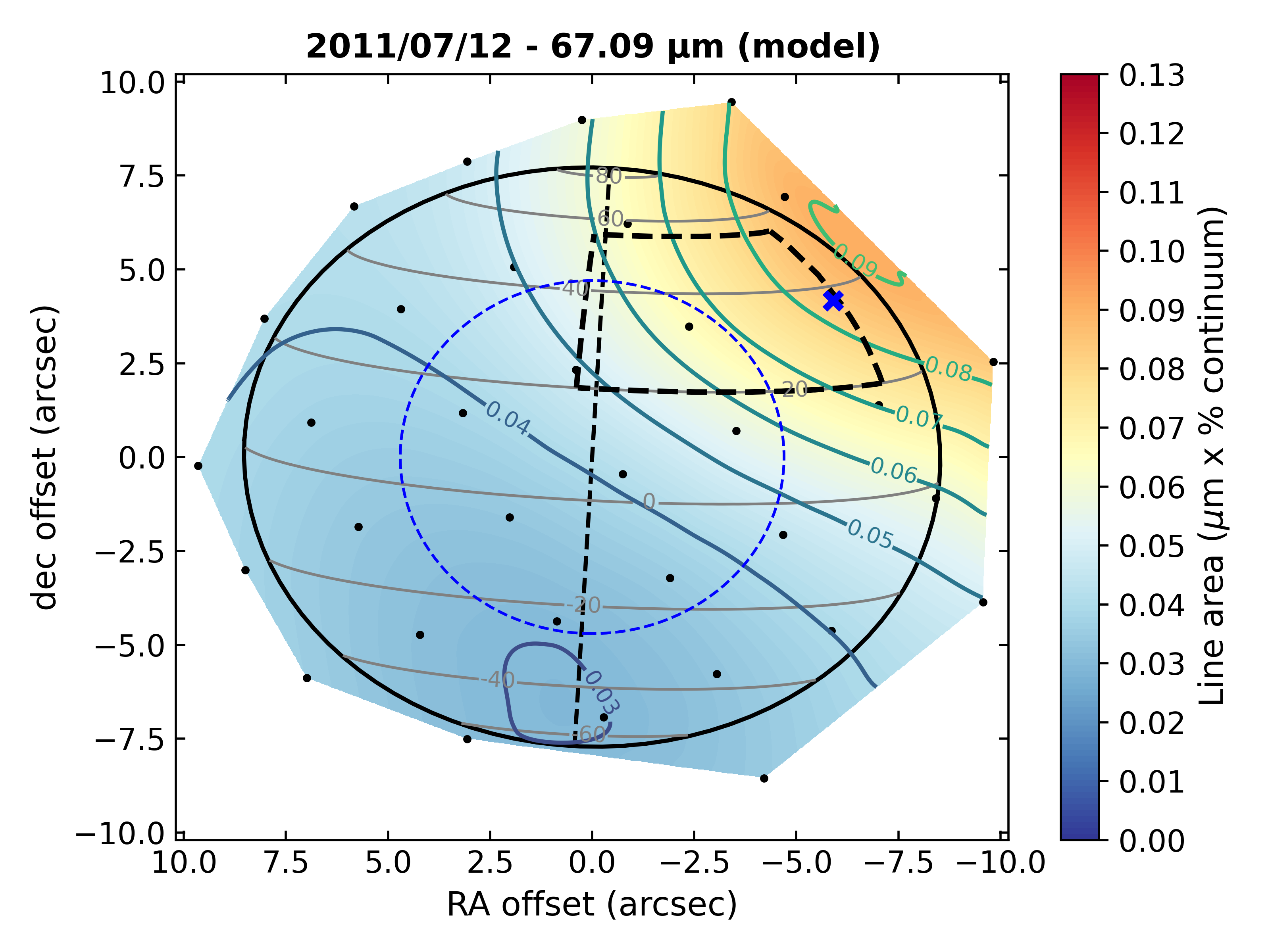}
 \includegraphics[width=0.47\textwidth]{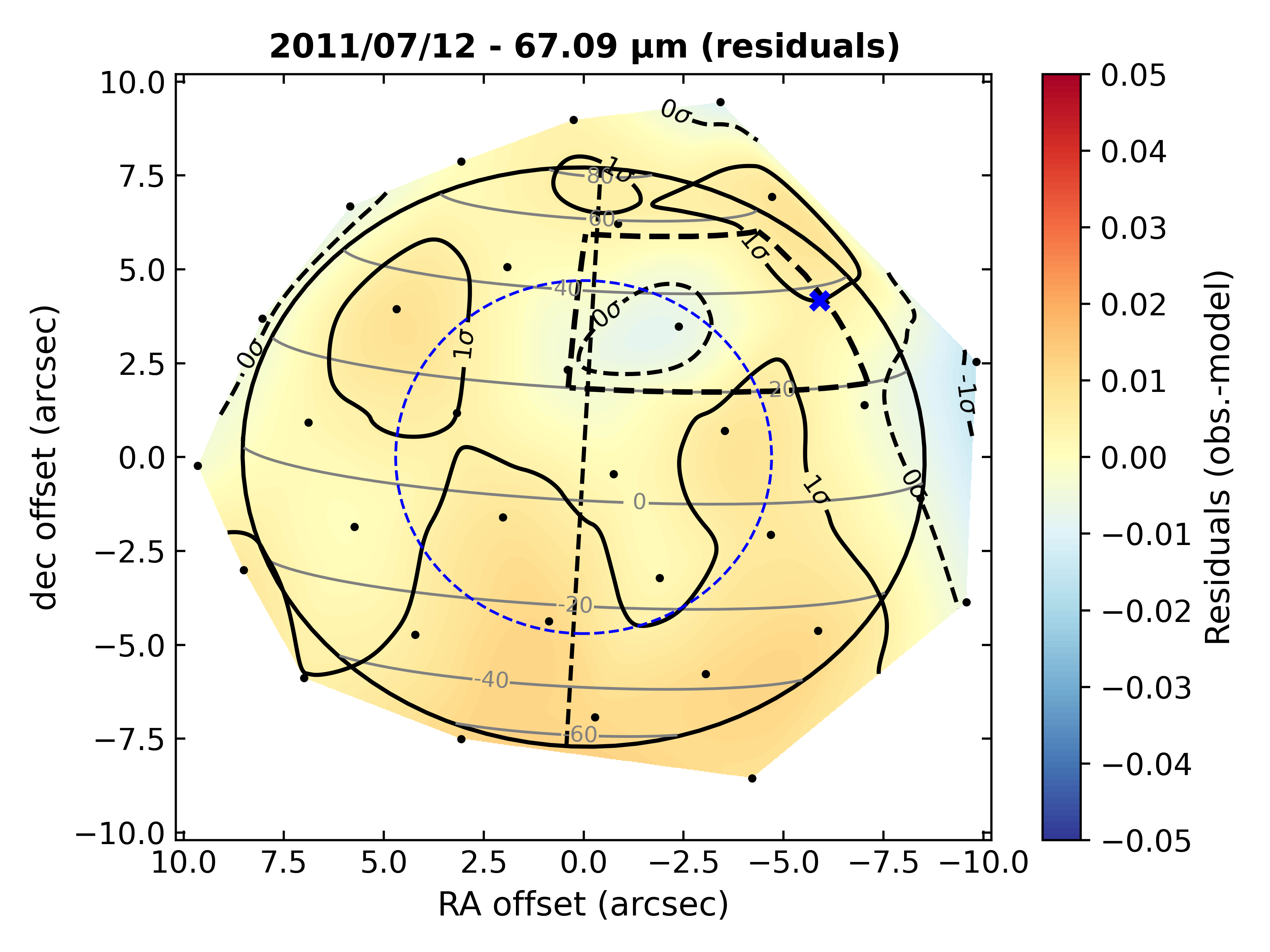}\\
 \caption{Modelled line area maps (\textit{left}) and residual maps (\textit{right}) for the first window (July 2011) and corresponding to the following models: (\textit{first row}) Step 1, in which we keep the Gaussian-latitude-dependent profiles of \citet{Cavalie2019} with no water vapour in the beacon new layers delimited by the pre-storm condensation level at 2 mbar and the new condensation level ; (\textit{second row}) Step 3, in which we set the water vapour abondance as a constant free parameter in the beacon to fit the observations. The best fit of July 2011 is presented here. The `x' markers in the maps at RA=-5.9\arcsec, dec=4.2\arcsec (i.e. in the beacon) correspond to the line of sight for which the observed and modelled spectra are shown in Fig. \ref{fig:Spectres_modeles}. The contours in the residuals are given in units of $\sigma$. The solid contours refer to positive residuals and the dashed contours indicate negative residuals. The results for the six other dates are presented in Fig. \ref{fig:Simu_T} for Step 1, and Fig. \ref{fig:Simu_best_fits} for Step 3.}
 \label{fig:Simu_resultats_ALL}
 \end{figure*}

\begin{figure}[!ht]
 \centering
 \includegraphics[width=0.7\textwidth]{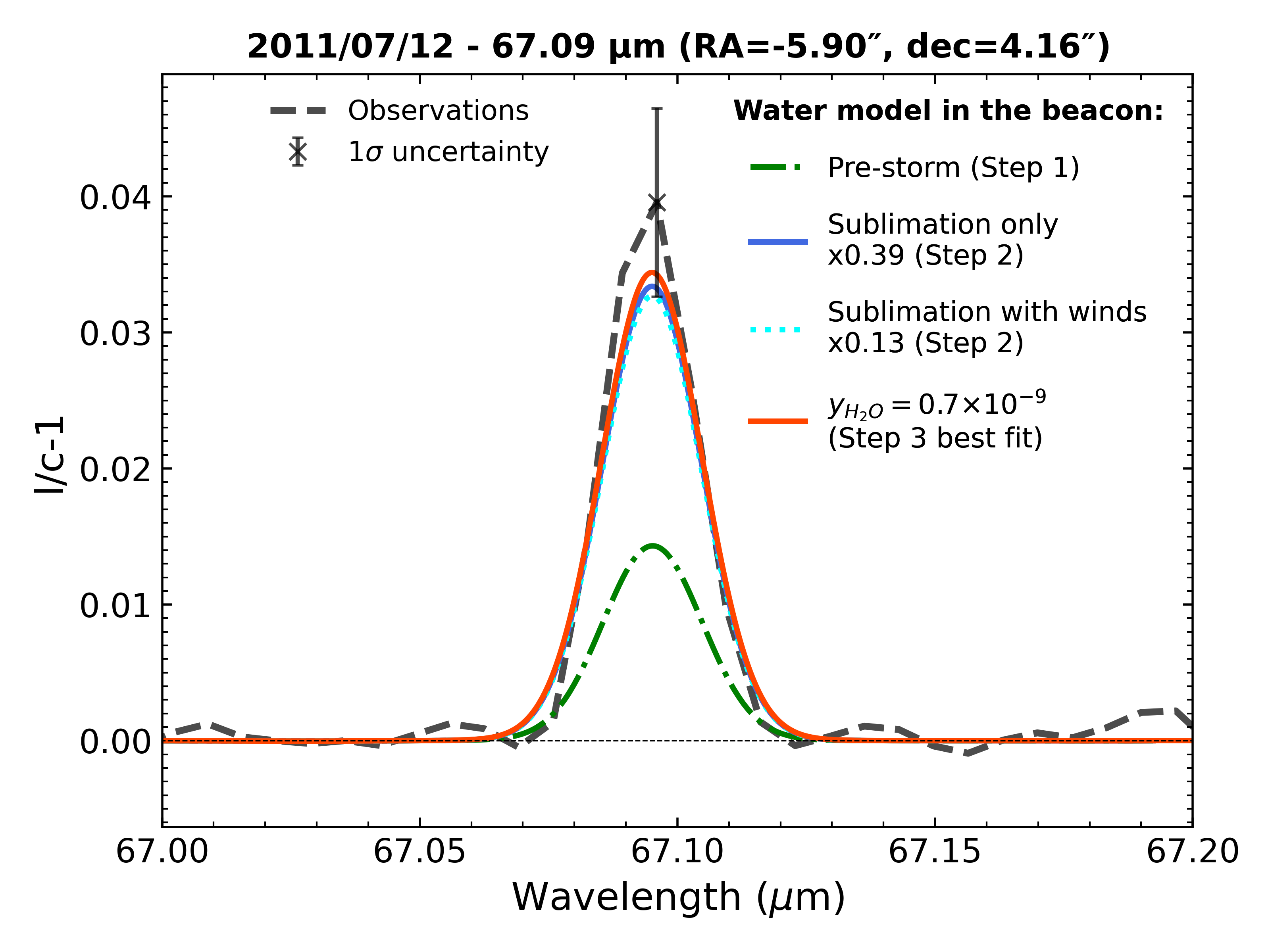}
 \caption{Example of observation and models of the water line at 67.09 $\mu$m at the centre of the beacon in July 2011, at the relative pointing coordinates RA=-5.90\arcsec, dec=4.16\arcsec (corresponding to the x-marker position in Fig. \ref{fig:Simu_resultats_ALL}). The spectrum is expressed in terms of continuum-substracted line-to-continuum (l/c-1). The observed spectrum is in dashed grey lines. It is obtained after the calibration and baseline removal steps presented in Section \ref{part:Data_reduction}. The dashed-dotted green line corresponds to a model in which only the temperature increase in the beacon is accounted for (Step 1 - see Section \ref{part:Step_1}). The solid blue line and dotted cyan line represent the rescaled photochemical models (Step 2 - see Section \ref{part:Step_2}) in which we consider only the haze sublimation (rescaled by a factor of 0.39), then we add the downward winds to the sublimation (rescaled by a factor of 0.13), respectively. The red line depicts a model in which the water mole fraction is fitted in the beacon as a single free parameter (Step 3 - see Section \ref{part:Step_3}). The black point with error bars corresponds to the $1\sigma$ line amplitude uncertainty of about 15-20\%, mostly caused by the baseline removal. This value is different from the line area uncertainty (see Section \ref{part:Data_reduction}).}
 \label{fig:Spectres_modeles}
 \end{figure}

	The modelled and the residual maps (which corresponds to the difference between the observed maps of Fig. \ref{fig:Line_area_maps} and the modelled maps of Fig. \ref{fig:Simu_resultats_ALL} top left panel) are shown in Fig. \ref{fig:Simu_resultats_ALL} (first row) as an example for the first window. All seven modelled and residual maps are presented in Fig. \ref{fig:Simu_T}. Water emission is enhanced locally around the beacon because of the higher temperatures, as seen on the modelled maps. The $\chi^2/N$ values, computed within a beam around the beacon region, range from 11 to 23; that is, $\sim3-5\sigma$. There is a strong lack of water emission around the beacon position within the seven modelled maps, up to 7$\sigma$ for the first window. To illustrate this point, Fig. \ref{fig:Spectres_modeles} shows the result for the spectrum located closest to the beacon centre for the first date. The location of this spectrum is indicated by a cross in Fig. \ref{fig:Simu_resultats_ALL}.
 
	The increase in temperature alone, coupled with the nominal water abundance field in quiescent conditions, thus fails to reproduce the observed water maps. The Saturn stratospheric water distribution results from the continuous infall of material from the Enceladus torus and the rings. \citet{Moses2000b} and \citet{Ollivier2000} estimated the flux of infalling oxygen in the upper atmosphere of Saturn at $\approx 10^{6}$ cm$^{-2}$.s$^{-1}$. More recent work from \citet{Hartogh2011a} found an infalling flux in Saturn’s atmosphere of $\approx 6 \times 10^{5}$ cm$^{-2}$.s$^{-1}$ from 2009-2010 \textit{Herschel} observations. After January 2011 and within the beacon, this flux would have continued to diffuse downward from the pre-storm condensation level to the deeper condensation level, enhancing locally the water column density. The typical water column density at the beacon latitude (35\degre N) in January 2011 is $\approx 0.3 \times 10^{15}$ cm$^{-2}$ and is not enough to reproduce the storm conditions. The photochemical model of \citet{Moses2015} predicts an increase in the water column by a factor of 8.5. It would take $\approx 130$ years to increase the nominal water column by a factor of 8.5 in the beacon with a constant influx rate of $6 \times 10^{5}$ cm$^{-2}$.s$^{-1}$. This timescale far exceeds the several months gap between the observations of January 2011 and our dataset. We can thus conclude that the enhanced water emission around the beacon position cannot be explained by the supplementary materiel coming from the continuous infall of water in the several months separating the January data and our dataset. Thus, the extra water abundance in the beacon must come from the stratospheric water haze sublimation and/or the effect of vertical downwelling winds. We could not think of other known mechanism that could significantly increase the water abundance in the stratosphere at the millibar level.

 \subsection{Step 2: Photochemical models}
 \label{part:Step_2}
 
    The exceptional increase in temperature alone is not sufficient to explain the observed line areas. We thus need to add an extra amount of water vapour to increase the water emission in our radiative transfer calculations. In this section, we explore two different water vertical profiles in the beacon based on the photochemical and transport coupled models of \citet{Moses2015}. They predict a factor-3 increase in water column density from the haze sublimation and a factor-2.8 increase from vertical transport (see Section \ref{part:H2O_profiles}) in the hot beacon core of May 4 2011 compared to their mid-2010 pre-storm conditions. A total increase in the water column by a factor of 8.5 is thus predicted by their model. Haze sublimation alone is tested first in the beacon in Section \ref{part:Step_2-1}, then the vertical downwelling winds are added in Section \ref{part:Step_2-2}. The vertical profiles, taken from \citet{Moses2015}, are presented in Fig. \ref{fig:Profils_Moses}. The general method to adjust these profiles in our radiative transfer calculations is discussed in Section \ref{part:Step_2-0}.

\subsubsection{Building the water vapour 3D-field}
\label{part:Step_2-0}

       The water vertical profiles can only be tested on our first observation window (July 12 2011) which is the closest in time to the hot beacon core stage of May 4 2011 on which the work of \citet{Moses2015} is based. The thermal field changed between the two dates, but the maximal temperature difference of 20 K is located at 2 mbar, a pressure level well above the new condensation level. This difference at 2 mbar has thus no impact on the water profile. The only important pressure level to consider is the condensation level, whose vertical position influences the vertical range where water can be in the vapour phase. According to \citet{Moses2015}, the water condensation level in the hot beacon core lies at $\approx9$ mbar. This level shifts to $\approx12$ mbar in July 2011 according to the \citet{Fletcher2012} profiles (see Fig. \ref{fig:Thermal_fields_profiles_CIRS_et_final} b). A deeper condensation level would eventually lead to an even higher water column density in the beacon if we extend their `sublimation' profile towards the condensation level of July 2011. The dynamic behaviour of the beacon during the 3 months separating the two dates is unknown, but if the vertical downwelling winds profile found in \citet{Moses2015} is still valid in July 2011, we can also expect even more water column in the beacon region from the continuous supply of the upper transport. In a first approximation, we use the profiles of May 2011 to evaluate the physical processes in the beacon in July 2011.
 
      The two water vertical profiles originate from a pre-storm profile obtained from a single disc-averaged observation of water in Saturn's stratosphere \citep{Moses2000b}. Our observations are latitude-dependent and therefore we first need to adapt the photochemical profiles to build a coherent 3D field as a function of latitude. The background 3D field is built in two steps, as follows: 1) the pre-storm profile is rescaled as a function of latitude to match the Gaussian distribution of equation \ref{eq:gaussian} at 1 mbar, then 2) this global latitude-dependent field is multiplied by a constant factor to fit the water map of January 2 2011. Details on the thermal field, on the observation strategy and geometry for this date are presented in \citet{Cavalie2019}. Radiative transfer calculations were performed in the same configuration as in \citet{Cavalie2019}, except for the 3D water field. We find a best global fitting factor of 2.1, which is equivalent to multiplying the pre-storm profile by a factor of 1.5 at the equator and by a factor of 0.5 at 35\degre N. The rescaled vertical profiles at the equator, 35\degre N and 50\degre N, are illustrated in Fig. \ref{fig:Profils_Moses}. Because we use a single vertical profile for all latitudes, the resulting 3D field does not follow the latitude dependency of the temperature field, which results in varying condensation levels, especially in the beacon where the condensation line was deeper in July 2011 than in May 2011. These vertical profiles are thus no longer physical in the condensation region. This constitutes a limitation to the method of building the background field.
      
\begin{figure}[!ht]
 \centering
 \includegraphics[width=0.48\textwidth]{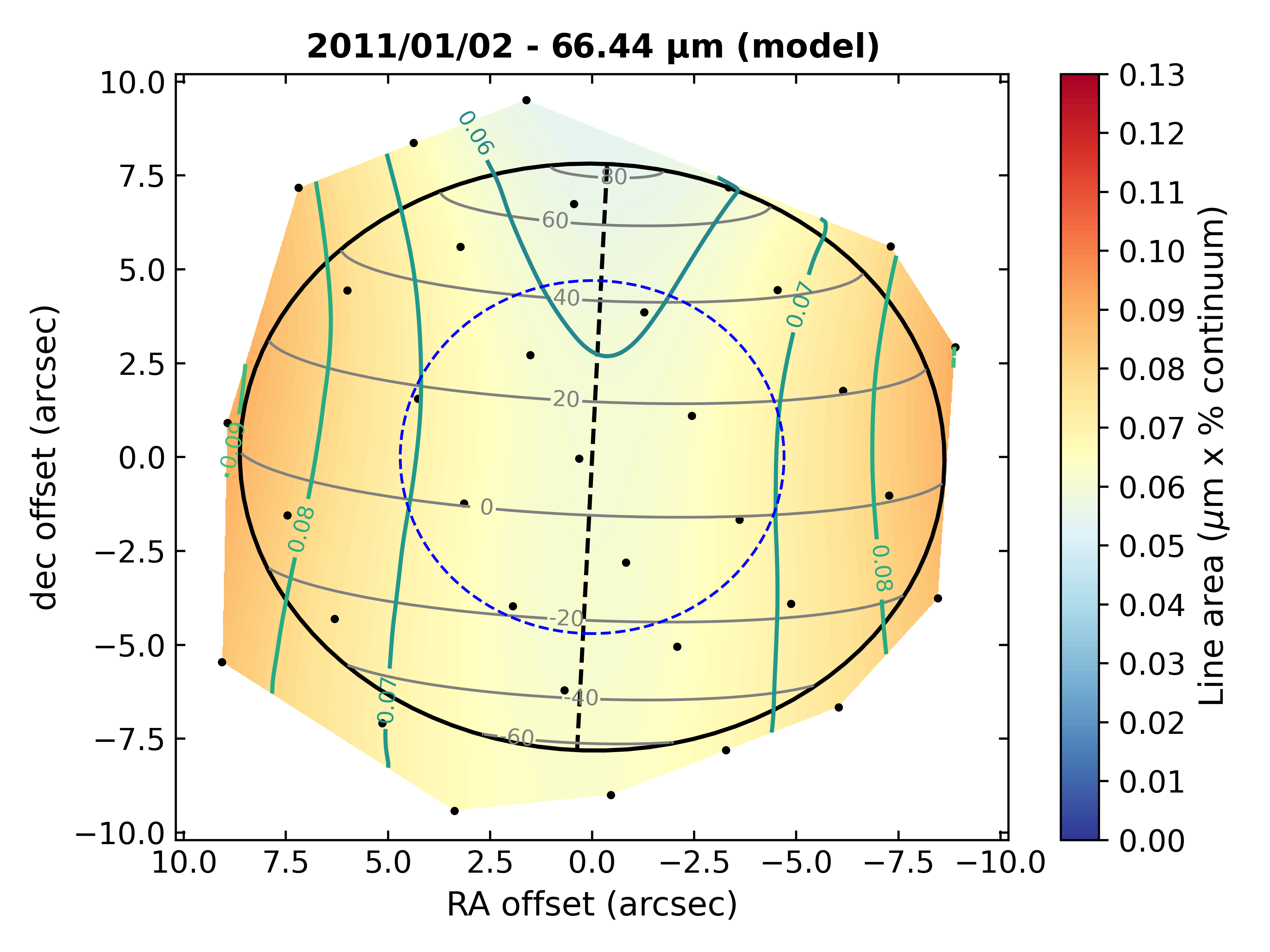}
 \includegraphics[width=0.48\textwidth]{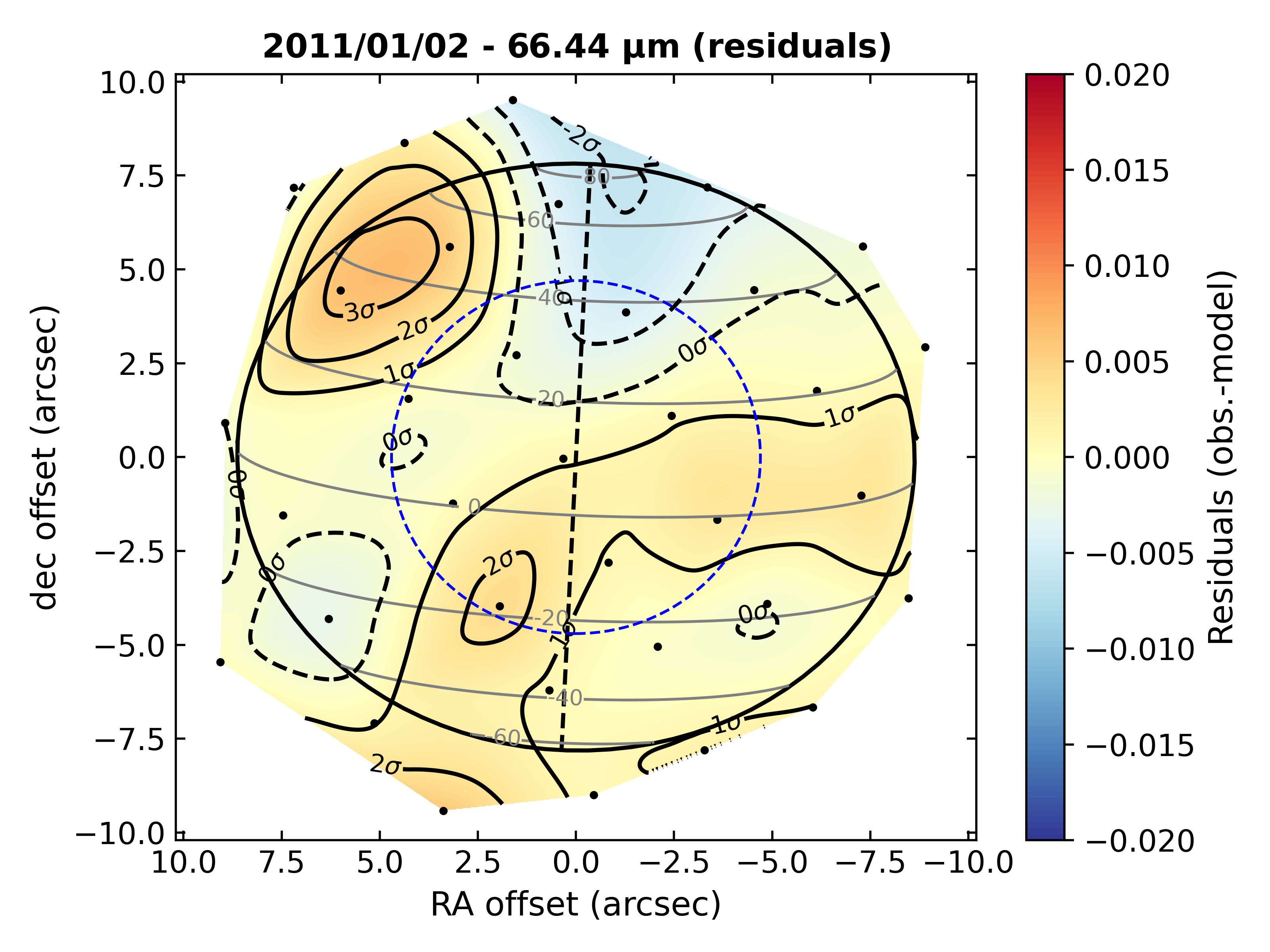}
 \caption{Modelled line area map (\textit{top}) and residual map (\textit{bottom}) for the best fit of January 2 2011 data \citep{Cavalie2019} with the pre-storm model of \citet{Moses2015}. The residual map corresponds to the difference between the observed map of \citet{Cavalie2019} (Fig. 2) and the modelled map. The contours are given in units of $\sigma$. The solid contours refer to positive residuals and the dashed contours indicate negative residuals. The overall description of the maps is the same as in Fig. \ref{fig:Line_area_maps}.}
 \label{fig:Fit_20110102_Moses}
 \end{figure}

      The water column calculated at the beacon latitude (35\degre N) is $0.91\pm0.07 \times 10^{15}$ cm$^{-2}$. Fig. \ref{fig:Fit_20110102_Moses} shows the best fit from radiative transfer calculations and the corresponding residuals. Even if the condensation levels are not well represented here as a function of latitude, a good agreement with the January 2011 data is found and thus the background field can be used for the next steps.
            
     The global field outside the beacon is set with this rescaled pre-storm profile for July 2011. Inside the beacon region delimited in latitude and longitude following data found in Table \ref{tab:Beacon_data}, we use the `sublimation-only' or `sublimation with vertical winds' models, and then we rescale the water field in the whole beacon with a fitting factor. We do not perform any Gaussian rescale as a function of latitude to remove the unknown latitude dependency in the beacon.

 \subsubsection{Sublimation only}
 \label{part:Step_2-1}
 
    We explore in this section the first model of \citet{Moses2015} accounting only for the haze sublimation, with the previous water vapour background 3D-field. The modelled water map as well as the residuals are shown in Fig. \ref{fig:Fit_20110712_Moses} (top two panels). The water line emission is enhanced around the beacon position in this model because of two related effects: 1) the increase in temperature in the millibar layers, as in Section \ref{part:Step_1}, and 2) the additional water vapour column density resulting from the haze sublimation at a few mbars. The enhanced emission is not centred on the beacon for geometrical reasons, and because of the large beam size and the planet's rotation during the observing time that spreads out the emission. An example of a modelled spectrum in the beacon is presented in Fig. \ref{fig:Spectres_modeles}, at the pointing depicted by a blue `x' symbol on Fig. \ref{fig:Fit_20110712_Moses}. The best fit is obtained for a scaling factor of $0.39^{+0.21}_{-0.16}$ with $1\sigma$ uncertainties. The corresponding water vertical profile is shown in Fig. \ref{fig:Profils_Moses} with the purple area. The $\chi^2/N$ value calculated within one beam around the beacon is 1.1. This model gives a good agreement with the observations after the proper rescaling. The column density at the beacon centre (35\degre N-55\degre W), calculated from 100 mbar to the top of the atmosphere, is 2.0$^{+1.0}_{-0.8} \times 10^{15}$ cm$^{-2}$ with this July 12 2011 `sublimation-only' model.

 \subsubsection{Sublimation with vertical winds}
 \label{part:Step_2-2}

\begin{figure*}[!ht]
 \centering
 \includegraphics[width=0.48\textwidth]{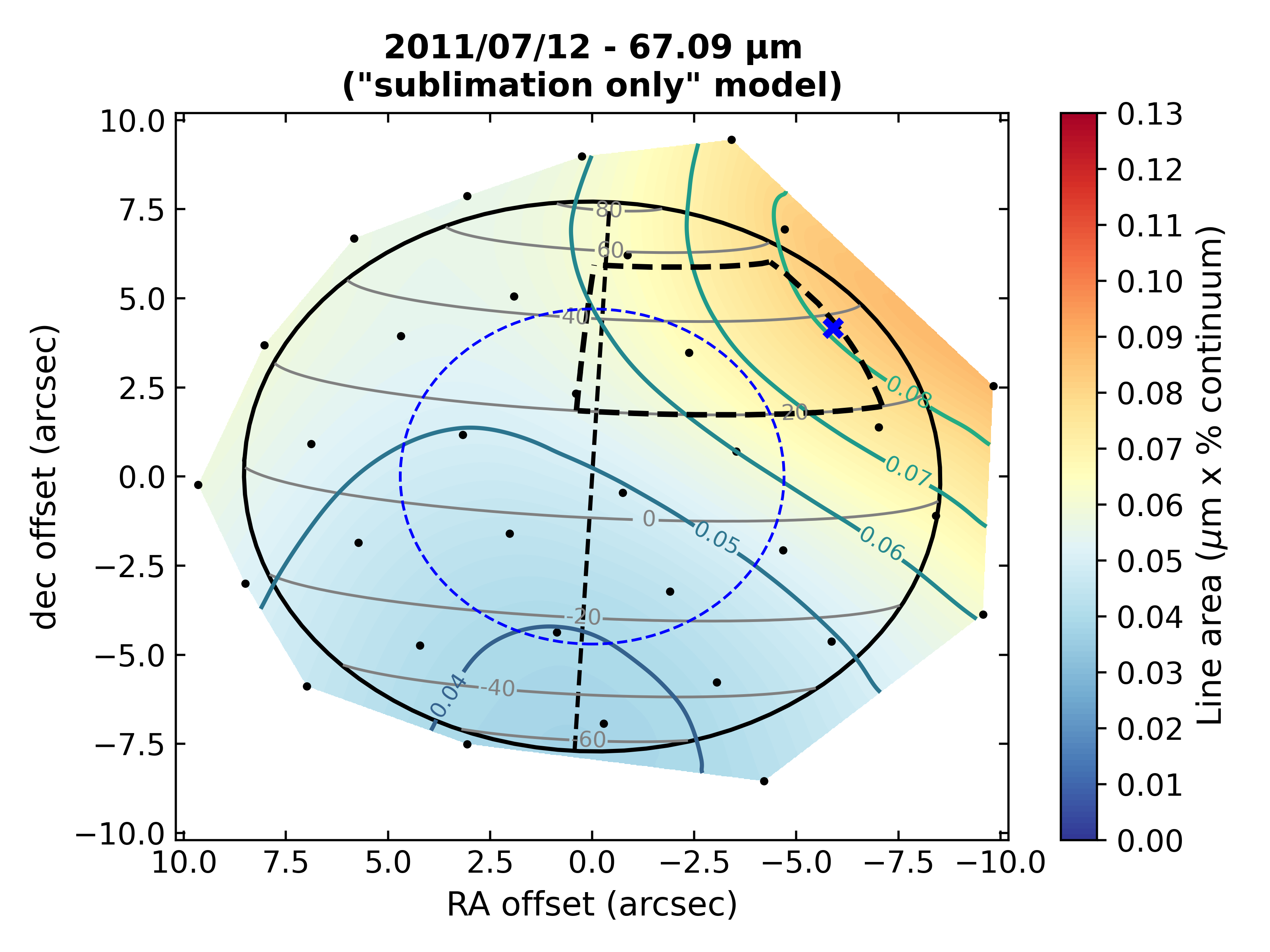}
 \includegraphics[width=0.48\textwidth]{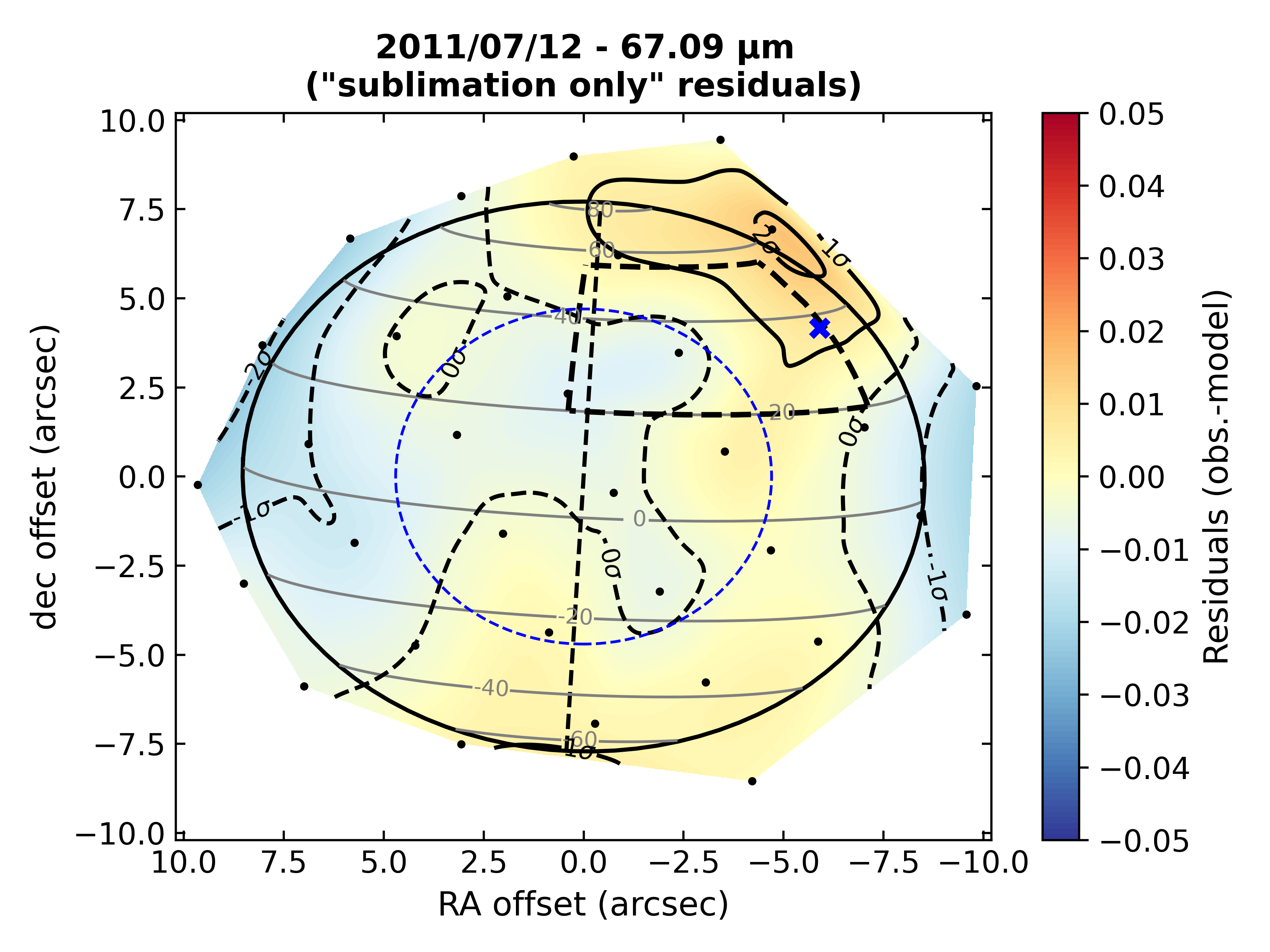}\\
 \includegraphics[width=0.48\textwidth]{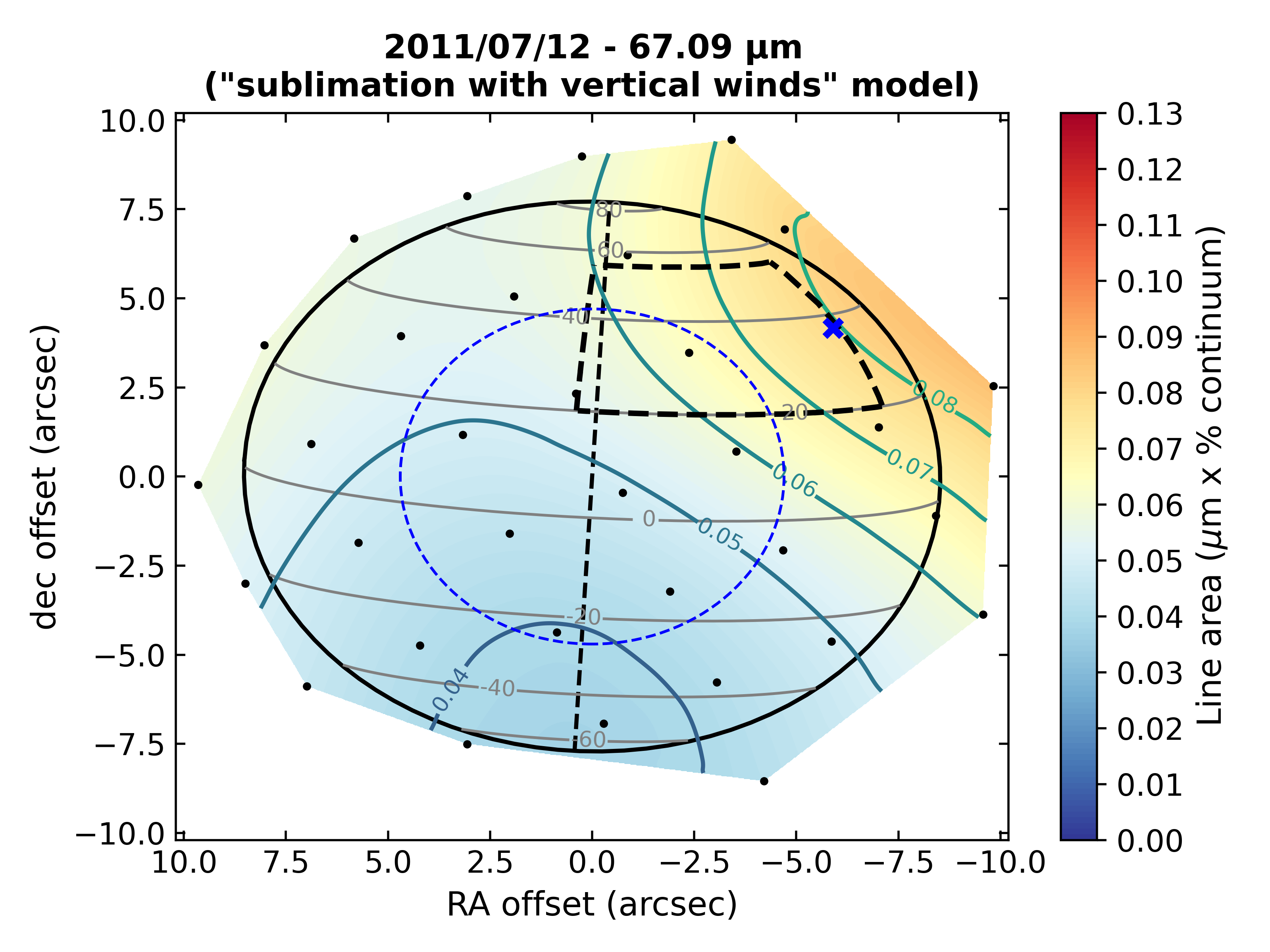}
 \includegraphics[width=0.48\textwidth]{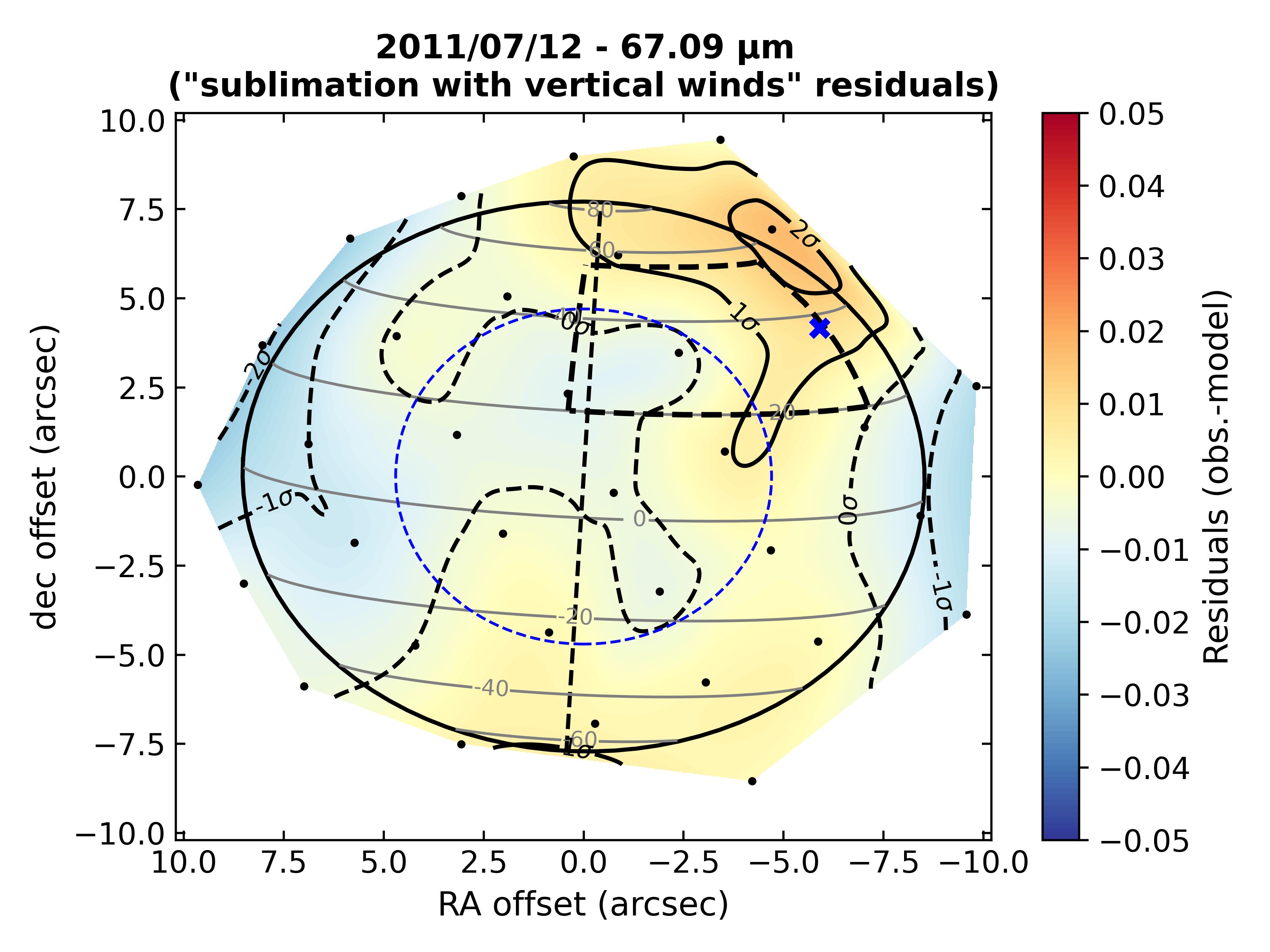}
 \caption{Modelled line area maps (\textit{left}) and residual maps (\textit{right}) for the two water models of \citet{Moses2015}: a first water profile considering only the sublimation of the water haze in the beacon (\textit{top}) and a second model in which the vertical downwelling winds are added on top of the haze sublimation (\textit{bottom}), both in the hot beacon core model of May 4 2011. The two profiles were used only on our first observation window (July 12 2011) which is the closest water observation to the hot beacon core model. The two profiles were copied uniformly in the beacon region and multiplied by a factor to fit the data. The background water field is derived from the fit of the January 2011 data with the pre-storm profile. The contours are given in units of $\sigma$. The solid contours refer to positive residuals and the dashed contours indicate negative residuals. The `x' markers at RA=-5.9\arcsec, dec=4.2\arcsec (i.e. in the beacon) correspond to the line of sight for which the observed and modelled spectra are shown in Fig. \ref{fig:Spectres_modeles}. The overall description of the maps is the same as in Fig. \ref{fig:Line_area_maps}.}
 \label{fig:Fit_20110712_Moses}
 \end{figure*}

   We now add the effect of vertical downwelling winds on the water vapour profile. Fig. \ref{fig:Fit_20110712_Moses} (bottom panels) presents the radiative transfer results with this second model. A spectrum corresponding to a line of sight in the beacon is given as an example in Fig. \ref{fig:Spectres_modeles}. The best fit corresponds to a scaling factor of $0.13^{+0.08}_{-0.06}$ with $1\sigma$ uncertainties and is obtained for a $\chi^2/N$ value of 1.1 around the beacon. The vertical profile associated with the best fit is illustrated in Fig. \ref{fig:Profils_Moses} with the grey area. The column density computed at the beacon centre is 1.9$^{+1.1}_{-0.8} \times 10^{15}$ cm$^{-2}$. It is consistent with the rescaled `sublimation-only' model column density. This was expected as we fit the right amount of water vapour in the sensitivity pressure region. The increase ratio between the storm fitted models and the pre-storm fitted model lies between 1.0 and 3.4.

   The previous results highlight that a solution is found from two different water vapour vertical profiles, which are derived from different physical assumptions (see Fig. \ref{fig:Fit_20110712_Moses}). The column densities inferred using the two vertical profiles are essentially the same, at about 2.0$ \times 10^{15}$ cm$^{-2}$. From Fig. \ref{fig:Profils_Moses}, we see that the fitted profiles from the two models overlap in the altitudinal range from which most of the line emission is formed (see the example of the July 2011 Step 3 sensitivity range in Fig. \ref{fig:Fonctions_contributions}, which is essentially the same here). This confirms that the right amount of water needed to reproduce the observations is fulfilled in this region. However, the two profiles do not match for levels lower than 0.1 mbar, and are also both incoherent with the rescaled background field at 35\degre N, but we are not sensitive at these high altitudes anyway with our observations.
   
   An extra amount of water vapour in the beacon is expected from vertical winds, as was observed for other molecules. All molecules should be affected similarly by the vertical winds, the only difference with other molecules is that water is a condensable species in the stratosphere, and some water vapour might be added on top of the winds contribution from the haze sublimation. These results show 1) the limitation of our dataset in terms of vertical sensitivity, and 2) the lack of knowledge on the beacon dynamical behaviour and chemical behaviour.
   
   Firstly, the lines are far from being spectrally resolved (the spectral resolution is about 100 times the natural line width). As was shown with the previous results, we cannot retrieve the water vertical profile in the beacon, and multiple solutions are acceptable. What really matters is the amount of water vapour put in the sensitivity region delimited in the altitudinal range from $\sim$1 mbar to $\sim$10 mbar (see the water contribution functions in Fig. \ref{fig:Fonctions_contributions}). The primary conclusion is that, due to the PACS limited spectral resolution, our sensitivity is restricted to a single parameter, the water column density within the beacon. We are also limited by the spatial resolution which is half the planet size, thus it is unnecessary to put a multi-parameter latitudinally dependent model of water in the beacon.
 
   Secondly, we are also limited by our lack of knowledge on the beacon which is a dynamic and time-evolving structure. Our observations are space and time dependent. Important physical variables such as the vertical mixing $K_{zz}$, are not constrained as a function of space and time in the beacon. Considering the vertical winds, the water excess transported from above is strongly dependent on the pre-storm water gradient in the stratosphere (see Fig. \ref{fig:Profils_Moses}), which in turn depends on $K_{zz}$. Moreover, the distribution of water ice particles in the stratospheric haze has never been observed before, which means that the quantity of water stored in this haze (and consequently the quantity released by the sublimation during the storm) can only be estimated with models. The ice distribution depends on the influx of water above the condensation level, on condensation, on the local distribution of aerosols on which water can stick, on coagulation, gravitational settling, evapouration from thermal fluctuations. Then, the evolution of the water vapour present in the new condensation region depends on an equilibrium between sublimation and re-condensation, and also depends on the vertical winds. All free parameters and unknowns of complex water models can therefore not be constrained with our spectrally and spatially limited dataset.
   
    In the next section, we propose a simpler approach, in which we only fit a single parameter, the water mole fraction (or, equivalently the water column density) in the beacon as a function of time. Finally, using a cloud model, we give a first estimation of the relative magnitude of sublimation versus vertical winds as the source of the extra water vapour seen in the beacon in Section \ref{part:cloud}.

 \subsection{Step 3: Fitting the water column density in the beacon}
 \label{part:Step_3}
 
   In these sets of empirical models, the water mole fraction inside the beacon is set as a free parameter, constant with altitude from the new local condensation level to the top of the atmosphere in the beacon. From the previous sections, we might expect an enhancement of water vapour from haze sublimation and/or vertical winds, which both tend to increase locally the water mole fraction above the new condensation level. We thus extend the empirical profile down to the new condensation level in the beacon. The corresponding contribution functions are displayed in Fig. \ref{fig:Fonctions_contributions}. They show the limited range of pressures ($\sim$1-5 mbar) from which the emission is produced; that is, just above the new condensation line in the beacon. The contribution to the line decreases drastically outside this pressure range. We can thus consider the value of the fitted water mole fraction as a fitting data point located at the pressure level where the contribution function peaks, and valid within a range of a few mbars around this peak. The water vertical profile corresponding to the water abundance in the beacon that best fits the first observation window is illustrated as an example in Fig. \ref{fig:Profils_eau}. The extension of this constant profile towards higher altitudes (i.e. typically for pressures lower than 0.1 mbar) has little impact on the water line, because the line is not so sensitive to these high altitudes.
 
  Uncertainties on the thermal field of $\pm 2$ K lead to an uncertainty on the location of the condensation level. Its exact position is important for two reasons: 1) we take a constant water mole fraction from the condensation level towards the top of the atmosphere, and 2) we still have some sensitivity down to this level, thus a deeper level means an extra column density contributing to the line. We estimated that the uncertainty on the thermal field leads to an uncertainty of about 30\% on the water mole fraction. This uncertainty was added quadratically to the fitting uncertainty.

   Radiative transfer calculations and residuals are presented in Fig. \ref{fig:Simu_best_fits} for the best fits for the seven water maps, and as an example for the first window in Fig. \ref{fig:Simu_resultats_ALL}. The spectrum which corresponds to the best fit is represented in Fig. \ref{fig:Spectres_modeles} for a line of sight in the beacon for the first window. The $\chi^2/N$ values, computed within a beam around the 2D extent of the beacon, are 0.6, 0.7, 1.2, 1.1, 0.4, 1.8 and 2.4, following the time order of Table \ref{tab:Obs_list} (July 2011 at 67.09 $\mu$m, February 2012 at 66.44 and 67.09 $\mu$m, July-August 2012 at 66.44 and 67.09 $\mu$m, January-February 2013 at 66.44 and 67.09 $\mu$m). The best-fitting water abundances are $0.7^{+0.3}_{-0.3}$, $1.1^{+0.5}_{-0.5}$, $1.2^{+0.7}_{-0.6}$, $1.4^{+0.8}_{-0.6}$, $2.1^{+1.2}_{-1.0}$, $1.1^{+0.5}_{-0.4}$, and $1.4^{+0.7}_{-0.6}$ $\times 10^{-9}$ ($1\sigma$ uncertainties), respectively. For a given time window, the observations at 66.44 $\mu$m and 67.09 $\mu$m give results that are compatible within error bars. The differences seen in the nominal values result from the difference in line amplitude. As the 67.09 $\mu$m line is 30\% less intense than the 66.44 $\mu$m one, water line area maps suffer from higher uncertainties at 67.09 $\mu$m and therefore a broader range of solutions is possible. The larger uncertainties of the July 2012 derived values are caused by the different observation geometry. The beacon was close to a pure nadir geometry, as opposed to the limb geometry of the other observations, and the observations are thus less sensitive to the water abundance.

   A good agreement between the model and the observations is found for each of the seven maps, but some details are important to notice. First, as mentioned above, the latitudinal extent of the beacon was not known for the last window (January-February 2013). The residual maps at 66.44 and 67.09 $\mu$m show a lack of emission in the model towards higher latitudes, but suffer from too strong an emission towards the equator. It probably results from our assumption on the latitudinal extent of the beacon at this point in time. Indeed, we do not have measurements of the temperature as a function of latitude in January-February 2013 and assumed the same latitudinal extent as in August 2012 (see Fig. \ref{fig:Thermal_fields_profiles_CIRS_ALL}). We may thus have over/under-estimated the temperature as a function of latitude at the northern and southern edges of the beacon and thus the size of the region in which we vary the water mole fraction. Being closer in time to the end of life of the beacon (see Fig. \ref{fig:Donnees_beacon_position_extension}), it may have been less extended in January-February 2013 than in August 2012. Reducing the latitudinal extension of the beacon and shifting its position towards northern latitudes improve the fits, however the water abundance needed to fit the observations stays relatively the same. It may indicate that the beacon moved towards higher latitudes at the end of its life, but we simply lack the observational constraints to conclude on this point.
 
   In addition, the water vertical profile within the beacon is certainly more complex than the constant water mole fraction profile we adopted in this paper, as shown by the photochemical profiles used in Section \ref{part:Step_2}. Our results still give a good estimate on the quantity of water present in the layers of sensitivity (at a few mbars in the beacon) through time. As is shown in Section \ref{part:Step_2}, it is unnecessary to make the model more complex with more free parameters because of the limited spatial and spectral resolutions of the observations. Even if the empirical and the photochemical water profiles are based on very different assumptions, the water mole fractions from the two approaches are consistent in the sensitivity layers. While we find a water vapour abundance from an empirical fit of $(0.7\pm0.3)\times 10^{-9}$ for July 2011, the two photochemical models overlaps precisely in this water mole fraction range at several mbars. This conclusion was, again, anticipated as we fill the sensitivity layers with the right amount of water vapour to fit the observations.
 
\begin{figure}[!ht]
 \centering
 \includegraphics[width=0.7\textwidth]{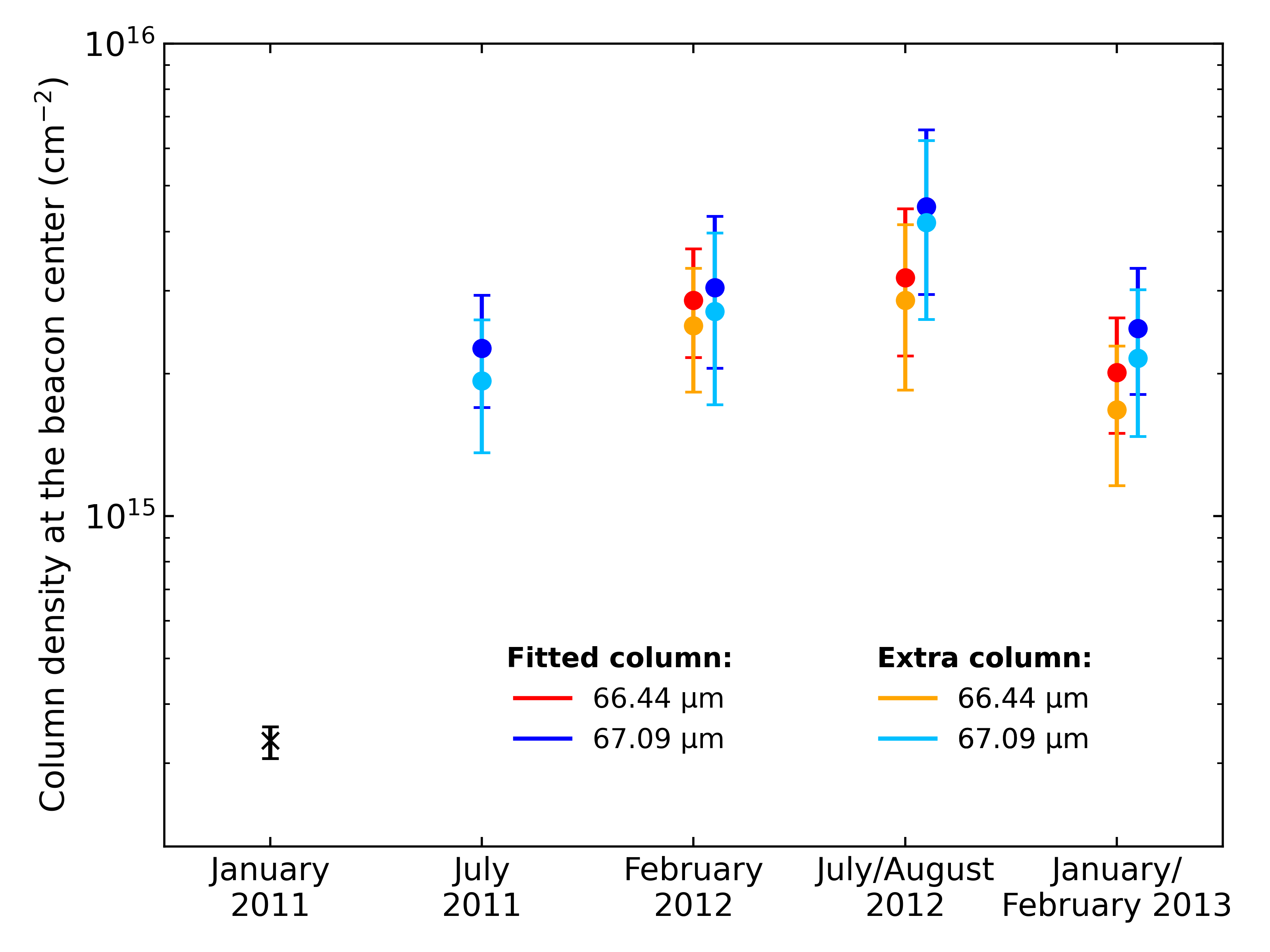}
 \caption{Column density at the beacon centre (35\degre N and central longitude) as a function of time. The column densities are integrated from $\sim$100 mbar towards the top of the atmosphere. The cross point in January 2011 is derived from the background water field of \citet{Cavalie2019} at 35\degre N. Darker coloured points relate to the fitting of the observations with a constant water mole fraction in the beacon. Light coloured points refer to the fitted water column density after subtracting the pre-storm column of January 2011, and therefore show the additional column density coming from a combination of vertical winds and sublimation. The red/orange points correspond to the water maps at 66.44 $\mu$m and the dark blue/light blue to those at 67.09 $\mu$m.}
 \label{fig:Densite_colonne}
 \end{figure}

   Column densities at the beacon centre are plotted over time in Fig. \ref{fig:Densite_colonne} from the previous fitting process. We find, respectively, $2.3^{+0.7}_{-0.6}$, $2.9^{+0.8}_{-0.7}$, $3.0^{+1.3}_{-1.0}$, $3.2^{+1.3}_{-1.0}$, $4.5^{+2.1}_{-1.6}$, $2.0^{+0.6}_{-0.5}$, and $2.5^{+0.9}_{-0.7}\times 10^{15}$ cm$^{-2}$, following the time order of Table \ref{tab:Obs_list}. It remains relatively constant (within the error bars) across the monitored time frame, at a mean of (2.5$\pm$0.3)$\times 10^{15}$ cm$^{-2}$. The pre-storm column found in January 2011 at 35\degre N with a similar method of empirical models is 0.34$^{+0.02}_{-0.03}\times 10^{15}$ cm$^{-2}$. Using similar empirical models for the water vertical distribution, we find that the water column is enhanced by a mean factor of (7.5$\pm$1.6) in the beacon compared to pre-storm conditions at the same latitude. The subtraction of the pre-storm column to the fitted column is also represented in Fig. \ref{fig:Densite_colonne} and shows the extra column density needed to reproduce the enhanced water emission seen in our \textit{Herschel}/PACS maps.

\section{Preliminary cloud modelling}
\label{part:cloud}
   
    Instrumental limitations prevent us from retrieving the water vertical profile and thus constraining directly the respective contributions of the vertical winds and the haze sublimation to the extra column density we observe in the beacon. But, cloud models, such as the PlanetCARMA (Community Aerosol and Radiation Model for Atmospheres) model \citep{Barth2020}, can be used to try to evaluate the contribution of sublimation. It couples aerosols microphysics and radiative transfer modelling to infer haze formation in a planetary atmosphere and compute, for example, the particle size and mass distributions. Important physical processes at play in hazes, like condensation, evapouration, coagulation, nucleation and sedimentation, are included in this model. More details can be found in \citet{Barth2020}.
    
    We have configured this model for this study in two cases prevailing for July 12 2011 conditions: (i) at the beacon mid-latitude but outside the beacon (35\degre N-115\degre W), and (ii) at the beacon centre (35\degre N-55\degre W). The external water vapour taken as an initial condition at the top of the atmosphere is set as the water mole fraction of the \citet{Cavalie2019} model at the central latitude of the beacon (35\degre N) in both cases (i) and (ii). This water vapour influx is kept constant until saturation is reached at some point. The water profile then follows the condensation curve. The two model runs are independent and are used here to give an idea of the water ice vertical distribution in quiescent and beacon-like conditions.
    
    Preliminary results obtained for a steady state are presented in Fig. \ref{fig:Profils_eau}. They show both the different altitudes and mass densities of the ice particles in the hazes inside and outside of the beacon at 35°N. The amount of water ice available for sublimation in the beacon is then simply obtained by integrating the difference between the beacon profile and the quiescent one from the condensation level in the beacon upwards; that is, from $\sim$12 mbar (beacon condensation level) to $\sim$2 mbar (quiescent condensation level). The resulting column density of sublimated water is 1.2$\times 10^{15}$ cm$^{-2}$. It corresponds to 45-85\% of the extra water vapour column density found at the beacon centre in July 12 2011, with a nominal value of 60\% and 1$\sigma$ uncertainties derived from the extra column calculations of Step 3 models. These preliminary PlanetCARMA results indicate that haze sublimation could be a significant source of water vapour, which could explain most of the extra water vapour we observe in the beacon.

\section{Discussion and conclusion}
\label{part:Discussion} 
 
In this paper, we have presented mapping observations of water vapour in the Saturn stratosphere, recorded by \textit{Herschel}/PACS between July 2011 and February 2013. These data, taken at a cadence of one dataset every 6 months, and thus covering 18 months, enabled us to monitor the water emission of the hot stratospheric vortex (the `beacon') that was formed in early 2011 as a consequence of Saturn's Great Storm of 2010-2011. The water vapour maps show significant a increase in the water emission in the regions around the location of the beacon. The peak of emission enhancement is not exactly at the beacon location because of geometrical effects in the build-up of line opacity. A priori knowledge of the temperature fields from independent \textit{Cassini}/CIRS measurements of \citet{Fletcher2012} pertaining to our observations enables the water abundance in the beacon to be derived from the \textit{Herschel} observations.

Using radiative transfer modelling coupled to the temperature retrievals, we find that the dramatic temperature increase measured in the 0.5-10\,mbar layers within the beacon by \citet{Fletcher2012} with the pre-storm water vapour distribution derived by \citet{Cavalie2019} cannot explain the enhanced water emission recorded by \textit{Herschel}/PACS. We thus demonstrate that we need to increase the abundance of water vapour locally in the beacon to reproduce the stronger water emission in the observations.

\citet{Moses2015} demonstrated that the local increase in hydrocarbon abundance in the beacon is caused by the presence of vertical transport, and we consider that water vapour was affected in a same way as the rest of the atmospheric species. By analogy to hydrocarbons, we may expect that a part of the additional water vapour needed to reproduce our water maps is caused by the vertical winds. This could constitute a first explanation for the higher emission observed in the beacon.

A possible second source of water at the millibar level is water haze sublimation. Indeed, a stratospheric water haze is expected to exist at pressure levels greater than $\sim$2 mbar from the condensation and accumulation over time of the external flux of water vapour \citep{Moses2000b,Cavalie2019} coming from Enceladus geysers \citep{Waite2006,Porco2006,Hansen2006}, as was predicted by models \citep{Moses2000b,Ollivier2000}. Another source of water vapour could then result from the partial sublimation of the local water haze caused by the important increase in the temperature in the millibar pressure range that implied a downward shift of the condensation level in the beacon. For instance, the condensation level shifted from 2 mbar in early 2011 (pre-storm conditions) to 12 mbar in July 2011 in the beacon. This corresponds to a shift in altitude of the order of 100 km. The total amount of water ice in the haze and the distribution of the ice particles has never been constrained by observations, and thus the amount that sublimated in the beacon is not known a priori.

Unfortunately, we cannot retrieve the water vertical profile in the vertical layers of interest because of the limitations of our dataset in terms of both spectral and spatial resolutions. As a consequence, it is difficult to evaluate the relative magnitude of those two sources. This is best illustrated by the equivalently good fits of July 2011 data obtained with the two rescaled models of \citet{Moses2015} that account for sublimation only, on the one hand, and sublimation and vertical winds, on the other hand.

Using a simpler empirical model, in which we fit a constant water mole fraction above the condensation level in the beacon, we find a relatively constant increase in the water column between July 2011 and February 2013. We derive a mean extra column density of (2.2$\pm$0.3)$\times 10^{15}$ cm$^{-2}$, corresponding to an increase over the full temporal coverage of (7.5$\pm$1.6) with respect to pre-storm conditions at the beacon latitude. It is worth noting that the temporal increase tentatively seen in the beacon centre column densities in Fig. \ref{fig:Densite_colonne} from July 2011 to July-August 2012 is not significant as the July-August 2012 window suffers from higher uncertainties due to the beacon's position in nadir geometry. We also remind the reader that the thermal field of the last window in January-February 2013 is not well constrained due to the lack of information about the beacon's latitudinal extent. The decrease seen in Fig. \ref{fig:Densite_colonne} for the last date is thus not well constrained.

When using physically based profiles following the work of \citet{Moses2015}, we find a water column increase of a factor of 1.0-3.4 at 35\degre N in July 2011. This factor is 2-4 times less than the one found with the empirical model. This is caused by the fact that neither the 66.44 nor the 67.09 $\mu$m lines are optically thin. The lines are thus not only abundance-dependent (and therefore column-density-dependent), but also temperature-dependent. As the opacity of the 66.44$\mu$m line is greater than the 67.09 $\mu$m line, the 67.09 $\mu$m line is more sensitive to the water abundance and column density. A good fit is obtained as long as the right amount of water vapour is put in the sensitivity altitudinal range. This explains the consistent water mole fractions found between the empirical model fitting and the two rescaled storm profiles of \citet{Moses2015} at $\sim$3 mbar ($\sim$contribution function peak) in the beacon for July 12 2011.

The fact that the 66.44 $\mu$m line has a stronger opacity than the 67.09 $\mu$m one even produces saturation at the line centre. Physical models with stronger gradients and lower abundances in the vicinity of the condensation layer will require more water at the higher-altitude levels to produce the same line area as the empirical model. At some point, the line centre saturates and obtaining a good fit requires one to add more water, in a non-linear way. Therefore, because of the saturation effect, multiple water columns can lead to the same line area. In addition, there is a discrepancy between the condensation level of the rescaled \citet{Moses2015} pre-storm profile tuned to mid-2010 and the one observed in January 2011. At the beacon latitude of 35\degre N taken as a reference, it results in too low a condensation level pressure ($\sim$1 mbar - see Fig. \ref{fig:Profils_Moses} - vs $\sim$2 mbar in \citealt{Cavalie2019}). A much higher fitting factor is then needed to compensate for the emission which is then missing from these deeper levels. Combining these two effects then causes a factor-of-3 discrepancy between the increase factors derived with the two methods (physical vs. empirical models). This point is directly illustrated by our January 2011 pre-storm fitting in Section \ref{part:Step_2-0}, in which we find a water column 3 times greater with the rescaled physical model of \citet{Moses2015} than the results of \citet{Cavalie2019} at 35\degre N that use an empirical model. The global fitting factor found in Section \ref{part:Step_2-0} is thus overestimated to compensate for the emission which is missing from these deeper levels and because of line saturation. With physical models more accurately considering the varying condensation levels with latitude, the line saturation issue at 66.44 $\mu$m should be attenuated and we should find a lower pre-storm column density. An increase factor closer to, yet probably not matching, the one found with empirical models should then be found.

The overall constant column seen in Fig. \ref{fig:Densite_colonne} should only be taken as indicative of the continued presence of water vapour in the warmer layers. The temporal evolution of the thermal vertical profile in the beacon after our first observations of July 2011 (see Fig. \ref{fig:Thermal_fields_profiles_CIRS_et_final} b) shows that the condensation level progressively moves upwards. One could then expect the water vapour of the lower levels to re-condense progressively, leading to a decrease in the column density. The rather constant column over time seen in Fig. \ref{fig:Densite_colonne} could then be caused by the continued effect of the downward winds within the beacon, adding more water vapour from higher levels to the observed ones given the abundance gradient found in the stratosphere in physical profiles such as those of \citet{Moses2015}. This effect is difficult to quantify, because the temporal evolution of the downward wind speeds is unconstrained, and from the PACS observations we cannot separate its contribution to the haze sublimation. Thus, further modelling of the 3D dynamical processes in the beacon and of the water haze vertical structure is needed to interpret this extra column density over time.

One solution is to look at cloud models to better constrain the contribution of sublimation. \citet{Moses2015} predicted a factor-3 increase in the water column density from the haze sublimation, which represents 27\% of their extra column. In this paper, we used the PlanetCARMA cloud model of \citet{Barth2020} to predict the water ice distribution in the stratospheric haze as a function of altitude in typical atmospheric conditions inside and outside the beacon in the temperature conditions prevailing in July 2011, with the water vapour influx presented in \citet{Cavalie2019}. From a preliminary study of the water ice distributions, we obtain as a first estimate that 45-85\% of the extra water column seen in July 2011 in the beacon centre could result from the haze sublimation, the rest being caused by the vertical transport of upper material to the millibar levels. The difference between the sublimation profile of \citet{Moses2015} and this study resides in how sublimation and evaporation are handled. The PlanetCARMA model accounts for aerosol microphysics and radiative transfer to produce a more complete picture of the processes occurring in the water condensation region than the parametrized technique used in \citet{Moses2015}. Nevertheless, both indicate that haze sublimation plays a significant role in explaining the water abundance increase we derived from the observations.

Three-dimensional time-dependent models of the storm, including the effects of vertical winds, haze sublimation, and recondensation, would help us to more precisely quantify the two counterparts at the origin of the extra column density of water vapour retrieved in the beacon over time. The quantity of water ice stored in the water haze results from an equilibrium between condensation and sedimentation to deeper levels in quiescent conditions. This, in turn, depends on the size and mass of the icy particles. Future work using aerosol microphysical models such as that of \citet{Barth2020} applied to the results of this paper may unveil some of the properties of Saturn's stratospheric water haze and, more generally, complete our understanding of stratospheric hazes \citep{Guerlet2015,Fletcher2023b}.

\section*{Acknowledgements}
C. Lefour and T. Cavali\'e were supported by the Programme National de Plan\'etologie (PNP) of CNRS/INSU and by the Centre National d'\'Etudes Spatiales (CNES). PACS has been developed by a consortium of institutes led by MPE (Germany) and including UVIE (Austria); KU Leuven, CSL, IMEC (Belgium); CEA, LAM (France); MPIA (Germany); INAF-IFSI/OAA/OAP/OAT, LENS, SISSA (Italy); IAC (Spain). This development has been supported by the funding agencies BMVIT (Austria), ESA-PRODEX (Belgium), CEA/CNES (France), DLR (Germany), ASI/INAF (Italy), and CICYT/MCYT (Spain). The \textit{Herschel} spacecraft was designed, built, tested, and launched under a contract to ESA managed by the \textit{Herschel}/Planck Project team by an industrial consortium under the overall responsibility of the prime contractor Thales Alenia Space (Cannes), and including Astrium (Friedrichshafen) responsible for the payload module and for system testing at spacecraft level, Thales Alenia Space (Turin) responsible for the service module, and Astrium (Toulouse) responsible for the telescope, with in excess of a hundred subcontractors. HCSS / HSpot / HIPE is a joint development (are joint developments) by the \textit{Herschel} Science Ground Segment Consortium, consisting of ESA, the NASA \textit{Herschel} Science Center, and the HIFI, PACS and SPIRE consortia. L. Fletcher was supported by STFC Consolidated Grant reference ST/W00089X/1. For the purpose of open access, the author has applied a Creative Commons Attribution (CC BY) licence to the Author Accepted Manuscript version arising from this submission. A portion of this work used the ALICE high performance computing facility at the University of Leicester. The authors thank J. Moses for her in-depth review of this paper and for providing the water profiles from her 2015 paper.


\begin{appendix}

\onecolumn
\section{All \textit{Cassini}/CIRS thermal fields used}

\begin{figure*}[!h]
\centering
\includegraphics[width=0.99\textwidth]{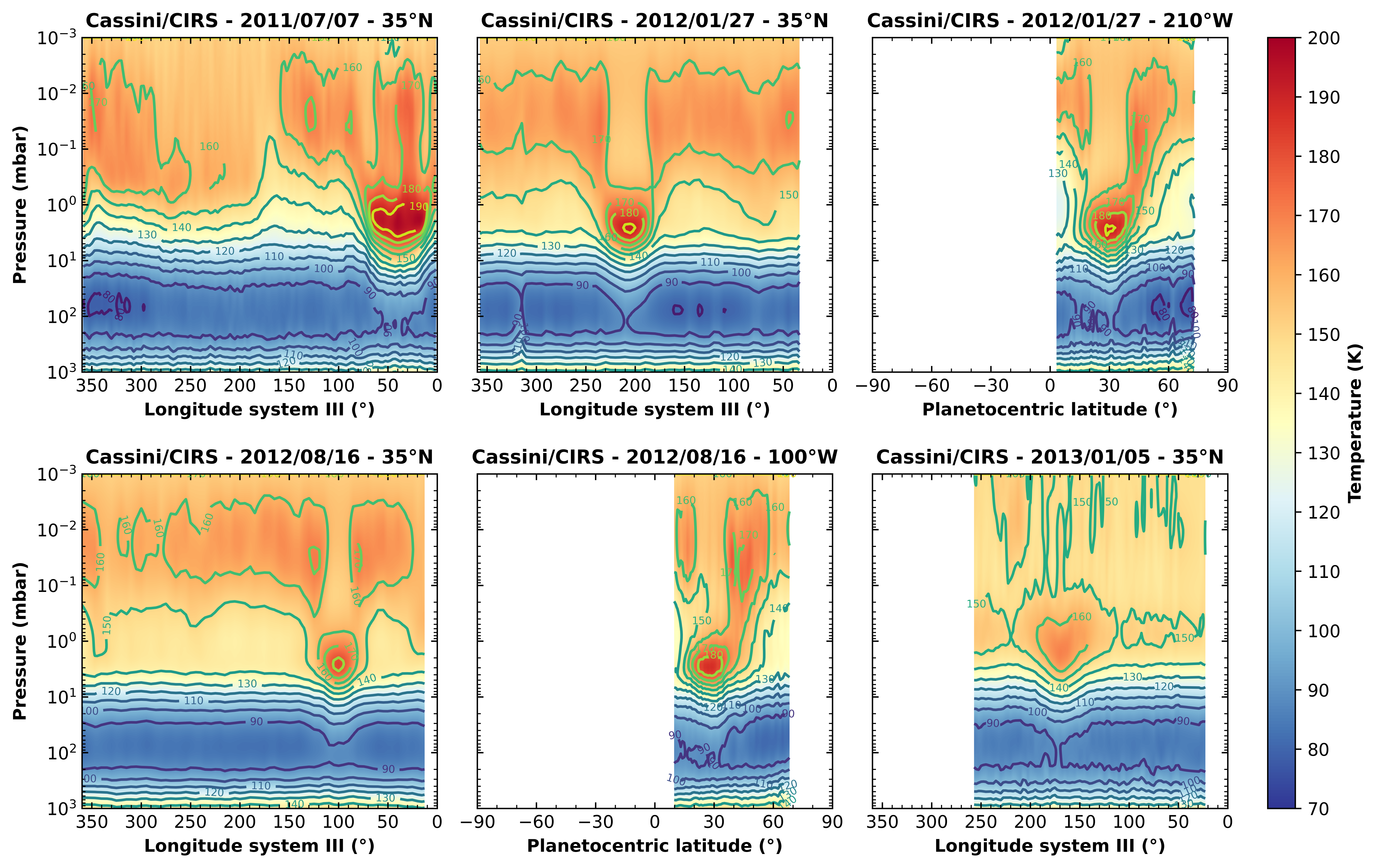}
\caption{Thermal fields retrieved from \textit{Cassini}/CIRS data \citep{Fletcher2012} and used to build the thermal fields for radiative transfer modelling. The fields shown here only represent the beacon region and its surroundings. They were derived from observations taken within our four \textit{Herschel}/PACS windows, except that they do not correspond to the same date of observation. The field dates are July 7 2011, January 27 2012, August 16 2012, and January 5 2013. Details on how these fields were retrieved can be found in Section \ref{part:Temperature_beacon}. The first and last dates have longitudinal coverage, and the two intermediate ones have also latitudinal coverage.}
\label{fig:Thermal_fields_profiles_CIRS_ALL}
\end{figure*}

The \textit{Cassini}/CIRS thermal 3D-fields used in this paper, at the beacon's central longitude and latitude (only for the two windows of 2012) are illustrated in Fig. \ref{fig:Thermal_fields_profiles_CIRS_ALL}.

\newpage
\section{Beacon position and extension from \textit{Cassini}/CIRS fields}

\begin{figure*}[!h]
\centering
\includegraphics[width=0.99\textwidth]{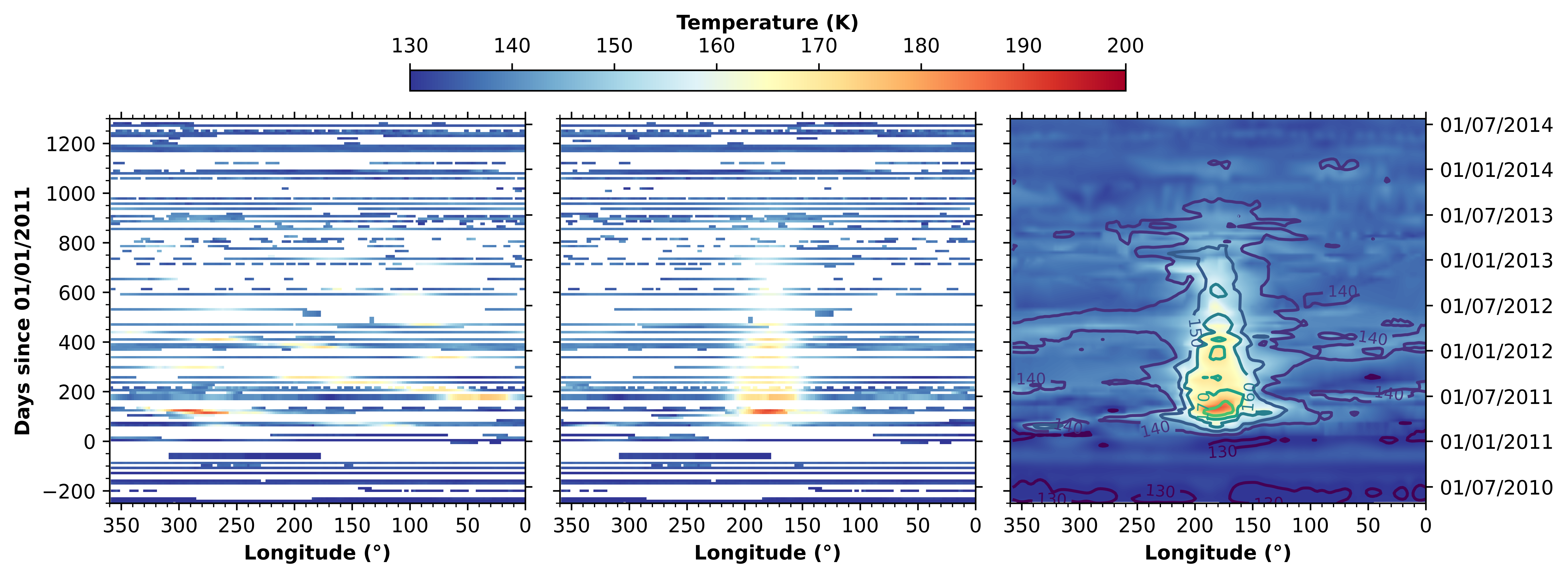}
\caption{\textit{Left:} Thermal fields retrieved from CH$_4$ emission from \textit{Cassini}/CIRS, averaged between 1280-1320 cm$^{-1}$, sensitive to the millibar levels in the stratosphere, as a function of longitude and time. The temperatures correspond to an average between 25-45\degre N in latitude. Data before 2012 are presented in \citet{Fletcher2012}, and data after the beginning of 2012 have not been published yet. \textit{Middle:} Same as the left figure, but after realigning the beacon centre at 180\degre W in longitude. The performed alignment results in the data points of Fig. \ref{fig:Beacon_position}. The interpolation and smoothing of this figure is illustrated in the \textit{right} panel to better see the extension of the beacon over time.}
\label{fig:Donnees_beacon_position_extension}
\end{figure*}

The \textit{Cassini}/CIRS thermal fields used to retrieve the beacon drift rate (Section \ref{part:Beacon_drift}) are presented in Fig. \ref{fig:Donnees_beacon_position_extension}. 

\newpage
\section{Step 1: The effect of temperature increase}

\begin{figure*}[!h]
\centering
\includegraphics[width=0.99\textwidth]{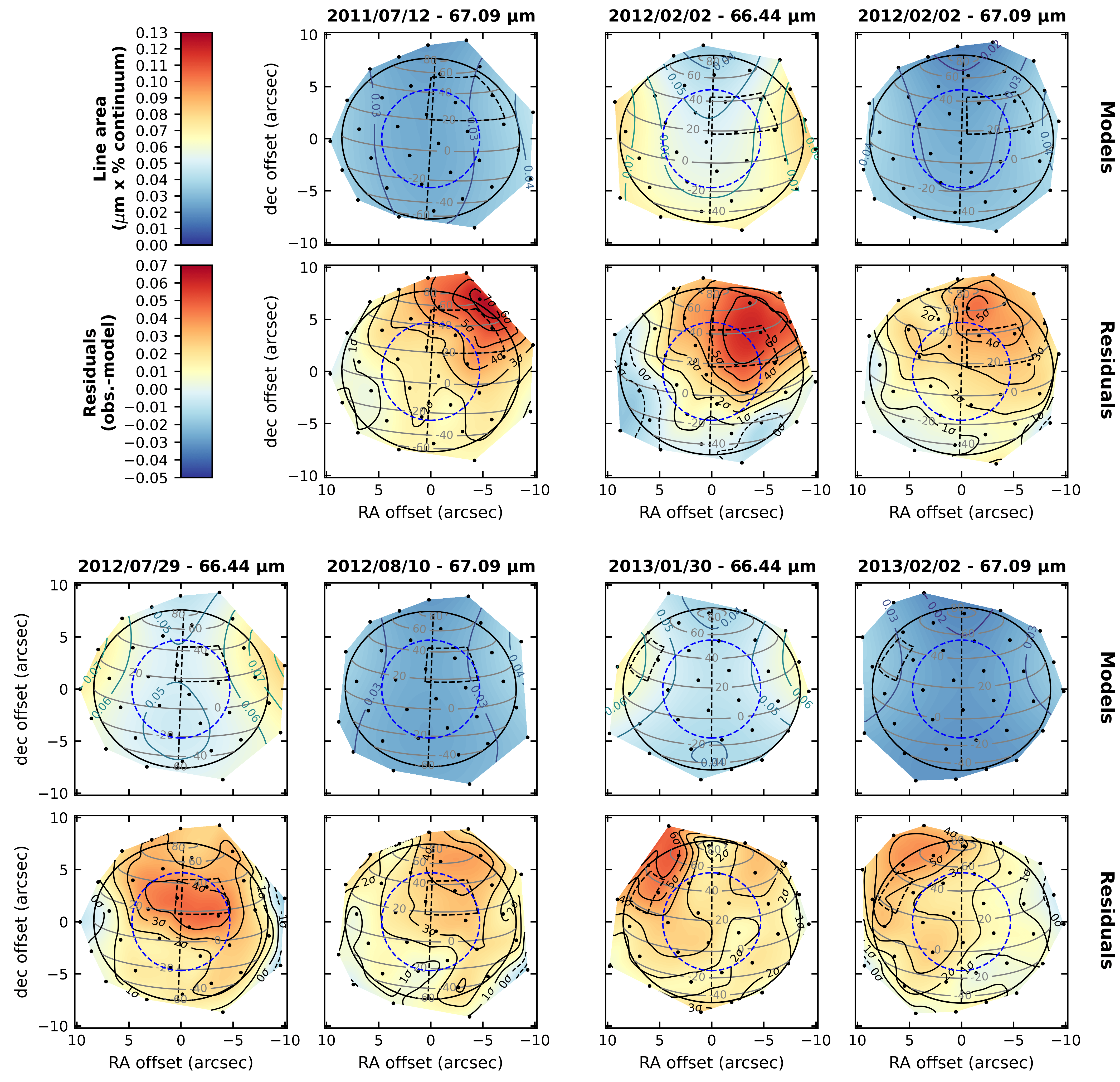}
\caption{Modelled H$_2$O water line area maps and residuals obtained within Step 1 (nominal water field of \citealt{Cavalie2019}), for the seven observation windows. The modelled line area maps are displayed in rows 1 and 3. The units are expressed as $\mu$m$\times$\% of continuum to directly compare to the observation maps of Fig. \ref{fig:Line_area_maps}. The corresponding residual maps are presented in rows 2 and 4 and correspond to the difference between the observed maps of Fig. \ref{fig:Line_area_maps} and the above modelled maps. The contours are given in units of $\sigma$. The solid contours refer to positive residuals and the dashed contours indicate negative residuals. The colour scales of the modelled and residual maps are shown on the top left corner. The overall description of the maps is the same as in Fig. \ref{fig:Line_area_maps}.}
\label{fig:Simu_T}
\end{figure*}

The results obtained with the water model of \citet{Cavalie2019} are presented in Fig. \ref{fig:Simu_T} for the seven observation maps.

\newpage
\section{Step 3: Fitting the water column density in the beacon}

\begin{figure*}[!h]
\centering
\includegraphics[width=0.99\textwidth]{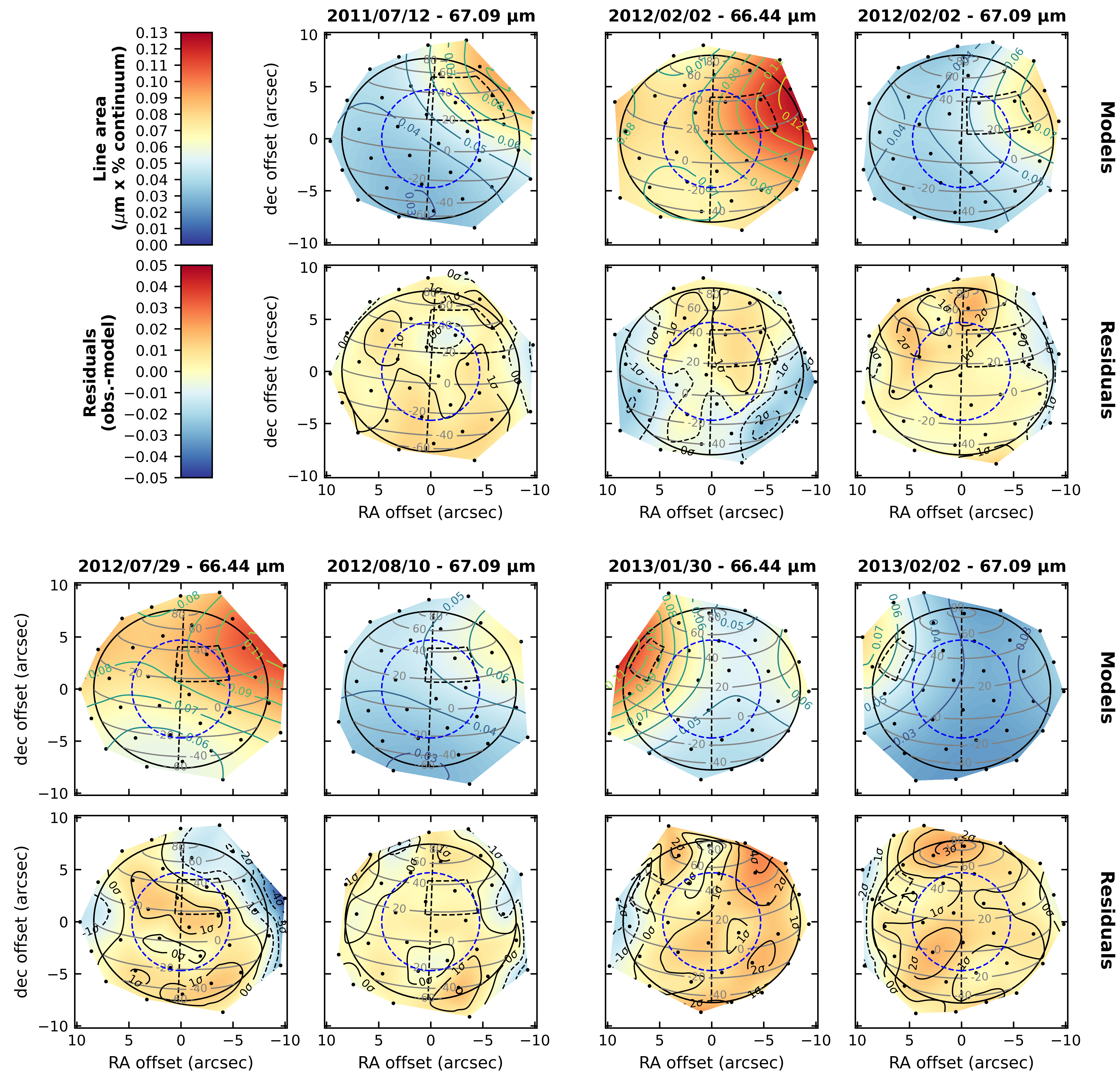}
\caption{Same as \ref{fig:Simu_T} but for Step 3 (water mole fraction fitting in the beacon).}
\label{fig:Simu_best_fits}
\end{figure*}

The results obtained with a model in which we fit the water abundance in the beacon as a constant free parameter above the condensation level, are shown in Fig. \ref{fig:Simu_best_fits} for the seven observation maps.

\end{appendix}


\begin{thebibliography}{55}

\bibitem[{{Barth}(2020)}]{Barth2020}
{Barth}, E. 2020, Atmosphere, 11, 1064

\bibitem[{{Benmahi} {et~al.}(2022){Benmahi}, {Cavali{\'e}}, {Fouchet},
  {Moreno}, {Lellouch}, {Bardet}, {Guerlet}, {Hue}, \& {Spiga}}]{Benmahi2022}
{Benmahi}, B., {Cavali{\'e}}, T., {Fouchet}, T., {et~al.} 2022, \aap, 666, A117

\bibitem[{{B{\'e}zard} {et~al.}(2002){B{\'e}zard}, {Lellouch}, {Strobel},
  {Maillard}, \& {Drossart}}]{Bezard2002}
{B{\'e}zard}, B., {Lellouch}, E., {Strobel}, D., {Maillard}, J.-P., \&
  {Drossart}, P. 2002, \icarus, 159, 95

\bibitem[{{Burgdorf} {et~al.}(2006){Burgdorf}, {Orton}, {van Cleve}, {Meadows},
  \& {Houck}}]{Burgdorf2006}
{Burgdorf}, M., {Orton}, G., {van Cleve}, J., {Meadows}, V., \& {Houck}, J.
  2006, \icarus, 184, 634

\bibitem[{{Cassidy} \& {Johnson}(2010)}]{Cassidy2010}
{Cassidy}, T.~A. \& {Johnson}, R.~E. 2010, \icarus, 209, 696

\bibitem[{{Cavali{\'e}} {et~al.}(2009){Cavali{\'e}}, {Billebaud}, {Dobrijevic},
  {Fouchet}, {Lellouch}, {Encrenaz}, {Brillet}, {Moriarty-Schieven},
  {Wouterloot}, \& {Hartogh}}]{Cavalie2009}
{Cavali{\'e}}, T., {Billebaud}, F., {Dobrijevic}, M., {et~al.} 2009, \icarus,
  203, 531
  
\bibitem[{{Cavali{\'e}} {et~al.}(2010){Cavali{\'e}}, {Hartogh}, {Billebaud},
  {Dobrijevic}, {Fouchet}, {Lellouch}, {Encrenaz}, {Brillet}, \&
  {Moriarty-Schieven}}]{Cavalie2010}
{Cavali{\'e}}, T., {Hartogh}, P., {Billebaud}, F., {et~al.} 2010, \aap, 510,
  A88
  
\bibitem[{{Cavali{\'e}} {et~al.}(2013){Cavali{\'e}}, {Feuchtgruber},
  {Lellouch}, {de Val-Borro}, {Jarchow}, {Moreno}, {Hartogh}, {Orton},
  {Greathouse}, {Billebaud}, {Dobrijevic}, {Lara}, {Gonz{\'a}lez}, \&
  {Sagawa}}]{Cavalie2013}
{Cavali{\'e}}, T., {Feuchtgruber}, H., {Lellouch}, E., {et~al.} 2013, \aap,
  553, A21  
  
\bibitem[{{Cavali{\'e}} {et~al.}(2014){Cavali{\'e}}, {Moreno}, {Lellouch},
  {Hartogh}, {Venot}, {Orton}, {Jarchow}, {Encrenaz}, {Selsis}, {Hersant}, \&
  {Fletcher}}]{Cavalie2014}
{Cavali{\'e}}, T., {Moreno}, R., {Lellouch}, E., {et~al.} 2014, \aap, 562, A33
  
\bibitem[{{Cavali{\'e}} {et~al.}(2015){Cavali{\'e}}, {Dobrijevic}, {Fletcher},
  {Loison}, {Hickson}, {Hue}, \& {Hartogh}}]{Cavalie2015}
{Cavali{\'e}}, T., {Dobrijevic}, M., {Fletcher}, L.~N., {et~al.} 2015, \aap,
  580, A55
  
\bibitem[{{Cavali{\'e}} {et~al.}(2019){Cavali{\'e}}, {Hue}, {Hartogh},
  {Moreno}, {Lellouch}, {Feuchtgruber}, {Jarchow}, {Cassidy}, {Fletcher},
  {Billebaud}, {Dobrijevic}, {Rezac}, {Orton}, {Rengel}, {Fouchet}, \&
  {Guerlet}}]{Cavalie2019}
{Cavali{\'e}}, T., {Hue}, V., {Hartogh}, P., {et~al.} 2019, \aap, 630, A87

\bibitem[{{Cavali{\'e}} {et~al.}(2024{\natexlab{a}}){Cavali{\'e}}, {Benmahi},
  {Lefour}, {Moreno}, {Fouchet}, {Lellouch}, {Gurwell}, {Ducreux}, {Gueth}, \&
  {Fletcher}}]{Cavalie2024b}
{Cavali{\'e}}, T., {Benmahi}, B., {Lefour}, C., {et~al.} 2024{\natexlab{a}}, in
  AAS/Division for Planetary Sciences Meeting Abstracts, Vol.~56, AAS/Division
  for Planetary Sciences Meeting Abstracts, 210.06

\bibitem[{{Cavali{\'e}} {et~al.}(2024{\natexlab{b}}){Cavali{\'e}}, {Lunine},
  {Mousis}, \& {Hueso}}]{Cavalie2024a}
{Cavali{\'e}}, T., {Lunine}, J., {Mousis}, O., \& {Hueso}, R.
  2024{\natexlab{b}}, \ssr, 220, 8

\bibitem[{{Coustenis} {et~al.}(1998){Coustenis}, {Salama}, {Lellouch},
  {Encrenaz}, {Bjoraker}, {Samuelson}, {de Graauw}, {Feuchtgruber}, \&
  {Kessler}}]{Coustenis1998}
{Coustenis}, A., {Salama}, A., {Lellouch}, E., {et~al.} 1998, \aap, 336, L85

\bibitem[{{Dobrijevic} {et~al.}(2014){Dobrijevic}, {H{\'e}brard}, {Loison}, \&
  {Hickson}}]{Dobrijevic2014}
{Dobrijevic}, M., {H{\'e}brard}, E., {Loison}, J.~C., \& {Hickson}, K.~M. 2014,
  \icarus, 228, 324

\bibitem[{{Feuchtgruber} {et~al.}(1997){Feuchtgruber}, {Lellouch}, {de Graauw},
  {B{\'e}zard}, {Encrenaz}, \& {Griffin}}]{Feuchtgruber1997}
{Feuchtgruber}, H., {Lellouch}, E., {de Graauw}, T., {et~al.} 1997, \nat, 389,
  159

\bibitem[{{Fischer} {et~al.}(2011){Fischer}, {Kurth}, {Gurnett}, {Zarka},
  {Dyudina}, {Ingersoll}, {Ewald}, {Porco}, {Wesley}, {Go}, \&
  {Delcroix}}]{Fischer2011}
{Fischer}, G., {Kurth}, W.~S., {Gurnett}, D.~A., {et~al.} 2011, \nat, 475, 75


\bibitem[{{Fletcher} {et~al.}(2011){Fletcher}, {Hesman}, {Irwin}, {Baines},
  {Momary}, {Sanchez-Lavega}, {Flasar}, {Read}, {Orton}, {Simon-Miller},
  {Hueso}, {Bjoraker}, {Mamoutkine}, {del Rio-Gaztelurrutia}, {Gomez},
  {Buratti}, {Clark}, {Nicholson}, \& {Sotin}}]{Fletcher2011}
{Fletcher}, L.~N., {Hesman}, B.~E., {Irwin}, P.~G.~J., {et~al.} 2011, \science,
  332, 1413
  
\bibitem[{{Fletcher} {et~al.}(2012){Fletcher}, {Swinyard}, {Salji},
  {Polehampton}, {Fulton}, {Sidher}, {Lellouch}, {Moreno}, {Orton},
  {Cavali{\'e}}, {Courtin}, {Rengel}, {Sagawa}, {Davis}, {Hartogh}, {Naylor},
  {Walker}, \& {Lim}}]{Fletcher2012}
{Fletcher}, L.~N., {Swinyard}, B., {Salji}, C., {et~al.} 2012, \aap, 539, A44

\bibitem[{{Fletcher} {et~al.}(2017){Fletcher}, {Guerlet}, {Orton}, {Cosentino},
  {Fouchet}, {Irwin}, {Li}, {Flasar}, {Gorius}, \&
  {Morales-Juber{\'{\i}}as}}]{Fletcher2017}
{Fletcher}, L.~N., {Guerlet}, S., {Orton}, G.~S., {et~al.} 2017, \natastron, 1,
  765

\bibitem[{{Fletcher} {et~al.}(2018){Fletcher}, {Orton}, {Sinclair}, {Guerlet},
  {Read}, {Antu{\~n}ano}, {Achterberg}, {Flasar}, {Irwin}, {Bjoraker},
  {Hurley}, {Hesman}, {Segura}, {Gorius}, {Mamoutkine}, \&
  {Calcutt}}]{Fletcher2018b}
{Fletcher}, L.~N., {Orton}, G.~S., {Sinclair}, J.~A., {et~al.} 2018,
  \natcommun, 9, 3564

\bibitem[{{Fletcher} {et~al.}(2023){Fletcher}, {King}, {Harkett}, {Hammel},
  {Roman}, {Melin}, {Hedman}, {Moses}, {Guerlet}, {Milam}, \&
  {Tiscareno}}]{Fletcher2023b}
{Fletcher}, L.~N., {King}, O. R.~T., {Harkett}, J., {et~al.} 2023, Journal of
  Geophysical Research (Planets), 128, e2023JE007924

\bibitem[{{Fouchet} {et~al.}(2016){Fouchet}, {Greathouse}, {Spiga}, {Fletcher},
  {Guerlet}, {Leconte}, \& {Orton}}]{Fouchet2016}
{Fouchet}, T., {Greathouse}, T.~K., {Spiga}, A., {et~al.} 2016, \icarus, 277,
  196

\bibitem[{{Fray} \& {Schmitt}(2009)}]{Fray2009}
{Fray}, N. \& {Schmitt}, B. 2009, \planss, 57, 2053

\bibitem[{{Guerlet} {et~al.}(2015){Guerlet}, {Fouchet}, {Vinatier}, {Simon},
  {Dartois}, \& {Spiga}}]{Guerlet2015}
{Guerlet}, S., {Fouchet}, T., {Vinatier}, S., {et~al.} 2015, \aap, 580, A89

\bibitem[{{Hansen} {et~al.}(2006){Hansen}, {Esposito}, {Stewart}, {Colwell},
  {Hendrix}, {Pryor}, {Shemansky}, \& {West}}]{Hansen2006}
{Hansen}, C.~J., {Esposito}, L., {Stewart}, A.~I.~F., {et~al.} 2006, \science,
  311, 1422

\bibitem[{{Hartogh} {et~al.}(2009){Hartogh}, {Lellouch}, {Crovisier},
  {Banaszkiewicz}, {Bensch}, {Bergin}, {Billebaud}, {Biver}, {Blake}, {Blecka},
  {Blommaert}, {Bockel{\'e}e-Morvan}, {Cavali{\'e}}, {Cernicharo}, {Courtin},
  {Davis}, {Decin}, {Encrenaz}, {Encrenaz}, {Gonz{\'a}lez}, {de Graauw},
  {Hutsem{\'e}kers}, {Jarchow}, {Jehin}, {Kidger}, {K{\"u}ppers}, {de Lange},
  {Lara}, {Lis}, {Lorente}, {Manfroid}, {Medvedev}, {Moreno}, {Naylor},
  {Orton}, {Portyankina}, {Rengel}, {Sagawa}, {S{\'a}nchez-Portal}, {Schieder},
  {Sidher}, {Stam}, {Swinyard}, {Szutowicz}, {Thomas}, {Thornhill},
  {Vandenbussche}, {Verdugo}, {Waelkens}, \& {Walker}}]{Hartogh2009}
{Hartogh}, P., {Lellouch}, E., {Crovisier}, J., {et~al.} 2009, \planss, 57,
  1596

\bibitem[{{Hartogh} {et~al.}(2011){Hartogh}, {Lellouch}, {Moreno},
  {Bockel{\'e}e-Morvan}, {Biver}, {Cassidy}, {Rengel}, {Jarchow},
  {Cavali{\'e}}, {Crovisier}, {Helmich}, \& {Kidger}}]{Hartogh2011a}
{Hartogh}, P., {Lellouch}, E., {Moreno}, R., {et~al.} 2011, \aap, 532, L2

\bibitem[{{Hesman} {et~al.}(2012){Hesman}, {Bjoraker}, {Sada}, {Achterberg},
  {Jennings}, {Romani}, {Lunsford}, {Fletcher}, {Boyle}, {Simon-Miller},
  {Nixon}, \& {Irwin}}]{Hesman2012}
{Hesman}, B.~E., {Bjoraker}, G.~L., {Sada}, P.~V., {et~al.} 2012, \apj, 760, 24

\bibitem[{{Hsu} {et~al.}(2018){Hsu}, {Schmidt}, {Kempf}, {Postberg},
  {Moragas-Klostermeyer}, {Sei{\ss}}, {Hoffmann}, {Burton}, {Ye}, {Kurth},
  {Hor{\'a}nyi}, {Khawaja}, {Spahn}, {Schirdewahn}, {O'Donoghue}, {Moore},
  {Cuzzi}, {Jones}, \& {Srama}}]{Hsu2018}
{Hsu}, H.-W., {Schmidt}, J., {Kempf}, S., {et~al.} 2018, \science, 362, aat3185

\bibitem[{{Lara} {et~al.}(2014){Lara}, {Lellouch}, {Gonz{\'a}lez}, {Moreno}, \&
  {Rengel}}]{Lara2014}
{Lara}, L.~M., {Lellouch}, E., {Gonz{\'a}lez}, M., {Moreno}, R., \& {Rengel},
  M. 2014, \aap, 566, A143

\bibitem[{{Lellouch} {et~al.}(1995){Lellouch}, {Paubert}, {Moreno}, {Festou},
  {Bezard}, {Bockelee-Morvan}, {Colom}, {Crovisier}, {Encrenaz}, {Gautier},
  {Marten}, {Despois}, {Strobel}, \& {Sievers}}]{Lellouch1995}
{Lellouch}, E., {Paubert}, G., {Moreno}, R., {et~al.} 1995, \nat, 373, 592

\bibitem[{{Lellouch} {et~al.}(2002){Lellouch}, {B{\'e}zard}, {Moses}, {Davis},
  {Drossart}, {Feuchtgruber}, {Bergin}, {Moreno}, \& {Encrenaz}}]{Lellouch2002}
{Lellouch}, E., {B{\'e}zard}, B., {Moses}, J.~I., {et~al.} 2002, \icarus, 159,
  112

\bibitem[{{Lellouch} {et~al.}(2005){Lellouch}, {Moreno}, \&
  {Paubert}}]{Lellouch2005}
{Lellouch}, E., {Moreno}, R., \& {Paubert}, G. 2005, \aap, 430, L37

\bibitem[{{Li} {et~al.}(2022){Li}, {Allison}, {Atreya}, {Fletcher}, {Galanti},
  {Guillot}, {Ingersoll}, {Kaspi}, {Li}, {Lunine}, {Orton}, {Oyafuso},
  {Steffes}, {Wong}, {Zhang}, {Levin}, \& {Bolton}}]{Li2022}
{Li}, C., {Allison}, M.~D., {Atreya}, S.~K., {et~al.} 2022, in AGU Fall Meeting
  Abstracts, Vol. 2022, P32C--1854

\bibitem[{{Mitchell} {et~al.}(2018){Mitchell}, {Perry}, {Hamilton}, {Westlake},
  {Kollmann}, {Smith}, {Carbary}, {Waite}, {Perryman}, {Hsu}, {Wahlund},
  {Morooka}, {Hadid}, {Persoon}, \& {Kurth}}]{Mitchell2018}
{Mitchell}, D.~G., {Perry}, M.~E., {Hamilton}, D.~C., {et~al.} 2018, \science,
  362, aat2236

\bibitem[{{Moreno} {et~al.}(2012){Moreno}, {Lellouch}, {Lara}, {Feuchtgruber},
  {Rengel}, {Hartogh}, \& {Courtin}}]{Moreno2012}
{Moreno}, R., {Lellouch}, E., {Lara}, L.~M., {et~al.} 2012, \icarus, 221, 753

\bibitem[{{Moreno} {et~al.}(2017){Moreno}, {Lellouch}, {Cavali{\'e}}, \&
  {Moullet}}]{Moreno2017}
{Moreno}, R., {Lellouch}, E., {Cavali{\'e}}, T., \& {Moullet}, A. 2017, \aap,
  608, L5

\bibitem[{{Moses} {et~al.}(2000){Moses}, {Lellouch}, {B{\'e}zard}, {Gladstone},
  {Feuchtgruber}, \& {Allen}}]{Moses2000b}
{Moses}, J.~I., {Lellouch}, E., {B{\'e}zard}, B., {et~al.} 2000, \icarus, 145,
  166
  
\bibitem[{{Moses} {et~al.}(2005){Moses}, {Fouchet}, {B{\'e}zard}, {Gladstone},
  {Lellouch}, \& {Feuchtgruber}}]{Moses2005}
{Moses}, J.~I., {Fouchet}, T., {B{\'e}zard}, B., {et~al.} 2005, \jgr, 110,
  E08001

\bibitem[{{Moses} {et~al.}(2015){Moses}, {Armstrong}, {Fletcher}, {Friedson},
  {Irwin}, {Sinclair}, \& {Hesman}}]{Moses2015}
{Moses}, J.~I., {Armstrong}, E.~S., {Fletcher}, L.~N., {et~al.} 2015, \icarus,
  261, 149

\bibitem[{{Moses} \& {Poppe}(2017)}]{Moses2017}
{Moses}, J.~I. \& {Poppe}, A.~R. 2017, \icarus, 297, 33

\bibitem[{{Ollivier} {et~al.}(2000){Ollivier}, {Dobrij{\'e}vic}, \&
  {Parisot}}]{Ollivier2000}
{Ollivier}, J.~L., {Dobrij{\'e}vic}, M., \& {Parisot}, J.~P. 2000, \planss, 48,
  699

\bibitem[{{Ott}(2010)}]{Ott2010}
{Ott}, S. 2010, in \pasp, Vol. 434, Astronomical Data Analysis Software and
  Systems XIX, ed. Y.~{Mizumoto}, K.-I. {Morita}, \& M.~{Ohishi}, 139

\bibitem[{{Pilbratt} {et~al.}(2010){Pilbratt}, {Riedinger}, {Passvogel},
  {Crone}, {Doyle}, {Gageur}, {Heras}, {Jewell}, {Metcalfe}, {Ott}, \&
  {Schmidt}}]{Pilbratt2010}
{Pilbratt}, G.~L., {Riedinger}, J.~R., {Passvogel}, T., {et~al.} 2010, \aap,
  518, L1

\bibitem[{{Poglitsch} {et~al.}(2010){Poglitsch}, {Waelkens}, {Geis},
  {Feuchtgruber}, {Vandenbussche}, {Rodriguez}, {Krause}, {Renotte}, {van
  Hoof}, {Saraceno}, {Cepa}, {Kerschbaum}, {Agn{\`e}se}, {Ali}, {Altieri},
  {Andreani}, {Augueres}, {Balog}, {Barl}, {Bauer}, {Belbachir}, {Benedettini},
  {Billot}, {Boulade}, {Bischof}, {Blommaert}, {Callut}, {Cara}, {Cerulli},
  {Cesarsky}, {Contursi}, {Creten}, {De Meester}, {Doublier}, {Doumayrou},
  {Duband}, {Exter}, {Genzel}, {Gillis}, {Gr{\"o}zinger}, {Henning},
  {Herreros}, {Huygen}, {Inguscio}, {Jakob}, {Jamar}, {Jean}, {de Jong},
  {Katterloher}, {Kiss}, {Klaas}, {Lemke}, {Lutz}, {Madden}, {Marquet},
  {Martignac}, {Mazy}, {Merken}, {Montfort}, {Morbidelli}, {M{\"u}ller},
  {Nielbock}, {Okumura}, {Orfei}, {Ottensamer}, {Pezzuto}, {Popesso},
  {Putzeys}, {Regibo}, {Reveret}, {Royer}, {Sauvage}, {Schreiber}, {Stegmaier},
  {Schmitt}, {Schubert}, {Sturm}, {Thiel}, {Tofani}, {Vavrek}, {Wetzstein},
  {Wieprecht}, \& {Wiezorrek}}]{Poglitsch2010}
{Poglitsch}, A., {Waelkens}, C., {Geis}, N., {et~al.} 2010, \aap, 518, L2

\bibitem[{{Porco} {et~al.}(2006){Porco}, {Helfenstein}, {Thomas}, {Ingersoll},
  {Wisdom}, {West}, {Neukum}, {Denk}, {Wagner}, {Roatsch}, {Kieffer}, {Turtle},
  {McEwen}, {Johnson}, {Rathbun}, {Veverka}, {Wilson}, {Perry}, {Spitale},
  {Brahic}, {Burns}, {Del Genio}, {Dones}, {Murray}, \& {Squyres}}]{Porco2006}
{Porco}, C.~C., {Helfenstein}, P., {Thomas}, P.~C., {et~al.} 2006, \science,
  311, 1393

\bibitem[{{Prather} {et~al.}(1978){Prather}, {Logan}, \&
  {McElroy}}]{Prather1978}
{Prather}, M.~J., {Logan}, J.~A., \& {McElroy}, M.~B. 1978, \apj, 223, 1072

\bibitem[{{S{\'a}nchez-Lavega} {et~al.}(2011){S{\'a}nchez-Lavega}, {del
  R{\'{\i}}o-Gaztelurrutia}, {Hueso}, {G{\'o}mez-Forrellad}, {Sanz-Requena},
  {Legarreta}, {Garc{\'{\i}}a-Melendo}, {Colas}, {Lecacheux}, {Fletcher},
  {Barrado y Navascu{\'e}s}, {Parker}, {International Outer Planet Watch Team},
  {Akutsu}, {Barry}, {Beltran}, {Buda}, {Combs}, {Carvalho}, {Casquinha},
  {Delcroix}, {Ghomizadeh}, {Go}, {Hotershall}, {Ikemura}, {Jolly}, {Kazemoto},
  {Kumamori}, {Lecompte}, {Maxson}, {Melillo}, {Milika}, {Morales}, {Peach},
  {Phillips}, {Poupeau}, {Sussenbach}, {Walker}, {Walker}, {Tranter}, {Wesley},
  {Wilson}, \& {Yunoki}}]{Sanchez-Lavega2011}
{S{\'a}nchez-Lavega}, A., {del R{\'{\i}}o-Gaztelurrutia}, T., {Hueso}, R.,
  {et~al.} 2011, \nat, 475, 71

\bibitem[{{S{\'a}nchez-Lavega} {et~al.}(2018){S{\'a}nchez-Lavega}, {Fischer},
  {Fletcher}, {Garcia-Melendo}, {Hesman}, {Perez-Hoyos}, {Sayanagi}, \&
  {Sromovsky}}]{Sanchez-Lavega2018}
{S{\'a}nchez-Lavega}, A., {Fischer}, G., {Fletcher}, L.~N., {et~al.} 2018, in
  Saturn in the 21st Century, ed. K.~H. {Baines}, F.~M. {Flasar}, N.~{Krupp},
  \& T.~{Stallard}, 377--416

\bibitem[{{Strobel} \& {Yung}(1979)}]{Strobel1979}
{Strobel}, D.~F. \& {Yung}, Y.~L. 1979, \icarus, 37, 256

\bibitem[{{Teanby} {et~al.}(2022){Teanby}, {Irwin}, {Sylvestre}, {Nixon}, \&
  {Cordiner}}]{Teanby2022}
{Teanby}, N.~A., {Irwin}, P.~G.~J., {Sylvestre}, M., {Nixon}, C.~A., \&
  {Cordiner}, M.~A. 2022, \psj, 3, 96

\bibitem[{{Venot} {et~al.}(2020){Venot}, {Cavali{\'e}}, {Bounaceur},
  {Tremblin}, {Brouillard}, \& {Lhoussaine Ben Brahim}}]{Venot2020}
{Venot}, O., {Cavali{\'e}}, T., {Bounaceur}, R., {et~al.} 2020, \aap, 634, A78

\bibitem[{{Waite} {et~al.}(2006){Waite}, {Combi}, {Ip}, {Cravens}, {McNutt},
  {Kasprzak}, {Yelle}, {Luhmann}, {Niemann}, {Gell}, {Magee}, {Fletcher},
  {Lunine}, \& {Tseng}}]{Waite2006}
{Waite}, J.~H., {Combi}, M.~R., {Ip}, W.-H., {et~al.} 2006, \science, 311, 1419

\bibitem[{{Waite} {et~al.}(2018){Waite}, {Perryman}, {Perry}, {Miller}, {Bell},
  {Glein}, {Grimes}, {Hedman}, {Cuzzi}, {Brockwell}, {Teolis}, {Moore},
  {Mitchell}, {Persoon}, {Kurth}, {Wahlund}, {Morooka}, {Hadid}, {Walker},
  {Nagy}, {Yelle}, {Ledvina}, {Johnson}, {Tseng}, {Tucker}, \&
  {Ip}}]{Waite2018}
{Waite}, J.~H., {Perryman}, R.~S., {Perry}, M.~E., {et~al.} 2018, \science,
  362, 51

\end{thebibliography}
\end{document}